\newcommand\DLB{{\cal D}_{\text{LB}}}
\newcommand\DNN{{\cal D}_{\text{NN}}}
\newcommand\Figref[1]{Figure~\ref{#1}}
\newcommand\Figsref[1]{Figures~\ref{#1}}
\newcommand\LQCD{\Lambda_{\text{QCD}}}
\newcommand\Secref[1]{Section~\ref{#1}}
\newcommand\Tabref[1]{Table~\ref{#1}}
\newcommand\aplan{{\cal{A}}}        %
\newcommand\bbar{\overline{b}}               %
\newcommand\brocket[1]{\left\langle #1 \right\rangle}
\newcommand\calD{{\cal D}}
\newcommand\chisq{\chi^2}         %
\newcommand\deteta{\eta_{\text{det}}} %
\newcommand\dzero{D\O}                         %
\newcommand\eqref[1]{Eq.~(\ref{#1})}
\newcommand\eqsref[1]{Eqs.~(\ref{#1})}
\newcommand\et{E_T}               %
\newcommand\etjet{\et^{\text{jet}}}
\newcommand\figref[1]{Fig.~\ref{#1}}
\newcommand\figsref[1]{Figs.~\ref{#1}}
\newcommand\gev{\ \text{GeV}}
\newcommand\gevc{\ \text{GeV} / c}
\newcommand\gevcc{\ \text{GeV} / c^2}
\newcommand\htran{H_T}            %
\newcommand\ipb{\ \text{pb}^{-1}}                     %
\newcommand\jet{\text{jet}}
\newcommand\jets{\text{jets}}
\newcommand\kt{k_T}               %
\def\lnlmin{-\ln L_{\text{min}}}
\newcommand\mG{{\bf G}}
\newcommand\mcorr{m_{\text{corr}}}
\newcommand\met{\mbox{${\hbox{$E$\kern-0.6em\lower-.1ex\hbox{/}}}_T$}}%
\newcommand\metcal{\mbox{${\hbox{$E$\kern-0.6em\lower-.1ex\hbox{/}}}_T^{\text{cal}}$}}%
\newcommand\mfit{m_{\text{fit}}}
\newcommand\mlb{m_{\text{LB}}}
\newcommand\mmin{m_{\text{min}}}
\newcommand\mnn{m_{\text{NN}}}
\newcommand\mt{m_t}                          %
\newcommand\mx{{\bf x}}
\def\mynocite#1{\relax}  %
\newcommand\pbar{\overline{p}}               %
\newcommand\progname[1]{{\sc\lowercase{#1}}}
\newcommand\pt{p_T}               %
\newcommand\qbar{\overline{q}}               %
\newcommand\secref[1]{Sec.~\ref{#1}}
\newcommand\slb{\sigma_{\text{LB}}}
\newcommand\smallbrocket[1]{\langle #1 \rangle}
\newcommand\snn{\sigma_{\text{NN}}}
\newcommand\tabref[1]{Table~\ref{#1}}
\newcommand\tabsref[1]{Tables~\ref{#1}}
\newcommand\tbar{\overline{t}}               %
\newcommand\tev{\ \text{TeV}}
\newcommand\ttbar{t\overline{t}}             %
\newcommand\ugev{\text{GeV}}
\newcommand\ugevc{\text{GeV} / c}
\newcommand\ugevcc{\text{GeV} / c^2}
\def\urlbreak#1{#1\discretionary{}{}{}}
\newcommand\vecmet{\vec{\met}} %


\makeatletter
\def \@addtotoporbot {%
\@bitor\@currtype\@dbldeferlist\if@test\else
   \@getfpsbit \tw@
   \ifodd \@tempcnta
     \@flsetnum \@topnum
     \ifnum \@topnum>\z@
       \@tempswafalse
       \@flcheckspace \@toproom \@toplist
       \if@tempswa
         \@bitor\@currtype{\@midlist\@botlist}%
         \if@test
         \else
          \@flupdates \@topnum \@toproom \@toplist
          \@inserttrue
         \fi
       \fi
     \fi
   \fi
   \if@insert
   \else
     \@addtobot
   \fi
\fi
}
\makeatother
\newif\ifmypreprint
\mypreprintfalse
\ifmypreprint
\documentstyle[prd,eqsecnum,aps,verbatim,floats,preprint]{revtex}
\else
\documentstyle[prd,eqsecnum,aps,verbatim,floats]{revtex}
\fi

\input psfig

\begin{document}
\draft

\preprint{xxxpreprints}

\title {Direct Measurement of the Top Quark Mass at \dzero}
\author{                                                                      
B.~Abbott,$^{30}$                                                             
M.~Abolins,$^{27}$                                                            
B.S.~Acharya,$^{45}$                                                          
I.~Adam,$^{12}$                                                               
D.L.~Adams,$^{39}$                                                            
M.~Adams,$^{17}$                                                              
S.~Ahn,$^{14}$                                                                
H.~Aihara,$^{23}$                                                             
G.A.~Alves,$^{10}$                                                            
N.~Amos,$^{26}$
% ...
\\                                                                            
\vskip 0.50cm                                                                 
\centerline{(D\O\ Collaboration)}                                             
\vskip 0.50cm                                                                 
}                                                                             
\address{                                                                     
\centerline{$^{1}$Universidad de los Andes, Bogot\'{a}, Colombia}             
\centerline{$^{2}$University of Arizona, Tucson, Arizona 85721}               
\centerline{$^{3}$Boston University, Boston, Massachusetts 02215}             
\centerline{$^{4}$Brookhaven National Laboratory, Upton, New York 11973}      
\centerline{$^{5}$Brown University, Providence, Rhode Island 02912}           
\centerline{$^{6}$Universidad de Buenos Aires, Buenos Aires, Argentina}       
\centerline{$^{7}$University of California, Davis, California 95616}          
\centerline{$^{8}$University of California, Irvine, California 92697}         
\centerline{$^{9}$University of California, Riverside, California 92521}      
\centerline{$^{10}$LAFEX, Centro Brasileiro de Pesquisas F{\'\i}sicas,        
                  Rio de Janeiro, Brazil}                                     
\centerline{$^{11}$CINVESTAV, Mexico City, Mexico}                            
\centerline{$^{12}$Columbia University, New York, New York 10027}             
\centerline{$^{13}$Delhi University, Delhi, India 110007}                     
\centerline{$^{14}$Fermi National Accelerator Laboratory, Batavia,            
                   Illinois 60510}                                            
\centerline{$^{15}$Florida State University, Tallahassee, Florida 32306}      
\centerline{$^{16}$University of Hawaii, Honolulu, Hawaii 96822}              
\centerline{$^{17}$University of Illinois at Chicago, Chicago,                
                   Illinois 60607}                                            
\centerline{$^{18}$Indiana University, Bloomington, Indiana 47405}            
\centerline{$^{19}$Iowa State University, Ames, Iowa 50011}                   
\centerline{$^{20}$Korea University, Seoul, Korea}                            
\centerline{$^{21}$Kyungsung University, Pusan, Korea}                        
\centerline{$^{22}$Institute of Nuclear Physics, Krak\'ow, Poland}            
\centerline{$^{23}$Lawrence Berkeley National Laboratory and University of    
                   California, Berkeley, California 94720}                    
\centerline{$^{24}$Louisiana Tech University, Ruston, Louisiana 71272}        
\centerline{$^{25}$University of Maryland, College Park, Maryland 20742}      
\centerline{$^{26}$University of Michigan, Ann Arbor, Michigan 48109}         
\centerline{$^{27}$Michigan State University, East Lansing, Michigan 48824}   
\centerline{$^{28}$Moscow State University, Moscow, Russia}                   
\centerline{$^{29}$University of Nebraska, Lincoln, Nebraska 68588}           
\centerline{$^{30}$New York University, New York, New York 10003}             
\centerline{$^{31}$Northeastern University, Boston, Massachusetts 02115}      
\centerline{$^{32}$Northern Illinois University, DeKalb, Illinois 60115}      
\centerline{$^{33}$Northwestern University, Evanston, Illinois 60208}         
\centerline{$^{34}$University of Notre Dame, Notre Dame, Indiana 46556}       
\centerline{$^{35}$University of Oklahoma, Norman, Oklahoma 73019}            
\centerline{$^{36}$University of Panjab, Chandigarh 16-00-14, India}          
\centerline{$^{37}$Institute for High Energy Physics, 142-284 Protvino,       
                   Russia}                                                    
\centerline{$^{38}$Purdue University, West Lafayette, Indiana 47907}          
\centerline{$^{39}$Rice University, Houston, Texas 77005}                     
\centerline{$^{40}$Universidade do Estado do Rio de Janeiro, Brazil}          
\centerline{$^{41}$University of Rochester, Rochester, New York 14627}        
\centerline{$^{42}$CEA, DAPNIA/Service de Physique des Particules,            
                   CE-SACLAY, Gif-sur-Yvette, France}                         
\centerline{$^{43}$Seoul National University, Seoul, Korea}                   
\centerline{$^{44}$State University of New York, Stony Brook,                 
                   New York 11794}                                            
\centerline{$^{45}$Tata Institute of Fundamental Research,                    
                   Colaba, Mumbai 400005, India}                              
\centerline{$^{46}$University of Texas, Arlington, Texas 76019}               
\centerline{$^{47}$Texas A\&M University, College Station, Texas 77843}       
}                                                                             
\date{\today}
\maketitle

\begin{abstract}
We determine the top quark mass $m_t$ using $t\tbar$ pairs produced
in the \dzero\ detector by $\sqrt{s} = 1.8\tev$ $p\pbar$ collisions
in a $125\ipb$ exposure at the Fermilab Tevatron.  We make a two
constraint fit to $m_t$ in $t\tbar \rightarrow b W^+\bbar W^-$ final
states with one $W$~boson decaying to $q\qbar$ and the other to $e\nu$ or
$\mu\nu$.  
Likelihood fits to the data yield
$m_t(l+\jets) = 173.3\pm 5.6\ \textrm{(stat)} \pm 5.5\ \textrm{(syst)}\gevcc$.
When this result is
combined with an analysis of events in which both $W$~bosons
decay into leptons,
we obtain
$m_t = 172.1\pm 5.2\ \textrm{(stat)} \pm 4.9\ \textrm{(syst)}\gevcc$.
An alternate analysis, using three constraint fits to fixed top quark
masses, gives
$m_t(l+\jets) = 176.0\pm 7.9\ \textrm{(stat)}\pm 4.8\ \textrm{(syst)}\gevcc$,
consistent with the above result.
Studies of kinematic distributions of the top quark
candidates are also presented.
\end{abstract}

\ifmypreprint\else
\twocolumn
\narrowtext
\fi

\newpage
\tableofcontents
\newpage

\section{Introduction}
\label{introduction}

The discovery of the top~quark by the
CDF~\cite{cdfdiscovery}
and
\dzero~\cite{d0discovery}
collaborations at the Fermilab Tevatron ended the search phase of top~quark
physics.  Since then, emphasis has shifted to determining
its properties --- especially its large mass (about
200 times that of a proton) and production cross section.  Reviews of
searches for and the initial observations of the top~quark are given
in Ref.~\cite{pdg}\mynocite{*franklin,*wimpenny}.
Details of the initial \dzero\ top~quark search
can be found in Ref.~\cite{d0topprd}.
This paper reports on the determination of the top~quark mass using all
the data
collected by the \dzero\ experiment during the 1992--1996 Tevatron runs.
This is more than
twice as much data as was available for the initial observation.
In addition,
improvements have been made in event selection, object reconstruction, and
mass analysis techniques.
The result is a reduction of the statistical and systematic errors
by nearly a factor of four.  A short paper giving results from this analysis
has been published~\cite{ljmassprl}.

The top~quark is one of the fundamental fermions
in the standard model of electroweak interactions and
is the weak-isospin partner of the bottom quark.
For a top~quark with mass substantially greater than that of
the $W$~boson, the standard model predicts it
to decay promptly (before hadronization) to a $W$~boson
plus a bottom quark with a branching fraction of nearly $100\%$.
A precision measurement of the top~quark mass,
along with the $W$~boson mass and
other electroweak data, can set constraints on the
mass of the standard model Higgs boson.
It may also be helpful in understanding the origin of quark masses.

In $p\pbar$ collisions at a $1.8\tev$ center of mass energy,
top~quarks are produced
primarily as $t\tbar$ pairs.  Each decays
into a $W$~boson plus a bottom~quark, resulting
in events having several jets and often a charged lepton.
Due to the large top~quark mass, these final state objects tend to have
large momenta transverse to the $p\pbar$ direction.
About $30\%$ of $t\tbar$ decays have a single electron or muon
(from the decay of
one of the $W$~bosons) with a large transverse momentum.  Typically,
the neutrino
that accompanies this electron or muon will
also have a large transverse momentum, producing significant
missing transverse energy.
These characteristics allow for the
selection of a sample of
``lepton + jets'' events with an enriched signal to background
ratio.  This sample is the basis for the top~quark mass analysis 
reported in this paper.  It also comprises a large portion of the
data sample used for the measurement of the
$p\pbar \rightarrow t\tbar$
production cross section~\cite{xsecprl}.
A similar mass analysis for the final state with two charged
leptons plus jets is described in Ref.~\cite{dilmassprl}.

Three methods have been used to determine the top~quark mass
in the lepton + jets channels.  Two of them
use constrained variable-mass kinematic fits to obtain a
best-fit mass value for each
event.  The top~quark mass is then extracted using a maximum likelihood fit to 
a two-dimensional distribution, with one axis being the best-fit mass,
and the other being a variable which discriminates $\ttbar$ events
from the expected backgrounds.
The difference between these two methods
is in the discriminant variable and the binning used.
The third method uses $\chisq$~values from fixed-mass kinematic
fits.  A cut is made using a top quark
discriminant to select a sample of events with low background.
The expected contribution from the background is subtracted from the
distribution of $\chisq$ versus mass, 
and the resulting background-subtracted distribution is fit near
the minimum to extract the top~quark mass.

This paper is organized as follows.
\Secref{d0detector} briefly describes aspects
of the \dzero\ detector essential for this analysis.
\Secref{event-selection} discusses event selection, including triggers, 
particle identification, and the criteria used to select the initial
event sample.
\Secref{jetcorrections} describes the jet energy corrections.
\Secref{mcsim} discusses the simulation of $\ttbar$ signal and
background events.
\Secref{likelihood} defines the two discriminants used to
separate top quark events from background.
\Secref{mass_fit} describes the variable-mass kinematic fits to individual
events and the likelihood fits used to extract the top~quark mass,
and gives results from these fits.
\Secref{pseudolikelihood}
describes the pseudo-likelihood method (which uses fixed-mass kinematic
fits), gives results from it, and compares these
results with those from the two likelihood methods.
\Secref{kinematics} examines some kinematic properties of top quark events.  
Finally, conclusions are presented in \secref{conclusion}.

\section{The \dzero\ Detector}
\label{d0detector}

\begin{figure}
\psfig{figure=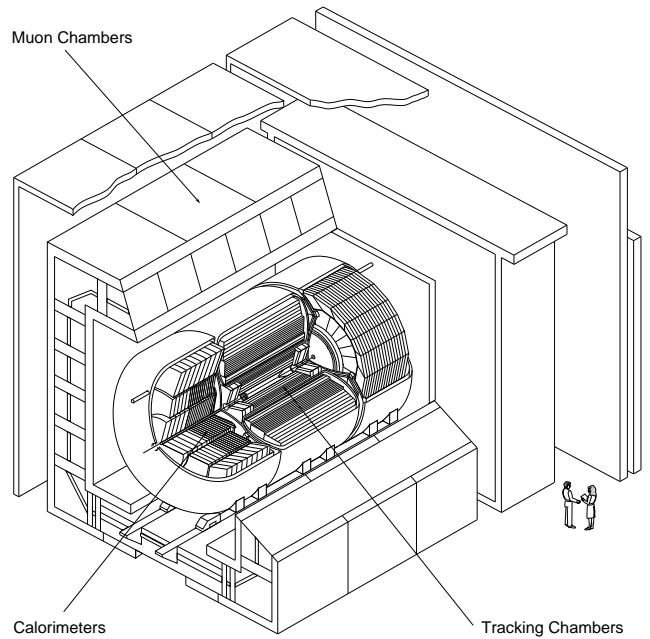,width=\hsize}
\caption{Cut away isometric view of the \dzero\ detector.}
\label{fg:figdet1}
\end{figure}

\dzero\ is a multipurpose detector designed to study  $p\pbar$ collisions at
the Fermilab Tevatron Collider. The detector was commissioned 
during  the summer of 1992. The work presented here is based
on approximately $125\ipb$ of accumulated data recorded during
the 1992--1996 collider runs.
A full description of the detector may be found in Ref.~\cite{d0nim}. 
Here, we
describe briefly the properties of the detector that are relevant for the top
quark mass measurement.

The detector was designed to have good electron and muon identification
capabilities, and to measure jets and missing transverse energy $\met$
with good resolution.
The detector consists of three major systems: a nonmagnetic central tracking
system, a hermetic uranium liquid-argon calorimeter,
and a muon spectrometer. A
cut away view of the detector is shown in \figref{fg:figdet1}.

The central detector (CD) consists of four tracking subsystems: a
vertex drift chamber, a transition radiation detector (not used
for this analysis), a
central drift chamber, and two forward drift chambers.
It measures the trajectories of charged particles and can discriminate
between single charged particles and $e^+e^-$ pairs from photon conversions
by measuring the ionization along their tracks.  It covers the region $|\eta|
< 3.2$ in pseudorapidity, where $\eta = \tanh^{-1}(\cos\theta)$.
(We define $\theta$ and $\phi$ to be the polar and azimuthal angles,
respectively.)

The calorimeter is divided into three parts: the central calorimeter
(CC) and the two end calorimeters (EC), which together cover
the pseudorapidity
range $|\eta| < 4.2$.  The inner electromagnetic (EM) portion
of the calorimeters is 21 radiation lengths deep, and is divided
into four longitudinal segments (layers).  The outer hadronic portions
are 7--9 nuclear interaction lengths deep, and are divided into
four (CC) or five (EC) layers.
The calorimeters are transversely segmented
into pseudoprojective towers with $\Delta\eta\times\Delta\phi$ = $0.1
\times 0.1$.  The third layer of the electromagnetic (EM) calorimeter,
in which the maximum of EM showers is expected, is segmented twice as
finely in both $\eta$ and $\phi$,
with cells of size $\Delta\eta\times\Delta\phi$ = $0.05 \times 0.05$.

Since muons from top quark decays populate predominantly the central
region, this work uses only the central portion of the \dzero\ muon
system, covering $|\eta| < 1.7$.  This system consists of four planes
of proportional drift tubes in front of magnetized iron toroids with a
magnetic field of 1.9~T and two groups of three planes each of
proportional drift tubes behind the toroids.  The magnetic field lines
and the wires in the drift tubes are oriented transversely to the beam
direction.  The muon momentum $p^\mu$ is measured from the muon's deflection
angle in the magnetic field of the toroid.

A separate synchrotron, the Main Ring,
lies above the Tevatron and passes through the outer region of the
\dzero\ calorimeter.  During data-taking, it is used to accelerate
protons for antiproton production.  Losses from the Main Ring may
deposit energy in the calorimeters, increasing the instrumental background.
We reject much of this background at the trigger level by not
accepting triggers during injection into the Main Ring, when losses
are large.  Some triggers are also disabled whenever a Main Ring bunch
passes through the detector or when losses are registered in scintillation
counters around the Main Ring.

\section{Event Selection}
\label{event-selection}

For the purposes of this analysis, we divide the lepton + jets final states
into electron and muon channels.  We further subdivide these channels
based on whether or not a muon consistent with $b\rightarrow\mu+X$
is present.  We thus have four channels, which will be denoted
$e+\jets$, $\mu+\jets$, $e+\jets/\mu$, and $\mu+\jets/\mu$.

The event sample used for determining the top quark mass is selected
using criteria similar to those used for the $\ttbar$ production cross
section measurement \cite{xsecprl}, with the exception of the cuts on
the event shape variables $\htran\equiv \sum \etjet$ and aplanarity.
The particle identification, trigger requirements, and event selection
cuts are summarized below.
More detailed information about
triggering, particle identification, and jet and $\met$ reconstruction
may be found in Ref.~\cite{d0topprd}.  (Note, however, that the
current electron and muon identification algorithms provide better
rejection of backgrounds and increased efficiencies than those used in
Ref.~\cite{d0topprd}.)

\subsection{Particle identification}
\subsubsection{Electrons}

Electron identification is based on a likelihood technique. 
Candidates are first identified by finding isolated clusters of energy
in the EM calorimeter with a matching track in the central
detector.  We then cut on a likelihood constructed from
the following four variables:
\begin{itemize}
\item The $\chisq$ from a covariance matrix which measures
      the consistency of the calorimeter cluster shape with
      that of an electron shower.
\item The electromagnetic energy fraction, defined as the ratio of the
      portion of the energy of the cluster found in the EM calorimeter
      to its total energy.
\item A measure of the consistency between the track position and the
      cluster centroid.
\item The ionization $dE/dx$ along the track.
\end{itemize}
To a good approximation, these four variables are 
independent of each other for electron candidates. 

Electrons from $W$~boson decay tend to be isolated, even in $\ttbar$ events.
Thus, we make the
additional cut
\begin{equation}
{E_{\text{tot}}(0.4) - E_{\text{EM}}(0.2) \over E_{\text{EM}}(0.2)} < 0.1,
\end{equation}
where $E_{\text{tot}}(0.4)$ is
the energy within $\Delta R < 0.4$ of the cluster centroid
($\Delta R = \sqrt{\Delta\eta^2 + \Delta\phi^2}$) and
$E_{\text{EM}}(0.2)$ is the energy in the EM calorimeter within
$\Delta R < 0.2$.

\subsubsection{Muons}

Two types of muon selection are used in this analysis. The first is used to
identify isolated muons from $W\rightarrow \mu\nu$ decay. 
The other is used to tag $b$-jets by identifying ``tag'' muons consistent with
originating from $b\rightarrow \mu + X$ decay.

Besides cuts on the muon track quality, both selections require that:
\begin{itemize}
\item The muon pseudorapidity $|\eta^\mu| \leq 1.7$.
\item The magnetic field integral $> 2.0\ \text{T}\cdot\text{m}$
(equivalent to a momentum change of $0.6\gevc$).
\item The energy deposited in the calorimeter along a muon track
      be at least that expected from a minimum ionizing particle.
\end{itemize}

For isolated muons, we apply the following additional selection
requirements:
\begin{itemize}
\item Transverse momentum $\pt \ge 20\gevc$.
\item The distance in the $\eta - \phi$ plane
      between the muon and the closest jet $\Delta R(\mu,j) > 0.5$.
\end{itemize}

For tag muons, we instead require:
\begin{itemize}
\item $\pt \ge 4\gevc$.
\item $\Delta R(\mu,j) < 0.5$.
\end{itemize}

\subsubsection{Jets and missing $\et$}

Jets are reconstructed in the calorimeter using a fixed-size cone
algorithm.  We use a cone size of $\Delta R = 0.5$.

Neutrinos are not detected directly.  Instead, their presence is
inferred from missing transverse energy $\met$.
Two different definitions of $\met$ are used in the event selection:
\begin{itemize}
\item $\metcal$, the calorimeter missing $\et$, obtained from the
      transverse  energy of all calorimeter cells.
\item $\met$, the muon corrected missing $\et$, obtained by subtracting the
      transverse momenta of identified muons from $\metcal$.
\end{itemize}

\subsection{Triggers}

The \dzero\ trigger system is responsible for reducing the event
rate from the beam crossing rate of 286~kHz to the approximately
3--4~Hz which can be recorded on tape.  The first stage of the
trigger (level~1) makes fast analog sums of the transverse energies
in calorimeter trigger towers.  These towers have a size of
$\Delta\eta\times\Delta\phi = 0.2 \times 0.2$ and are
segmented longitudinally into electromagnetic and hadronic sections.
The level 1 trigger operates on these sums along
with patterns of hits in the muon spectrometer.  It can make
a trigger decision within the space of a single beam crossing
(unless a level 1.5 decision is required; see below).
After level~1 accepts an event, the complete event is digitized
and sent to the level~2 trigger, which consists of a farm of
48~general-purpose processors.  Software filters running in these
processors make the final trigger decision.

The triggers used are defined in terms of combinations of specific
objects (electron, muon, jet, $\met$) required in the level~1 and
level~2 triggers.  These elements are summarized below.  For more
information on the \dzero\ trigger system, see Refs.~\cite{d0topprd,d0nim}.

To trigger on electrons, level~1 requires that the transverse energy
in the EM section of a trigger tower be above a programmed threshold.
The level~2 electron algorithm examines the regions around the level~1
towers which are above threshold, and uses the full segmentation of the
EM calorimeter to identify showers with shapes consistent with 
those of electrons.
The level~2 algorithm can also apply an isolation requirement or demand
that there be an associated track in the central detector.

For the latter portion of the run, a ``level~1.5''
processor was also available for electron triggering.  The $\et$
of each EM trigger tower
above the level~1 threshold is summed with the neighboring tower
with the most energy.  A cut is then made on this sum.
The hadronic portions of the two towers are also
summed, and the ratio of EM transverse energy to total transverse energy
in the two towers is required to be above 0.85.
The use of a level~1.5 electron trigger is indicated
in the tables below as an ``EX'' tower.

The level~1 muon trigger uses the pattern of drift tubes with hits
to provide the number of muon candidates in different regions of the
muon spectrometer.  A level~1.5 processor may optionally be used
to put a $\pt$ requirement on the candidates (at the expense of slightly
increased dead time).  In level~2, the full digitized data are available,
and the first stage of the full event reconstruction is performed.
The level~2 muon algorithm can optionally require the presence
of an energy deposit in the calorimeter consistent with that from a muon;
this is indicated in the tables below by ``cal confirm''.

For a jet trigger, level~1 requires that the sum of the transverse
energies in the EM and hadronic
sections of a trigger tower be above a programmed threshold.  Alternatively,
level~1 can sum the transverse energies within ``large tiles'' of size
$0.8\times 1.6$ in $\eta\times\phi$ and cut on these sums.
Level~2
then sums calorimeter cells around the identified towers
(or around the $\et$-weighted centroids of the large tiles)
in cones of a specified
radius $\Delta R$, and imposes a cut on the total transverse energy.

The $\met$ in the calorimeter can also be computed in both level~1
and level~2.  The $z$ position used for the interaction vertex in
level 2 is determined
from the relative timing of hits in scintillation counters located
in front of each EC (level~0).

The trigger requirements used for this analysis are summarized
in \tabsref{tb:trigger-1a}--\ref{tb:trigger-1c}.  These tables
are divided according to the three major running periods.  Run~1a
was from 1992--1993, run~1b was from 1994--1995, and run~1c was during
the winter of 1995--1996.  Note that not all the triggers listed
were active simultaneously, and that differing requirements
were used to veto possible Main Ring events.
In addition, some of the triggers
were prescaled at high luminosity.  The ``exposure'' column in the
tables takes these factors into account.

\widetext
\begin{table*}
\caption{Triggers used during run~1a (1992--1993).
``Exposure'' gives the effective integrated luminosity for each trigger,
taking into account any prescaling.}
\squeezetable  %
\begin{tabular}{l|d|c|c|l}
Name & Exposure & Level 1 & Level 2 & Used by \\
     & ($\ipb$) & & & \\
\tableline
\progname{ele-high} & 11.0 & 1 EM tower, $\et > 10\gev$
                           & 1 isolated $e$, $\et > 20\gev$
                           & $e+\jets$ \\
                    &&&    & $e+\jets/\mu$ \\
\tableline
\progname{ele-jet}  & 14.4 & 1 EM tower, $\et > 10\gev$, $|\eta| < 2.6$
                           & 1 $e$, $\et > 15\gev$, $|\eta| < 2.5$
                           & $e+\jets$ \\
                    &      & 2 jet towers, $\et > 5\gev$
                           & 2 jets ($\Delta R=0.3$), $\et > 10\gev$, $|\eta| < 2.5$
                           & $e+\jets/\mu$ \\
                    &      &
                           & $\metcal > 10\gev$ \\
\tableline
\progname{mu-jet-high}
                    & 10.2 & 1 $\mu$, $|\eta| < 2.4$
                           & 1 $\mu$, $\pt > 8\gevc$
                           & $\mu+\jets$ \\
                    &      & 1 jet tower, $\et > 5\gev$    
                           & 1 jet ($\Delta R=0.7$), $\et > 15\gev$ 
                           & $\mu+\jets/\mu$ \\
\end{tabular}
\label{tb:trigger-1a}
\end{table*}
\narrowtext

\widetext
\begin{table*}
\caption{Same as \tabref{tb:trigger-1a} for run~1b (1994--1995).}
\squeezetable  %
\begin{tabular}{l|d|c|c|l}
Name & Exposure & Level 1 & Level 2 & Used by \\
     & ($\ipb$) & & & \\
\tableline
\progname{em1-eistrkcc-ms}
                    & 93.4 & 1 EM tower, $\et > 10\gev$          
                           & 1 isolated $e$ w/track, $\et > 20\gev$
                           & $e+\jets$ \\
                    &      & 1 EX tower, $\et > 15$\gev\tablenotemark[1]
                           & $\metcal > 15\gev$
                           & $e+\jets/\mu$ \\
\tableline
\progname{ele-jet-high}
                    & 98.0 & 1 EM tower, $\et > 12\gev$, $|\eta| < 2.6$ 
                           & 1 $e$, $\et > 15\gev$, $|\eta| < 2.5$
                           & $e+\jets$ \\
                    &      & 2 jet towers, $\et > 5\gev$, $|\eta| < 2.0$   
                           & 2 jets ($\Delta R=0.3$), $\et > 10\gev$, $|\eta| < 2.5$
                           & $e+\jets/\mu$ \\
                    &      &
                           & $\metcal > 14\gev$ \\
\tableline
\progname{mu-jet-high}
                    & 66.4 & 1 $\mu$, $\pt > 7\gevc$\tablenotemark[1], $|\eta| < 1.7$
                           & 1 $\mu$, $\pt > 10\gevc$, $|\eta| < 1.7$
                           & $\mu+\jets$ \\
                    &      & 1 jet tower, $\et > 5\gev$, $|\eta| < 2.0$\tablenotemark[1]
                           & 1 jet ($\Delta R=0.7$), $\et > 15\gev$, $|\eta| < 2.5$
                           & $\mu+\jets/\mu$ \\
\tableline
\progname{mu-jet-cal}
                    & 88.0 & 1 $\mu$, $\pt > 7\gevc$\tablenotemark[1], $|\eta| < 1.7$             
                           & 1 $\mu$, $\pt > 10\gevc$, $|\eta| < 1.7$, cal confirm
                           & $\mu+\jets$ \\
                    &      & 1 jet tower, $\et > 5\gev$, $|\eta| < 2.0$\tablenotemark[1]
                           & 1 jet ($\Delta R=0.7$), $\et > 15\gev$, $|\eta| < 2.5$
                           & $\mu+\jets/\mu$ \\
\tableline
\progname{mu-jet-cent}
                    & 48.5 & 1 $\mu$, $|\eta| < 1.0$                
                           & 1 $\mu$, $\pt > 10\gevc$, $|\eta| < 1.0$
                           & $\mu+\jets$ \\
                    &      & 1 jet tower, $\et > 5\gev$, $|\eta| < 2.0$   
                           & 1 jet ($\Delta R=0.7$), $\et > 15\gev$, $|\eta| < 2.5$
                           & $\mu+\jets/\mu$ \\
\tableline
\progname{mu-jet-cencal}
                    & 51.2 & 1 $\mu$, $|\eta| < 1.0$                             
                           & 1 $\mu$, $\pt > 10\gevc$, $|\eta| < 1.0$, cal confirm
                           & $\mu+\jets$ \\
                    &      & 1 jet tower, $\et > 5\gev$, $|\eta| < 2.0$   
                           & 1 jet ($\Delta R=0.7$), $\et > 15\gev$, $|\eta| < 2.5$
                           & $\mu+\jets/\mu$ \\
\tableline
\progname{jet-3-mu}
                    & 11.9 & 3 jet towers, $\et > 5\gev$                 
                           & 3 jets ($\Delta R=0.7$), $\et > 15\gev$, $|\eta| < 2.5$
                           & $\mu+\jets$ \\
                    &      & $\metcal > 20\gev$
                           & $\metcal > 17\gev$
                           & $\mu+\jets/\mu$ \\
\tableline
\progname{jet-3-miss-low}
                    & 57.8 & 3 large tiles, $\et > 15$, $|\eta| < 2.4$       
                           & 3 jets ($\Delta R=0.5$), $\et > 15\gev$, $|\eta| < 2.5$
                           & $\mu+\jets$ \\
                    &      & 3 jet towers, $\et > 7\gev$, $|\eta| < 2.6$
                           & $\metcal > 17\gev$
                           & $\mu+\jets/\mu$ \\
\tableline
\progname{jet-3-l2mu}
                    & 25.8 & 3 large tiles, $\et > 15$, $|\eta| < 2.4$           
                           & 1 $\mu$, $\pt > 6\gevc$, $|\eta| < 1.7$, cal confirm
                           & $\mu+\jets$ \\
                    &      & 3 jet towers, $\et > 7\gev$, $|\eta| < 2.6$   
                           & 3 jets ($\Delta R=0.5$), $\et > 15\gev$, $|\eta| < 2.5$
                           & $\mu+\jets/\mu$ \\
                    &      &
                           & $\metcal > 17\gev$ \\
\end{tabular}
\tablenotetext[1]{This cut was looser than indicated during early
portions of the run.}
\label{tb:trigger-1b}
\end{table*}
\narrowtext

\widetext
\begin{table*}
\caption{Same as \tabref{tb:trigger-1a} for run~1c (1995--1996).}
\squeezetable  %
\begin{tabular}{l|d|c|c|l}
Name & Exposure & Level 1 & Level 2 & Used by \\
     & ($\ipb$) & & & \\
\tableline
\progname{ele-jet-high}
                    &  1.9 & 1 EM tower, $\et > 12\gev$, $|\eta| < 2.6$
                           & 1 $e$, $\et > 15\gev$, $|\eta| < 2.5$
                           & $e+\jets$ \\
                    &      & 2 jet towers, $\et > 5\gev$, $|\eta| < 2.0$   
                           & 2 jets ($\Delta R=0.3$), $\et > 10\gev$, $|\eta| < 2.5$
                           & $e+\jets/\mu$ \\
                    &      &
                           & $\metcal > 14\gev$ \\
\tableline
\progname{ele-jet-higha}
                    & 11.0 & 1 EM tower, $\et > 12\gev$, $|\eta| < 2.6$
                           & 1 $e$, $\et > 17\gev$, $|\eta| < 2.5$
                           & $e+\jets$ \\
                    &      & 2 jet towers, $\et > 5\gev$, $|\eta| < 2.0$   
                           & 2 jets ($\Delta R=0.3$), $\et > 10\gev$, $|\eta| < 2.5$
                           & $e+\jets/\mu$ \\
                    &      & 1 EX tower, $\et > 15\gev$
                           & $\metcal > 14\gev$ \\
\tableline
\progname{mu-jet-cent}
                    &  8.9 & 1 $\mu$, $|\eta| < 1.0$                
                           & 1 $\mu$, $\pt > 12\gevc$, $|\eta| < 1.0$
                           & $\mu+\jets$ \\
                    &      & 1 jet tower, $\et > 5\gev$, $|\eta| < 2.0$    
                           & 1 jet ($\Delta R=0.7$), $\et > 15\gev$, $|\eta| < 2.5$
                           & $\mu+\jets/\mu$ \\
                    &      & 2 jet towers, $\et > 3\gev$
                           & \\
\tableline
\progname{mu-jet-cencal}
                    & 11.4 & 1 $\mu$, $|\eta| < 1.0$                             
                           & 1 $\mu$, $\pt > 12\gevc$, $|\eta| < 1.0$, cal confirm
                           & $\mu+\jets$ \\
                    &      & 1 jet tower, $\et > 5\gev$, $|\eta| < 2.0$    
                           & 1 jet ($\Delta R=0.7$), $\et > 15\gev$, $|\eta| < 2.5$
                           & $\mu+\jets/\mu$ \\
                    &      & 2 jet towers, $\et > 3\gev$
                           & \\
\tableline
\progname{jet-3-l2mu}
                    & 11.3 & 3 large tiles, $\et > 15$, $|\eta| < 2.4$           
                           & 1 $\mu$, $\pt > 8\gevc$, $|\eta| < 1.7$, cal confirm
                           & $\mu+\jets$ \\
                    &      & 3 jet towers, $\et > 5\gev$, $|\eta| < 2.0$   
                           & 3 jets ($\Delta R=0.5$), $\et > 15\gev$, $|\eta| < 2.5$
                           & $\mu+\jets/\mu$ \\
                    &      & 4 jet towers, $\et > 3\gev$
                           & $\metcal > 17\gev$ \\
\end{tabular}
\label{tb:trigger-1c}
\end{table*}
\narrowtext

\subsection{Event selection}
\label{event-selection-sub}

The first set of cuts used to define the sample for mass analysis
is very similar to that used for the cross section analysis
\cite{xsecprl}:
\begin{itemize}
\item An isolated electron or muon with $\et > 20 \gev$.
\item $ |\eta^e| < 2.0$ or $|\eta^\mu| < 1.7$.
\item At least 4 jets with $\et > 15 \gev$ and $ |\eta^{\jet}| < 2.0$.
\item $\metcal > 25 \gev $ for $e$+jets (untagged) or 
      $\metcal > 20 \gev$ for $ \mu$+jets (both tagged and untagged).
\item $\met > 20 \gev$.
\end{itemize}
We reject events which contain photons --- isolated clusters in the
EM calorimeter with shapes consistent with an EM shower and with a poor
match to any track in the central detector, and satisfying
$\et > 15\gev$ and $|\eta| < 2$.  Three such events are rejected.
We also reject events which contain extra isolated high-$\pt$
electrons or which fail additional cuts to remove calorimeter noise
and Main Ring effects.  

After these cuts, the remaining background is primarily $W+\jets$,
with a small ($\approx 20\%$) admixture of QCD multijet events in which
a jet is misidentified as a lepton.

If a candidate has a tag muon, we require it to pass additional cuts
on the direction of the $\met$ vector.  For the $e+\jets/\mu$ channel,
we require
\begin{itemize}
\item $\met > 35\gev$, if $\Delta\phi(\met,\mu) < 25^\circ$,
\end{itemize}
while for the $\mu+\jets/\mu$ channel, we require that the
highest-$\pt$ muon satisfy
\begin{itemize}
\item $\Delta\phi(\met,\mu) < 170^\circ$ and
\item $|\Delta\phi(\met,\mu) - 90^\circ|/90^\circ < \met/(45\gev)$.
\end{itemize}
These cuts remove QCD multijet background events which appear to have a large
$\met$ due to a mismeasurement of the muon momentum.

For the remaining, untagged, events, we require:
\begin{itemize}
\item $\et^W \equiv |\et^{\text{lep}}| + |\met| > 60 \gev$.
\item $|\eta^W| < 2.0$.
\end{itemize}
For the purpose of these two cuts, we define $\eta^W$ by assuming that
the entire $\met$ of the event is due to the neutrino from the decay of
the $W$~boson.  The longitudinal component of the neutrino momentum
$p_z^\nu$ is
found by using the $W$~boson mass $M_W$ as a constraint.  If the transverse
mass of the lepton and neutrino $M_T(l\nu)$ is less than $M_W$, there are two real
solutions; the one with the smallest absolute value of $p_z^\nu$
is used.  Monte
Carlo studies show that this is the correct solution about $80\%$ of
the time.  If $M_T(l\nu) > M_W$ there are no real solutions.  In this
case, the $\met$ is scaled so that $M_T(l\nu) = M_W$.  This scaled
$\met$ is also used for the $\et^W$ cut (but not for the previous
cuts on $\met$ alone).

\begin{figure}
\psfig{file=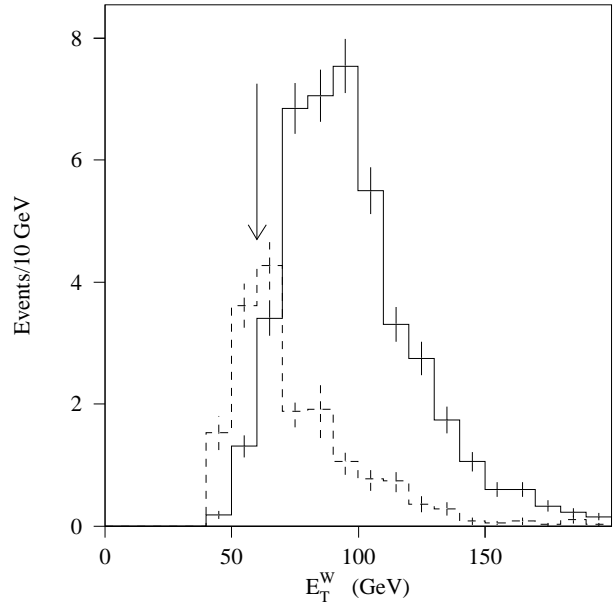,width=\hsize}
\caption {$\et^W$ distribution for
Monte Carlo $W$+jets events (solid histogram)
and for QCD multijet background data (dashed histogram).
All selection cuts are applied except for the $\et^W$ cut.
The arrow shows the cut value.
(The normalizations are taken from the result of the LB
fit to the data, as described in \secref{fits-to-data}, with
channels combined as described in \secref{vbls-and-binning}.  The
models used to simulate the data are described in \secref{mcsim}.)}
\label{fg:et_l}
\end{figure}

This cut on $\et^W$ removes a portion of the QCD multijet background.
Figure \ref{fg:et_l} compares the $\et^W$ distribution for this
background to that from Monte Carlo $W+\jets$ events.

\begin{figure}
\psfig{file=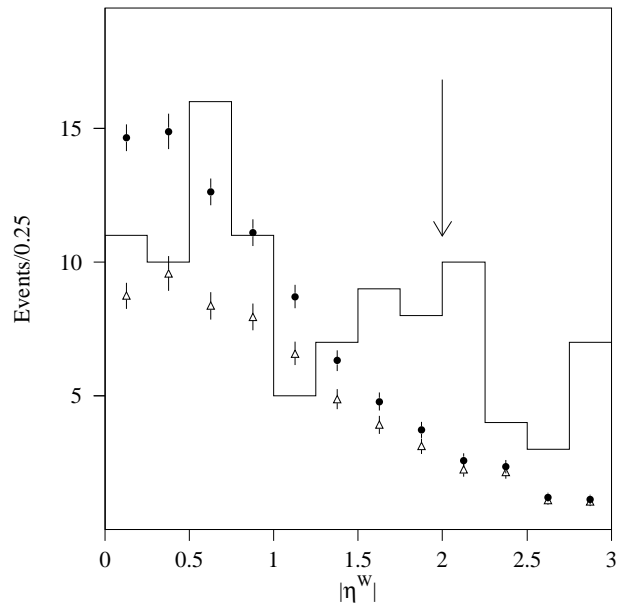,width=\hsize}
\caption {$|\eta^W|$ distribution for data (histogram), predicted
signal plus background (filled circles), and background alone (open triangles).
All selection cuts are applied except for the $\eta^W$ cut.
The arrow shows the cut value.
(The normalizations are as in \figref{fg:et_l}.)}
\label{fg:eta_w}
\end{figure}

We show in \figref{fg:eta_w} the distributions of $|\eta^W|$ for our
data and for the Monte Carlo prediction.  The data are seen to
significantly exceed the prediction of the \progname{vecbos} Monte
Carlo (described in \secref{mcsim})
in the far forward region.  The amount of $\ttbar$ signal
with $|\eta^W| > 2$ is only a few percent
($\approx 3\%$ for $m_t = 175\gevcc$).
In addition, a check of the
$W$~boson transverse mass and $\met$ distributions shows that the QCD
multijet background plays no unusually prominent role at high
$|\eta^W|$.  We note that the \progname{vecbos} Monte Carlo, while
the best currently available, is only a tree-level calculation of the
$W+\jets$ process.  Particularly in the forward direction, one would
expect higher order corrections to play a larger role.  To mitigate
the effects of this discrepancy, and to further reduce the background,
we require $|\eta^W| < 2$.  Once
this cut is made, the $\chisq_L$ between the data and prediction is
12.2 for 7 d.o.f., giving a $9\%$ probability.
($\chisq_L \equiv 2\sum_i \left[y_i - N_i + N_i \ln (N_i/y_i)\right]$, where
$N$ is the number of observed events and $y$ is the total number expected from
Monte Carlo.  This form is appropriate for low statistics
\cite{baker}.)  The contribution of this effect to the systematic
error will be discussed in \secref{massfit-gendep} (and is found to be
negligible).

These event selection cuts are summarized in \tabref{tbl:selection-cuts}.
When applied to the approximately $125\ipb$ of data from the
1992--1996 collider runs, 91 events are selected~\cite{ev10822-note},
seven of which have a tag muon.  This sample will be referred to as the
``precut'' sample, and the set of cuts as the ``PR'' cuts.
One additional
cut is made to define the final sample.  This is based on the $\chisq$ of
a kinematic fit to the $\ttbar$ decay hypothesis ($\chisq < 10$),
and is described
in \secref{mass_fit}.  This final cut reduces the sample to
77~candidate events, of which five are tagged.

\widetext
\begin{table*}
\caption{Summary of event selection cuts.}
\begin{tabular}{cllll}
 Channel  & $e+\jets$ & $\mu+\jets$ & $e+\jets/\mu$ & $\mu+\jets/\mu$ \\
\hline
Lepton & $\et^e > 20\gev$      & $\pt^\mu > 20\gevc$   &
         $\et^e > 20\gev$      & $\pt^\mu > 20\gevc$   \\
       & $|\eta^e| < 2$        & $|\eta^\mu| < 1.7$    &
         $|\eta^e| < 2$        & $|\eta^\mu| < 1.7$    \\
\hline
$\met$ & $\met > 20\gev$       & $\met > 20\gev$       &
         $\met > 20\gev$       & $\met > 20\gev$       \\
       & $\metcal > 25\gev$    & $\metcal > 20\gev$    &
                               & $\metcal > 20\gev$    \\
\hline
Jets   & $\ge 4\ \jets$        & $\ge 4\ \jets$        &
         $\ge 4\ \jets$        & $\ge 4\ \jets$        \\
       & $\et^{\jet} > 15\gev$ & $\et^{\jet} > 15\gev$ &
         $\et^{\jet} > 15\gev$ & $\et^{\jet} > 15\gev$ \\
       & $|\eta^{\jet}| < 2.0$ & $|\eta^{\jet}| < 2.0$ &
         $|\eta^{\jet}| < 2.0$ & $|\eta^{\jet}| < 2.0$ \\
\hline
$\mu$ Tag & No tag             & No tag                &
         Tag required          & Tag required          \\
\hline
Other  & $\et^W > 60\gev$      & $\et^W > 60\gev$      &
         $\met > 35\gev$ &
         $\Delta\phi(\met,\mu) < 170^\circ$ \\
       & $|\eta^W| < 2.0$    & $|\eta^W| < 2.0$    & 
         \quad if $\Delta\phi(\met,\mu) < 25^\circ$  &
         $|\Delta\phi(\met,\mu) - 90^\circ|/90^\circ <$  \\
       & & & & \quad $\met/(45\gev)$ \\
\hline
\hline
Events passing cuts& 43 & 41 & 4 & 3 \\
With $\chisq < 10$ & 35 & 37 & 2 & 3 \\
\end{tabular}
\label{tbl:selection-cuts}
\end{table*}
\narrowtext

\section{Jet Corrections and Energy Scale Error}
\label{jetcorrections}

To calibrate the energy scale so that data and
Monte Carlo (MC) are on an equal footing, we apply a series of energy
corrections to the measured objects.  These corrections are carried
out in three steps.  The first of these corrections is done
before events are selected and is used by most \dzero\ analyses; the
other two corrections are applied during the kinematic fit and are
specific to the top quark mass analysis.

\subsection{Standard corrections}
\label{standard-corrections}

For the standard corrections, electromagnetic objects are first
scaled by a factor which was chosen to make the
invariant mass peak from dielectron events match the $Z$ boson mass as
measured by the LEP experiments.  (This factor is determined separately
for each of the three cryostats of the calorimeter.)
Next, jet energies are corrected
using
\begin{equation}
E(\text{corrected}) = {E(\text{measured})-O \over R (1-S)}.
\end{equation}
Here, $R$ is the calorimeter response; it is found using
$\et$ balance (as determined from the total $\met$)
in $\gamma + \jets$ events.  This determination is done
separately and symmetrically for both data and Monte Carlo.  $O$ is the
offset due to the underlying event, multiple interactions, and noise
from the natural radioactivity of the uranium absorber.  It is
determined by comparing data in which a hard interaction is required
to data in which that requirement is relaxed, and by comparing data
taken at different luminosities.  The term $S$ is the fractional
shower leakage outside the jet cone in the calorimeter.  It is
determined by using single particle showers measured in the test beam
to construct simulated showers from MC jets; this leakage is
approximately $3\%$ for a $50\gev$ jet 
($\Delta R = 0.5$) in the central calorimeter.
Further details about these corrections may be found in Ref.~\cite{cafix}.

\subsection{Parton-level corrections}

The procedure of the previous section corrects for the portions of
showers in the calorimeter which spread outside of the jet cone,
but not for any
radiation outside of the cone.  Thus, the corrected jet energies are
systematically lower than the corresponding parton-level energies
(i.e., before QCD evolution or fragmentation in the MC).  We make a
correction to match the scale of the jet energies to that of the
unfragmented partons in the MC.

 \begin{figure}
 \psfig{file=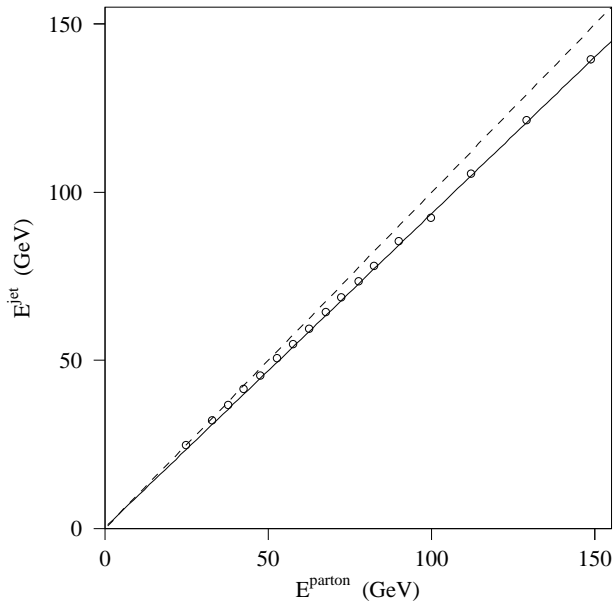,width=\hsize}
 \caption{The measured jet energies for quarks from
   $W\rightarrow q\qbar$ in $t\tbar$ MC are plotted
   against the corresponding parton energies.
   Radiation outside of the jet cone
   causes the measured jet energy to be
   lower than the energy at the parton level.  The dashed line
   is drawn along the diagonal, and
   the solid line is a linear fit to the points.   This plot is based on
   \progname{HERWIG} fragmentation with $|\deteta^{\jet}| < 0.2$.}
 \label{fg:out-of-cone}
 \end{figure}

To derive this correction, we use \progname{herwig}~\cite{her}
$t\tbar$ Monte
Carlo and match reconstructed jets to the partons from top~quark
decay.  Their energies are then plotted against each other, as in
\figref{fg:out-of-cone}.  This relation is observed to be nearly
linear.  We fit it separately for
light quark jets and for untagged $b$~quark jets.  The results are given in
\tabref{tb:oocparms} for different regions in $\deteta$
($\deteta$ $\equiv$ `detector-$\eta$' $\equiv$ the pseudorapidity
corresponding to a particle coming from the geometric center of the
detector, rather than from the interaction vertex).
Separating the $b$~quark jets allows us to
correct, on average, for the neutrinos from $b$~decays.  This
correction is observed not to depend strongly on the MC top quark
mass.

\begin{table}
\caption{Parameters for parton-level jet corrections.
$E(\text{corrected}) = (E - A) / B$.}
\begin{tabular}{ldddd}
 & \multicolumn{2}{c}{Light quark jets}
 & \multicolumn{2}{c}{Untagged $b$~jets} \\
\cline{2-3}
\cline{4-5}
$\eta$ region           & $A$ ($\ugev$)   & $B$   & $A$ ($\ugev$)    & $B$   \\
\tableline
$0.0 < |\deteta| < 0.2$ & 0.322 & 0.933 & -0.672 & 0.907 \\
$0.2 < |\deteta| < 0.6$ & 0.635 & 0.930 & -1.34  & 0.914 \\
$0.6 < |\deteta| < 0.9$ & 1.86  & 0.883 &  0.002 & 0.868 \\
$0.9 < |\deteta| < 1.3$ & 1.70  & 0.933 & -0.548 & 0.904 \\
$1.3 < |\deteta|$\hfill & 4.50  & 0.882 &  2.46  & 0.859 \\
\end{tabular}
\label{tb:oocparms}
\end{table}

\begin{figure}
\psfig{file=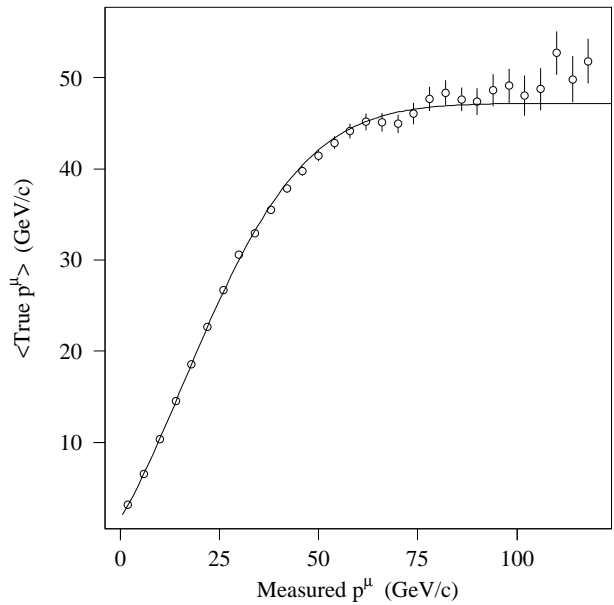,width=\hsize}
\caption{Correlation between the measured momentum and the
true momentum of the tag muon
in Monte Carlo $t\tbar$ events.  The curve is the result of an empirical fit,
$%
 47.19 [1 - \exp (-0.03398 - 0.01593 p^\mu - 0.0005554 (p^\mu)^2)]$.}
\label{fg:mu-unsmear}
\end{figure}

\begin{figure}
\psfig{file=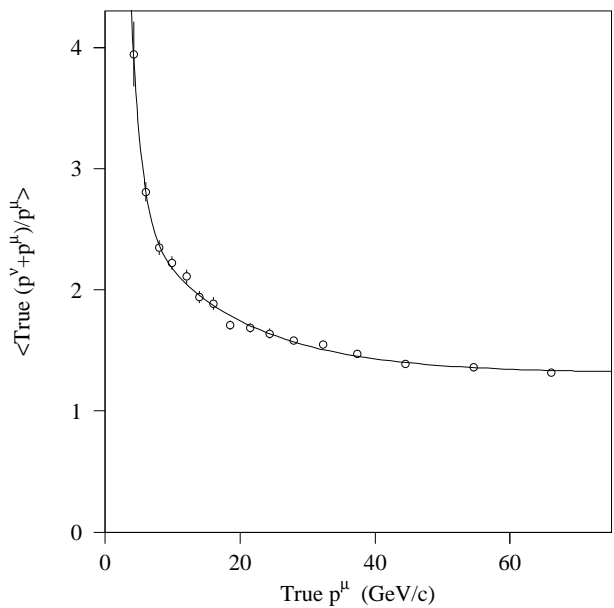,width=\hsize}
\caption{Correlation between the tag muon momentum and the total leptonic
energy from $b$~quark decay in \hbox{MC} $t\tbar$ events.  The curve is
the result of an empirical fit,
$%
 1.313 + \exp (3.101 - 0.6528 p^\mu) + \exp (0.4622 - 0.06514 p^\mu)$.}
\label{fg:lepcorr}
\end{figure}

For tagged $b$~quark jets, we have additional information from the tag
muon.  However, the momentum spectrum of muons from $b$~quark decay in
$t\tbar$ events is rather steeply falling; furthermore, the resolution
of the muon system is more nearly Gaussian in the inverse momentum $1/p$
than in $p$.
Thus, measurement errors will cause the measured momentum of a tag muon
to be biased upwards.  We correct for this bias using $t\tbar$ MC, as
illustrated in \figref{fg:mu-unsmear}.  We then further scale the muon
momentum to account for the unobserved neutrino, as shown in
\figref{fg:lepcorr}.  The jet itself is corrected using the light
quark corrections; the estimated leptonic energy is then added to this
corrected jet energy.

\subsection{$\eta$-dependent adjustment and energy scale error}

\begin{figure}
\psfig{file=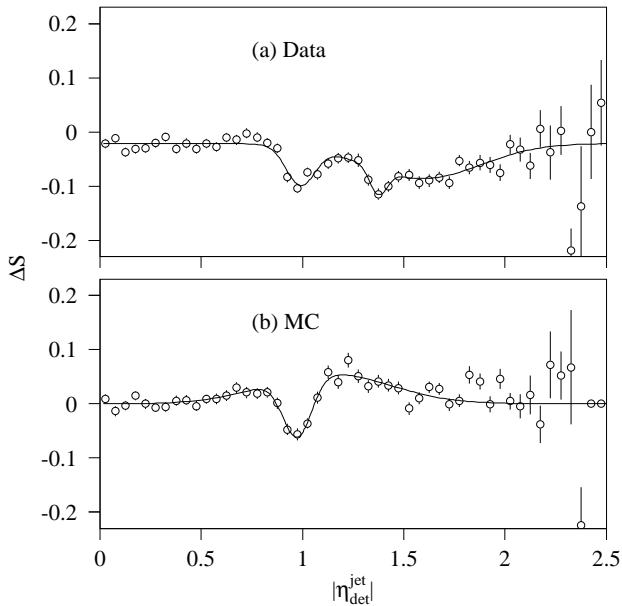,width=\hsize}
\caption{The energy scale deviation $\Delta S$ as a function
of $\deteta^{\jet}$ for (a) data and (b) Monte Carlo.  The curves are
empirical multigaussian fits to the points.}
\label{fg:deltas}
\end{figure}

For the final corrections, we study the response of the detector to
$\gamma + 1\ \jet$ events, using both data and Monte Carlo.  We select
events containing exactly one photon with $\et^{\gamma} > 20\gev$,
$|\deteta^{\gamma}| < 1.0$
or $1.6 < |\deteta^{\gamma}| < 2.5$, and exactly one reconstructed jet
of any energy (excluding the photon).  We
require that the jet satisfy $\et > 15\gev$, $|\eta| < 2$, and
$|\pi - \Delta\phi(j, \gamma)| < 0.2\ \text{rad}$.  We reject events
with Main Ring activity and those which are likely to be multiple
interactions.  To reject $W$~boson decays, we further require that 
$\met / \et^\gamma < 1.2$ if $\et^\gamma < 25\gev$, or
$\met / \et^\gamma < 0.65$ otherwise.  With this selection, we
compute
\begin{equation}
\Delta S = \brocket{\et^{\jet} - \et^\gamma \over \et^\gamma}
\end{equation}
and plot it as a function of $\deteta^{\jet}$.  The result is shown in
\figref{fg:deltas}.  This reveals detector inhomogeneities in the
transition region between the central and end calorimeters~\cite{cafix-note}.
The curve from Monte Carlo is also seen to have a somewhat different
shape than that from data.
To remove these effects,
we smooth the $\Delta S$ distributions by fitting them to the sum
of several Gaussians, and scale each jet by $1/(1+\Delta S(\deteta^{\jet}))$.
This is done separately for data and for Monte Carlo.

\begin{figure}
 \psfig{file=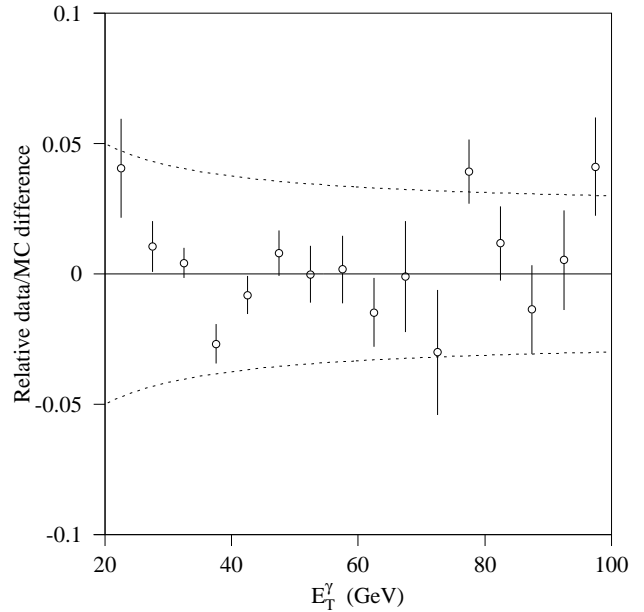,width=\hsize}
 \caption{The relative energy scale difference between data and MC as a
  function of photon $\et$ after all jet corrections are applied.
  The curves are the error band 
  $\pm(2.5 \% + {0.5\gev})$.}
 \label{fg:scale_diff}
 \end{figure}

To estimate the uncertainty in the relative scale between data and
Monte Carlo after all corrections, we derive $\Delta S$ as a function of
$\et^\gamma$ (averaging over $\deteta^{\jet}$)
for both data and MC after all corrections have been
applied.  The difference of the two is plotted in
\figref{fg:scale_diff}, along with a band of
$\pm(2.5\% + 0.5\gev)$, which we use as our estimate of the
systematic error of the jet energy calibration.  (It is the relative
data-MC difference that is relevant,
rather than the absolute error, since the final
mass is extracted by comparing the data to MC generated with known
top quark masses.)

\begin{figure}
 \psfig{file=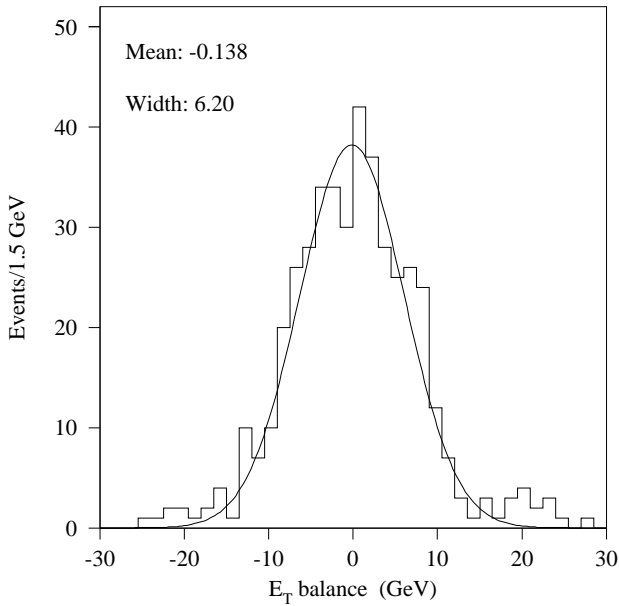,width=\hsize}
 \caption{Transverse energy balance for $(Z\rightarrow ee)+\jets$
    events.  The vector $\vec{p}_T^{\,Z} + 
    \sum_{\jets} \vec{E}_T^{\jet}$
    is projected onto the angle bisector of the two electrons.  All
    jet corrections are applied.  The curve is a Gaussian fit to the
    histogram.}
 \label{fg:Z_jets}
 \end{figure}

\begin{figure}
\psfig{file=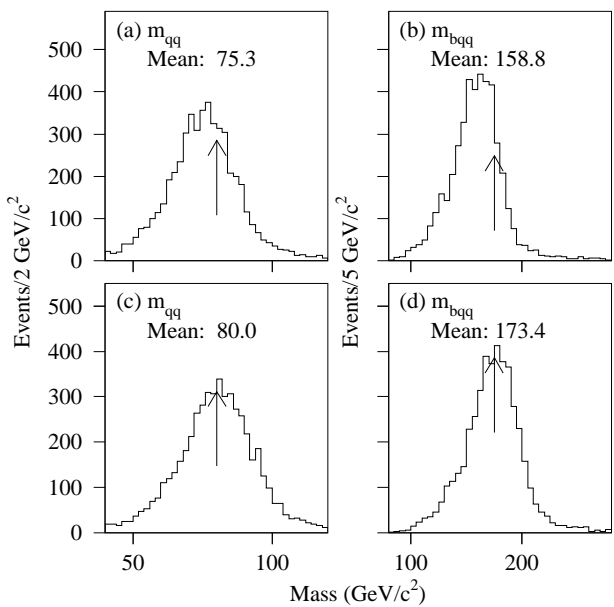,width=\hsize}
\caption{Masses of $W\rightarrow q\qbar$ and $t\rightarrow bq\qbar$
in $t\tbar$ MC with $m_t = 175\gevcc$, both (a), (b) with standard
corrections only and (c), (d) with all jet corrections.  The arrows
locate the input $W$~boson and top~quark masses.}
\label{fg:jetcorr-mccheck}
\end{figure}

A cross-check of these corrections is provided by
$(Z\rightarrow ee) + \jets$ events.  As shown in \figref{fg:Z_jets},
the corrected jets satisfactorily balance the $Z$~boson.  We also show
in \figref{fg:jetcorr-mccheck} the $W\rightarrow q\qbar$ and
$t\rightarrow bq\qbar$ masses from $t\tbar$ MC before and after the final
two corrections.  It is seen that the proper masses are recovered.

The accuracy of these corrections depends on how well the
Monte Carlo models jet widths.
Studies of jets in \dzero\ data show that \progname{herwig} models
the transverse energy distribution within jets to
within 5--$10\%$~\cite{d0jetshape}.  
Note, however, that since the determination of the
response is done separately for data and for Monte Carlo, any
disagreements would, to first order, be removed from the energy scale
determination.  There can still be second-order effects: for example,
if jets in \progname{herwig} were slightly too narrow, and if two jets
were to overlap slightly, then the perturbation to the apparent jet
energies due to that overlap would be slightly underestimated in the
Monte Carlo.  For this situation, we calculate that the fraction of
the energy of a jet between $R=0.5$ and $R=1.0$ of the jet axis which
leaks into the nearest jet is about $10\%$.  We further find that this
region in $R$ contains about $10\%$ of the total energy of a
\progname{herwig} jet.  Thus, the leakage of energy from a jet to a
neighbor is on the order of $1\%$.  If the fraction of the
jet energy outside of $R=0.5$ is substantially larger in data than in
\progname{herwig}, e.g., $20\%$, a $1\%$ miscalibration would
result.  This is well within the errors we assign for moderate $\et$
jets.

\section{Event Simulation}
\label{mcsim}

Monte Carlo simulation is used to model the final states expected
from top quark decays and their principal physics backgrounds.  Although the
overall background normalization is estimated using the observed data,
the simulation is essential to determine the expected shapes of kinematic
distributions.

\subsection{Signal events}

Our primary model for $t\tbar$ production is the
\progname{HERWIG} generator, version 5.7, with
CTEQ3M~\cite{cteq} parton distribution functions.  \progname{HERWIG}
models $t\tbar$ production starting with the elementary hard
process, choosing the parton momenta according to matrix element
calculations.  Initial and final state gluon emission is modeled using
leading log QCD evolution~\cite{dglap}.  Each top quark is then
decayed to a $W$ boson and a $b$ quark, and final state partons are
hadronized into jets. Underlying spectator interactions are also
included in the model.

For this analysis, samples are generated with top quark
masses between 110 and $230\gevcc$.  To increase the
efficiency in the processing of lepton plus jets events, one of the
$W$~bosons is forced to decay to one of the three lepton families.
Events with no final state electrons or muons are vetoed, and half of
the events in which both $W$~bosons decayed leptonically are
discarded in order to preserve the proper branching ratios.  The
generated events are run through the \progname{d\o geant} detector
simulation~\cite{d0geant,geant}
and the \dzero\ event reconstruction program.

Additional samples are made using the \progname{isajet}~\cite{isa}
generator to allow for cross-checks.

\subsection{W+jets background}

The background due to the production of a $W$~boson along
with multiple jets is modeled using the
\progname{VECBOS}~\cite{vecbos} event generator. \progname{VECBOS} supplies
final state partons as a result of a leading order calculation which
incorporates the exact tree level matrix elements for $W$~and
$Z$~boson production with up to four additional partons.  To include the
effects of additional radiation and the underlying processes, and to model
the hadronization of final state partons, the output of
\progname{vecbos} is passed through \progname{herwig}'s
QCD evolution and fragmentation stages.  Since \progname{HERWIG}
requires information about the color labels of its input partons,
it and \progname{vecbos} were modified to assign color and flavor to
the generated partons.  Flavors are assigned probabilistically by
keeping track of the relative weights of each diagram contributing to
the process.  Color labels are simply assigned randomly.
To estimate systematic errors, we also
generate samples which use \progname{isajet} instead of
\progname{herwig} to fragment the \progname{vecbos} partons.  
We test the reliability of the \progname{HERWIG} and \progname{ISAJET}
simulations of higher order processes by comparing $W+$ four jet events
generated using the \progname{vecbos} $W+$ four jet process to those
generated using the $W+$ three jet process.

Events are generated using the same parton distribution functions
assumed for the signal sample.  The dynamical scale of the process is
set to be the average jet $\pt$.  Systematic uncertainties arising
from this choice are estimated by changing the scale to the mass of
the $W$~boson in a second sample of events.  The background samples
are processed through the detector simulation, reconstruction, and
event selection in the same manner as for the signal samples.

\subsection{QCD multijet background}

The non-$W$ QCD multijet background is estimated, both for the electron and
the muon channels, using background-enriched data samples. In
the former channels, the sample consists of events containing highly
electromagnetic jets failing the electron identification cuts. In the
latter, events are selected containing a muon which fails the
isolation requirement, but which otherwise passes the muon identification
cuts.

\section{Top Discriminants}
\label{likelihood}

The key feature that distinguishes top quark events from the $W$+jets
and QCD multijet backgrounds is the fitted mass $\mfit$ obtained from kinematic
fits of the events to the top quark decay hypothesis. Since the top
quark is heavy, the fitted mass tends to be larger for top quark
events than for the backgrounds.  Therefore, if both the signal to
background ratio and the signal are large enough, we should see a clear
signal peak in the $\mfit$ distribution.  However, there is a caveat:
this is true only if the cuts to enhance the signal to noise ratio do
not significantly distort the fitted mass distributions. Unfortunately,
powerful selection variables such as $\htran \equiv \sum \et^{\jet}$ tend to be
highly correlated with the fitted mass.  Cuts on them thus introduce
severe distortions in $\mfit$ which reduce the differences between
the distributions for $\ttbar$ signal and background, and between
the distributions for $\ttbar$ signal at different top~quark masses,
thus impairing the mass measurement.

This distortion of the $\mfit$ distribution can be avoided by using
variables which are only weakly correlated with the fitted mass. The
challenge is to find variables that also provide a useful measure of
discrimination between signal and background.  After an extensive
search of variables that exploit the expected qualitative differences
between the kinematics of top quark events and the backgrounds, we have
succeeded in finding four variables $x_1$--$x_4$ with the desired
properties.

This success, however, comes at a price: the discrimination afforded
by these variables tends to be weaker than that provided by variables,
like $\htran$, that are mass dependent.  But by treating these
variables collectively, rather than applying a cut on each separately,
we can compensate for their weaker discrimination.  It is most
effective to combine the variables into a multivariate discriminant
$\calD({\bf x})$ with the general form
\begin{equation}
\calD({\bf x}) \equiv \frac{f_{s}({\bf x})}{f_{s}({\bf x})+f_{b}({\bf x})},
\label{eq:discrim-def}
\end{equation}
where ${\bf x}$ denotes the 4-tuple of mass-insensitive variables and
$f_s({\bf x})$ and $f_b({\bf x})$ are functions that pertain to the
signal and background, respectively. We choose the functions $f_s$ and
$f_b$ so that $\calD({\bf x})$ is concentrated near zero for the
background and near unity for the signal.

In the following sections we describe the variables $x_1$--$x_4$
and the two complementary forms we have used for the functions
$f_s({\bf x})$ and $f_b({\bf x})$.

\subsection{Variables}

The four variables $\{x_1,x_2,x_3,x_4\} \equiv {\bf x}$ are defined as
follows:
\begin{eqnarray}
    x_1 & \equiv &  \met        \nonumber \\
    x_2 & \equiv &  \aplan      \\
    x_3 & \equiv &  H_{T2}/H_z  \nonumber \\
    x_4 & \equiv &  \Delta R^{\text{min}}_{jj} \et^{\text{min}} /
                    \et^{W}. \nonumber
\end{eqnarray}
Our use of the variable $x_1$ is motivated by the fact that top quark
events have substantial missing transverse energy, due to the neutrino
from the leptonically-decaying $W$~boson,
while QCD multijet background events
do not. Variable $x_2$ is the aplanarity $\aplan$~\cite{aplan},
which is defined in terms of the normalized momentum tensor of the
jets and the $W$~boson:
\begin{equation}
M_{ab} = \sum_i p_{ia} p_{ib} / \sum_i p_i^2,
\end{equation}
where $\vec p_i$ is the three-momentum of the $i$th object in the
laboratory frame, and $a, b$ run over $x$, $y$, and $z$.
(For this and the remaining two variables, we use all jets
satisfying $\et^{\jet} > 15\gev$ and $|\eta^{\jet}| < 2$.)
The $W$~boson momentum is defined by the sum of the lepton
and neutrino momentum vectors, where the $z$-component of the
neutrino momentum is determined as described in
\secref{event-selection-sub}.
If the
three eigenvalues of $M_{ab}$ are denoted $Q_j$ such that
\begin{equation}
Q_1 \le Q_2 \le Q_3,
\end{equation}
then
\begin{equation}
\aplan = {3\over 2} Q_1.
\end{equation}
This variable is a
measure of the degree to which the final state particles lie out of a
plane. In $W+\jets$ events, a high $\pt$ $W$~boson recoils against a
hadronic system that is typically dominated by a single high $\pt$
jet. In QCD multijet events, two jets, perturbed by gluon radiation, recoil
against each other. The signal, by contrast, has a momentum flow that
is more spherical.  It therefore has a larger aplanarity than do the
backgrounds, which have more longitudinal topologies.
(The aplanarity for top quark events is expected to decrease
with increasing $\mt$ due to the $W$~boson decay products becoming more
collimated.  This effect, however, is very small for $\mt < 200\gevcc$.)

\begin{figure}
\psfig{file=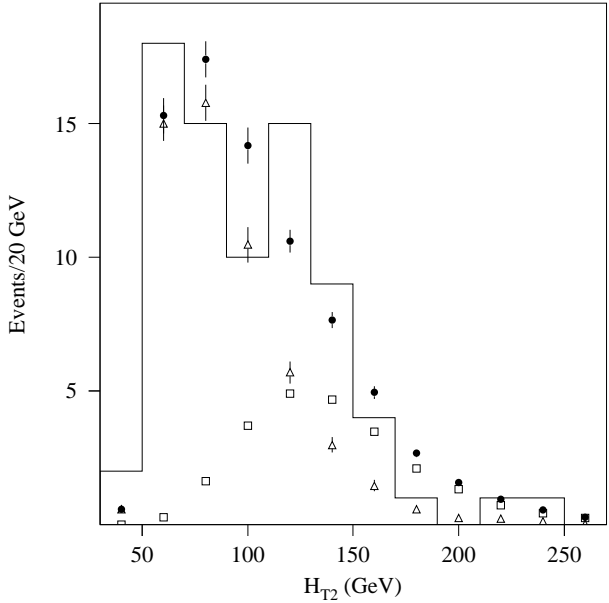,width=\hsize}
\caption{Plot of $H_{T2}$ for the 77-event candidate sample, compared
with the expectation for $m_t=175\gevcc$ signal plus background
(filled circles), signal alone (open squares),
and background alone (open triangles).
(The normalizations are as in \figref{fg:et_l}.)}
\label{fg:ht2}
\end{figure}

The variable $\htran$, as noted above, is a powerful discriminant
between signal and background. But, since both the signal and
background tend to have at least one high $\pt$ jet, we can improve the
discrimination somewhat by removing the highest $\pt$ jet from
$\htran$, yielding $H_{T2}$.  A plot of this variable is shown in
\figref{fg:ht2}.  This variable, however, is correlated with the
fitted mass.  Therefore, we divide by another mass-sensitive variable,
namely $H_z$ (equal to the sum of $|p_z|$ of the lepton, neutrino, and
the jets), in order to reduce that correlation.  The longitudinal component of
the neutrino momentum is found by the same method used to define
$\eta^W$.  We thus arrive at variable $x_3$, which measures the
centrality of the events --- top quark events being more central than
the backgrounds.

The last variable, $x_4$, is motivated by the observation that the
four highest $\et$ jets in top quark events have a different origin
than the jets in $W$+jets and QCD multijet events. For $\ttbar$ events, 
the four highest $\et$~jets are mostly from the decay of the $t\tbar$
system. These jets tend to be widely separated in $\eta-\phi$ space.
For the backgrounds, usually at least one jet is the result
of gluon radiation and is therefore somewhat closer to another jet, on
average, than the jets in $\ttbar$ events.  Therefore, we are led to
consider the six~possible pairs of the four~highest $\et$ jets and
take the pair with the minimum separation
$\Delta R^{\text{min}}_{jj}$ in $\eta-\phi$ space. We then multiply
this minimum separation by the $\et$ of the lesser jet of the pair,
thus constructing a variable akin to the $\pt$ of one jet relative to
another. Again, to reduce the correlation with mass, we divide by
another mass-sensitive variable,
$\et^W \equiv |\et^{\text{lep}}| + |\met|$.

\begin{figure}
\psfig{file=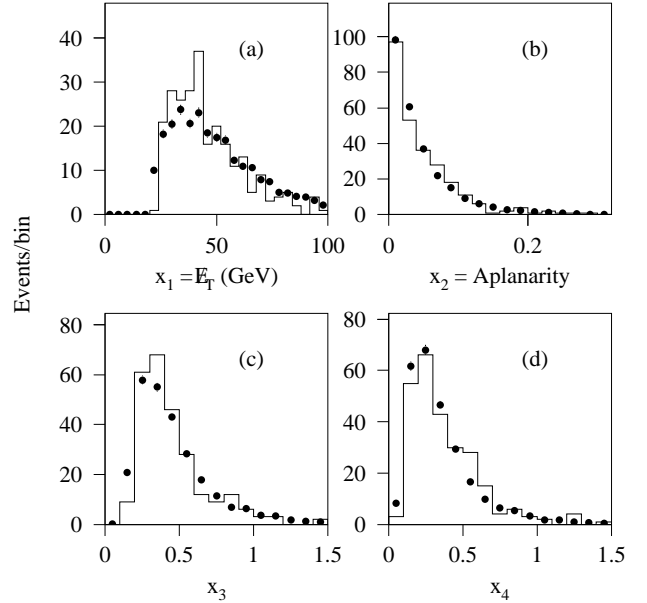,width=\hsize}
\caption{The variables $x_1 \ldots x_4$ used as input to the
top quark discriminants, for $W+3$ jet control samples.  Histograms
are data, and the circles are the expected signal + background mixture.}
\label{fg:w3j}
\end{figure}

\begin{figure}
\psfig{file=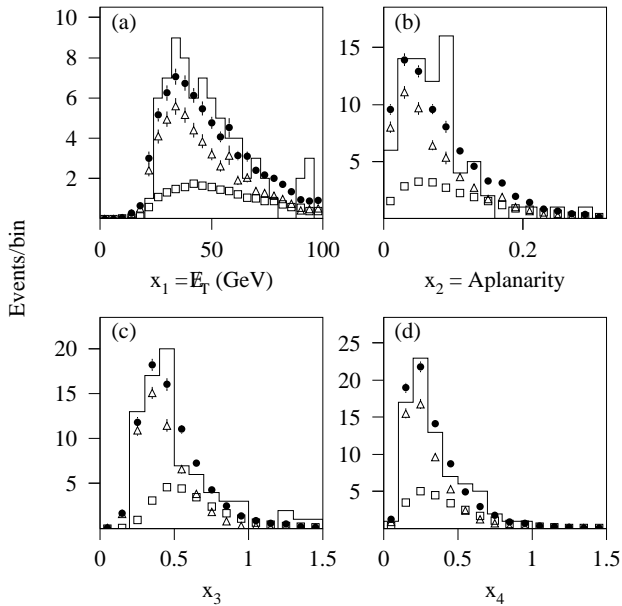,width=\hsize}
\caption{The variables $x_1 \ldots x_4$ used as input to the
top quark discriminants, for the 77-event candidate sample (histogram),
$\ttbar$ signal plus background for $m_t = 175\gevcc$ (filled circles),
signal alone (open squares), and background alone
(open triangles).
(The normalizations are as in \figref{fg:et_l}.)}
\label{fg:discrim-vars}
\end{figure}

We have verified that the variables $x_1$--$x_4$ are well modeled by our
Monte Carlo calculations. \Figref{fg:w3j} shows the observed distributions of
these variables compared with the Monte Carlo predictions for a sample
of $W$+3 jet events, which is dominated by background.  In addition,
\figref{fg:discrim-vars} shows the distributions of these variables
for the 77-event candidate sample, compared with Monte Carlo expectations.
The Monte Carlo models the data well.  We thus use
these variables for the multivariate
discriminants we now describe.

\subsection{Likelihood discriminant}
\label{lb-discrim}

The correlations among the variables
$x_1$--$x_4$ are small. Although we may not conclude that the
variables are, as a consequence, independent, experience shows that it
is frequently true that weakly correlated variables are also nearly
independent. We assume this to be true for $x_1$--$x_4$ and write the
functions $f_s$ and $f_b$ as
\begin{eqnarray}
f_s({\bf x}) & \equiv & \prod_{i=1}^{4} s^{w_i}_{i}(x_i),
\label{eq:dlb-funcs} \\ \nonumber
f_b({\bf x}) & \equiv & \prod_{i=1}^{4} b^{w_i}_{i}(x_i),
\end{eqnarray}
where $s_i(x_i)$ and $b_i(x_i)$ are the normalized distributions of
variable $x_i$ for signal and background, respectively. These forms
reduce to the usual likelihood function for strictly independent
variables when the weights $w_i = 1$. With the weights adjusted
slightly away from unity, we can nullify the correlation between $\mfit$
and the discriminant $\DLB({\bf x})$ formed from \eqsref{eq:discrim-def}
and~(\ref{eq:dlb-funcs}), while maintaining maximal
discrimination between high-mass ($> 170 \gevcc$) top events and the
background.  The subscript ``LB'' (= ``low bias'') denotes the fact
that cuts on $\DLB$ introduce negligible bias (that is, distortion)
in the $\mfit$ distributions.

We have found it useful to have a parameterized form for the discriminant
$\DLB$. Rather than directly parameterizing the functions $f_s$ and
$f_b$, it is simpler to parameterize the ratio ${\cal{L}} \equiv f_s/f_b$
by using polynomial fits to the four
functions ${\cal{L}}_i \equiv s_i(x_i)/b_i(x_i)$ and then computing
${\cal{L}} \equiv \exp \sum_i w_i \ln {\cal{L}}_i$~\cite{vbl-note}.
We then find
$\DLB = {\cal{L}}/(1 + {\cal{L}})$.

We also make use of cuts based on $\DLB$ and $H_{T2}$.
All tagged events pass this
``LB selection''; for untagged events, we require:
\begin{itemize}
 \item $\DLB > 0.43$ and
 \item $H_{T2} > 90\gev$.
\end{itemize}
This selection is used in several places to separate the sample into
signal-rich and background-rich portions.  The cut $\DLB > 0.43$
was chosen to minimize the error on the top quark mass when analyzing
Monte Carlo samples.  The $H_{T2}$ cut removes very little signal
for the top quark masses of interest (see \figref{fg:ht2}), but provides
an easy way of further reducing the background.

\subsection{Neural network discriminant}

\begin{figure}
\psfig{file=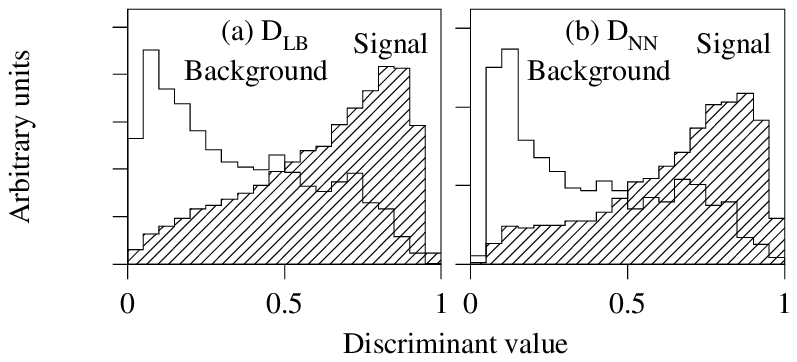,width=\hsize}
\caption{The discriminant variables (a)~$\DLB$ and (b)~$\DNN$ plotted
for the $m_t=175\gevcc$ $\ttbar$ (hatched) sample and
the simulated background (unhatched).
All histograms are normalized to unity.}
\label{fg:dlb-and-dnn}
\end{figure}

\begin{figure}
\psfig{file=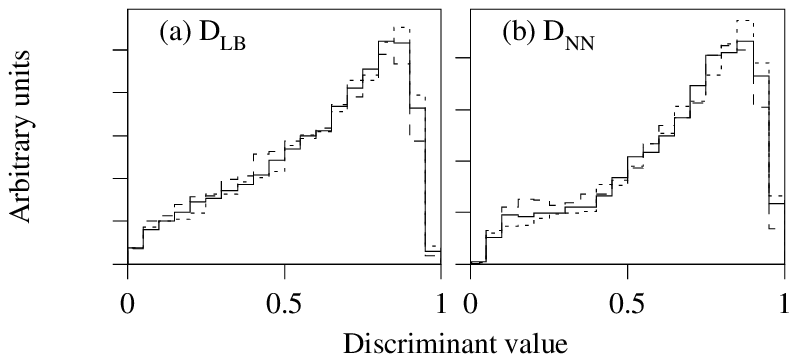,width=\hsize}
\caption{The discriminant variables (a)~$\DLB$ and (b)~$\DNN$ 
for $t\tbar$ Monte Carlo with
$m_t = 150\gevcc$ (dashed lines),
$m_t = 175\gevcc$ (solid lines), and
$m_t = 200\gevcc$ (dotted lines).
All histograms are normalized to unity.}
\label{fg:discrim-vs-mass}
\end{figure}

\begin{figure}
\psfig{file=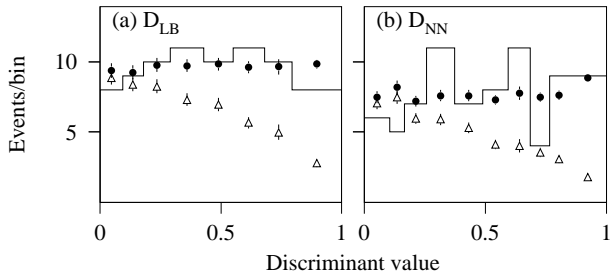,width=\hsize}
\caption{The discriminant variables (a)~$\DLB$ and (b)~$\DNN$ for
the 77-event candidate sample (histogram), $\ttbar$ signal plus
background (filled circles), and background alone (open triangles).
The binnings were chosen such that the predicted signal plus background
distribution would be approximately flat.}
\label{fg:discrim-data}
\end{figure}

The variables $x_1$--$x_4$ were chosen to have minimal correlations
with the fitted mass.
We therefore
consider a second, complementary, discriminant in which no attempt
is made to nullify the correlation between the discriminant and the
fitted mass.  We do attempt, however, to account for the small correlations
that exist among the variables $x_1$--$x_4$.  This discriminant,
denoted by $\DNN$, is calculated with a neural network (NN) having
four input nodes, three hidden nodes, and a single output node, whose
value is $\DNN$. The network is trained using the back-propagation
algorithm provided in the program \progname{JETNET} V3.0 \cite{jetnet}
using the default training parameters.  We use \progname{herwig}
$t\tbar$ Monte Carlo with $m_t = 170\gevcc$ as the signal, and
\progname{vecbos} $W+\jets$ events as the background (equal numbers of
each).  During training, the target outputs are set to unity for the
signal and zero for the background.  Under these conditions, the
network output approximates the ratio $s({\bf x})/[s({\bf x})+ b({\bf
x})]$ \cite{network1}\mynocite{*network2},
where $s({\bf x})$ is the normalized density for
the signal and $b({\bf x})$ is the normalized density for the
background. Since the correlations among $x_1\ldots x_4$ are small,
as are the correlations with
the fitted mass, we should anticipate that the discriminants $\DLB$ and
$\DNN$ will provide comparable levels of signal to background
discrimination. That this is true is evident, qualitatively, from
\figref{fg:dlb-and-dnn} which compares the distributions of $\DLB$ and
$\DNN$ for top quark events and for the mixture of
$W$+jets and QCD multijet events appropriate for the precuts discussed
earlier.  The dependence of the discriminants on the top quark
mass is indeed small, as shown in \figref{fg:discrim-vs-mass}.
In \figref{fg:discrim-data}, we compare the distributions of
the two discriminants obtained from the candidate sample to those
predicted from Monte Carlo; the agreement is quite good.

Analogous to the LB selection, we will also make use of a cut on $\DNN$.
This ``NN selection'' is defined by $\DNN > 0.6$.  This cut value yields
roughly the same discrimination as the LB selection.

\section{Variable-Mass Fit}
\label{mass_fit}

\subsection{Introduction}

The method used can be summarized as follows.  For each event in the
precut sample, we perform a constrained kinematic fit to the hypothesis
$t\tbar\rightarrow l + \jets$ to arrive at a ``fitted mass''
$\mfit$.  Events which fit poorly are discarded.  For
each event, we also compute a top quark discriminant $\calD$ (either $\DLB$
or $\DNN$).  The events are then entered into a two-dimensional
histogram in the $(\calD, \mfit)$ plane.  Similar histograms are also
constructed for a sample of background events and for signal Monte
Carlo at various top quark masses.  For each of these MC masses, we
fit a sum of the signal and background
histograms to the data histogram.  This fit yields a background fraction and a
corresponding likelihood value.  These likelihood values are then
plotted as a function of the top quark mass, and the final result
extracted by fitting a quadratic function to their logarithms.

\subsection{Kinematic fit}

The goal of the kinematic fit is to constrain a measured event to the
hypothesis
\begin{equation}
p\pbar \rightarrow t\tbar + X \rightarrow (W^+b)(W^-\bbar) + X
\rightarrow (l\nu b)(q\qbar\bbar) + X
\end{equation}
(or the charge conjugate)
and thus arrive at an estimate $\mfit$ of the top quark mass.  There
is a complication, however, in that when reconstructing the event, we
do not know \textit{a priori} which observed jet corresponds to which
parton.  In fact, due to QCD radiative effects, jet merging and
splitting during reconstruction, and jet reconstruction
inefficiencies, the observed jets may have no one-to-one
correspondence with the unfragmented partons from the $t\tbar$ decay.
Nevertheless, the fitted mass $\mfit$ constructed from the observed jets
is correlated with the true top
quark mass and can thus be used for a measurement; however, $\mfit$ should
not be thought of as ``the top quark mass'' for a particular event.

The inputs to the fit are the kinematic parameters
of the lepton, the jets, and the missing
transverse energy vector $\vecmet$.  Only the four jets with the largest $\et$
within $|\eta| < 2.5$
are used in the fit (any additional jets are assumed to be due
to initial state radiation).  We parameterize electrons and jets in
terms of energy~$E$, azimuthal angle~$\phi$, and
pseudorapidity~$\eta$.  For muons, we parameterize the momentum in
terms of $k = 1/p$, since the resolution is more nearly Gaussian in
that variable.  The muon direction is also represented as
$(\phi,\,\eta)$.  Leptons and light quarks are fixed to zero mass;
$b$ quarks are fixed to a mass of $5\gevcc$.  The transverse momentum
of the neutrino is taken to be $\vecmet$.  However, we do not use $\vecmet$
directly in the fit, as it is correlated with all the other objects in
the event.  Instead, we use the $x$ and $y$ components of
\begin{equation}
\vec \kt = \vecmet + \vec{E}_T^{\text{lep}} + \sum_{4\ \jets} \vec{E}_T^{\jet}.
\end{equation}
This can be thought of as the transverse momentum of the
$t\tbar$ pair.  Note that this is not necessarily a small quantity if
the event has more than four jets.  One additional variable is
needed to uniquely define the event kinematics: we take that to be
the $z$-component of the neutrino momentum $p_z^\nu$.  This variable
is not measured, but is determined by the fit.  This gives a total of
18~variables.

With this parameterization, there are three kinematic constraints
which can be applied:
\begin{eqnarray}
\label{eq:constraints}
m(t\rightarrow l\nu b) &=& m(\tbar\rightarrow q\qbar\bbar)\nonumber\\
m(l\nu) &=& M_W\\
m(q\qbar) &=& M_W.\nonumber
\end{eqnarray}
Three constraints and one unmeasured variable allow for a 2C
fit.

Since we do not know the correspondence
between jets and partons, we try all twelve distinct assignments of
the four jets to the partons $(b\bbar qq)$.  (But if the event has a
$b$-tag, only the six~permutations in which the tagged jet is used as
a $b$ quark are considered.)  Once a permutation is chosen, we apply
the parton-level and $\eta$-dependent jet corrections described in
\secref{jetcorrections}.  We apply a loose cut on the hadronic $W$
boson mass before the fit: $40 < m(q\qbar) < 140\gevcc$.  Permutations
failing this cut are rejected without being fit in order to speed up
the computation.  We arrange the measured variables into a vector
$\mx^m$ and form the $\chisq$
\begin{equation}
\chisq = (\mx - \mx^m)^T \mG (\mx - \mx^m) ,
\label{eq:chisq}
\end{equation}
where $\mG$ is the inverse error matrix.  This $\chisq$ is then
minimized subject to the kinematic constraints of~\eqref{eq:constraints}.
The minimization algorithm uses the method of Lagrange multipliers;
the nonlinear constraint equations are solved using an iterative
technique.  (The algorithm used is very similar to that of the
\progname{squaw} kinematic fitting program~\cite{squaw1}\mynocite{*squaw2}; a
detailed description may be found in Ref.~\cite{sssthesis}.)  If this
minimization does not converge, the permutation is rejected.  A
permutation is also rejected if $\chisq > 10$.  For each surviving
permutation, this method gives a fitted mass $\mfit$ and a $\chisq$.
We pick the $\mfit$ value corresponding to the smallest $\chisq$ as
$\mfit$ for the event.

There is one additional wrinkle to the above procedure.  In order to
start each fit, we must specify an initial value for the unmeasured
variable $p_z^\nu$.  We choose it so that the two top quarks are assigned
equal mass.  This yields a quadratic equation for $p_z^\nu$.  If the
solutions are complex, the real part is used.  Otherwise, there are
two real solutions.  Both are tried, and the fit which gives the
smaller $\chisq$ is retained.  Note, however, that since $p_z^\nu$
does not enter into the $\chisq$ (its measurement error is effectively
infinite), the only effect its initial value can have on the final
result is to influence which local minimum the fit will find, should there
happen to be more than one.
In the majority of cases, two distinct neutrino solutions yield
nearly the same fit result.

The error matrix $\mG^{-1}$ is taken to be diagonal.  The resolutions
used are given in \tabref{tb:objref}.  (The lepton angular resolutions
are much smaller than the other resolutions, and can be taken to be
effectively zero.)
In most cases, these resolutions were derived
from $t\tbar$ Monte Carlo events by comparing reconstructed objects to
generator-level objects.

\widetext
\begin{table*}
\caption{Object resolutions.  The operator $\oplus$ denotes a sum
in quadrature.}
\begin{tabular}{lccc}
 & Energy resolution & $\sigma(\phi)$ & $\sigma(\eta)$ \\
\tableline
Electrons  &  ${\sigma(\et)/\et} =
                   0.0157 \oplus
                   {0.072\gev^{1/2}/\sqrt{\et}} \oplus
                   {0.66\gev/\et}$
           & %
           & %
             \\
Muons      & $\sigma(1/p) =
               C
\tablenote{$C = 0.0045/(\ugevc)$ if the muon track could be matched
           with a track
           in the central detector; $C = 0.01/(\ugevc)$ otherwise.}
                \oplus {0.2/p}$
           & %
           & %
             \\
Jets&&&\\
\quad $0 < |\deteta| < 0.8$
           & ${\sigma(E)/ E} = 0.036 \oplus
                                       {1.145\gev^{1/2}/\sqrt{E}}$
           & $0.04\>\text{rad}$
           & $0.04$ \\
\quad $0.8 < |\deteta| < 1.4$
           & ${\sigma(E)/ E} = 0.082 \oplus
                                        {1.264\gev^{1/2}/\sqrt{E}}$
           & $0.05\>\text{rad}$
           & $0.05$ \\
\quad $1.4 < |\deteta| < 2.0$
           & ${\sigma(E)/ E} = 0.046 \oplus
                                        {1.305\gev^{1/2}/\sqrt{E}}$
           & $0.05\>\text{rad}$
           & $0.05$ \\
$\kt$      & $\sigma({\kt}_x) = \sigma({\kt}_y) = 12\gev$\\
\end{tabular}
\label{tb:objref}
\end{table*}
\narrowtext

\begin{figure}
\psfig{file=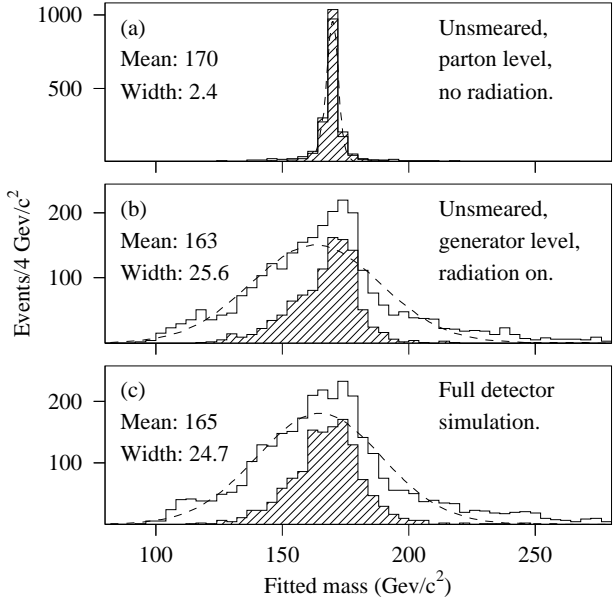,width=\hsize}
\caption{Tests of kinematic fit method on $t\tbar$ Monte Carlo
samples ($m_t = 170\gevcc$, $e+\jets$ channel).  (a) Using
\progname{herwig} partons directly.  (b) Final state Monte Carlo
particles, after clustering into $R=0.5$ cones.  (c) After full
detector simulation and reconstruction.  The hatched plots show the
results for the correct jet permutation (regardless of whether or not it
has the lowest $\chisq$).
Displayed means and widths
are from a Gaussian fit, shown by the dashed curve.}
\label{fg:mc-tests}
\end{figure}

Results of this procedure on Monte Carlo $t\tbar$ samples are shown in
\figref{fg:mc-tests}.  \Figref{fg:mc-tests}(a) shows results using
the \progname{herwig} partons directly, before any
QCD~evolution has taken place.  A rather sharp peak is seen; further,
about $80\%$ of the time, the permutation with the lowest $\chisq$ is
the one which is actually correct.  The residual width seen in the
plot is due mainly to the non-zero widths of the $W$ bosons.
\Figref{fg:mc-tests}(b) shows results from the same sample, but after
QCD~evolution and jet fragmentation.  The final state particles are
clustered together in cones of width
$\Delta R=0.5$ in order to simulate the action of
the jet reconstruction algorithm.  This distribution is
considerably broader.  There are fewer events in the
hatched plot because it is not always possible to uniquely define the
correct permutation.  Due to splitting and merging effects, jet
finding inefficiencies, and jets falling below the selection
threshold, the correct permutation can be uniquely identified in
only about $50\%$ of events.  In that case, the correct permutation is the
lowest $\chisq$ permutation about $40\%$ of the time.  Finally,
\figref{fg:mc-tests}(c) shows results for a sample which has been
through the full detector simulation and reconstruction.  The
resulting distribution has essentially the same width as that of
\figref{fg:mc-tests}(b); this indicates that the dominant contribution
to the width of this distribution comes from QCD radiation and jet
combinatoric effects, and not from the detector resolution.

\begin{figure}
\psfig{file=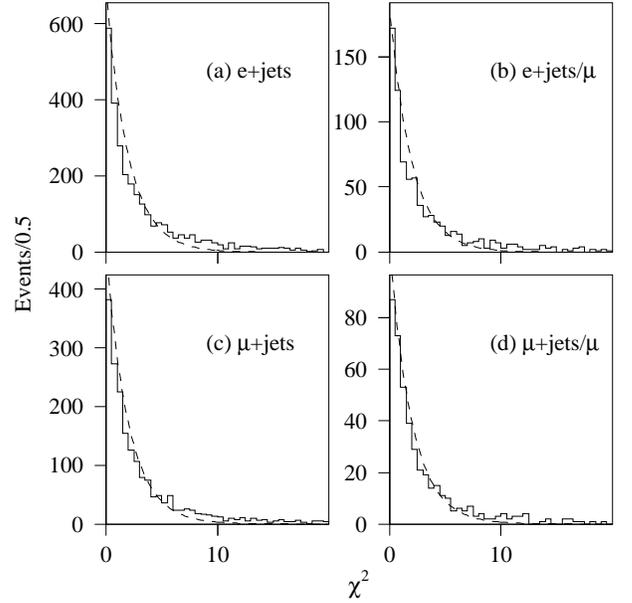,width=\hsize}
\caption{Fit $\chisq$ distributions for the correct jet permutation
for $t\tbar$ Monte Carlo samples ($m_t = 170\gevcc$).  The dashed curve
is the $\chisq$ distribution for two degrees of freedom, normalized to
the area of the histogram.}
\label{fg:chisq-corr}
\end{figure}

\begin{figure}
\psfig{file=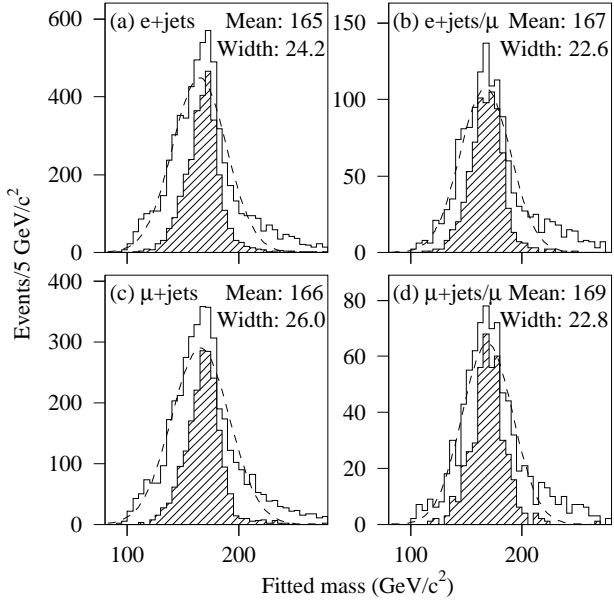,width=\hsize}
\caption{Fitted mass distributions
for $t\tbar$ Monte Carlo samples ($m_t = 170\gevcc$)
for the jet permutation with the lowest $\chisq$.
Hatched
histograms show the results for the correct jet permutation
(regardless of whether or not it has the lowest $\chisq$).
Displayed means and widths
are from a Gaussian fit, shown by the dashed curve.}
\label{fg:mt-vs-channel}
\end{figure}

The (MC) fit $\chisq$ distributions resulting from the fit to the correct
jet permutation are shown in \figref{fg:chisq-corr}.  The
distributions agree reasonably well with the expectations for a two
degree-of-freedom $\chisq$, except for a tail at the high end due to
non-Gaussian tails in the resolutions.  The (MC) $\mfit$ distributions for
the four channels are shown in \figref{fg:mt-vs-channel}.

\begin{figure}
\psfig{file=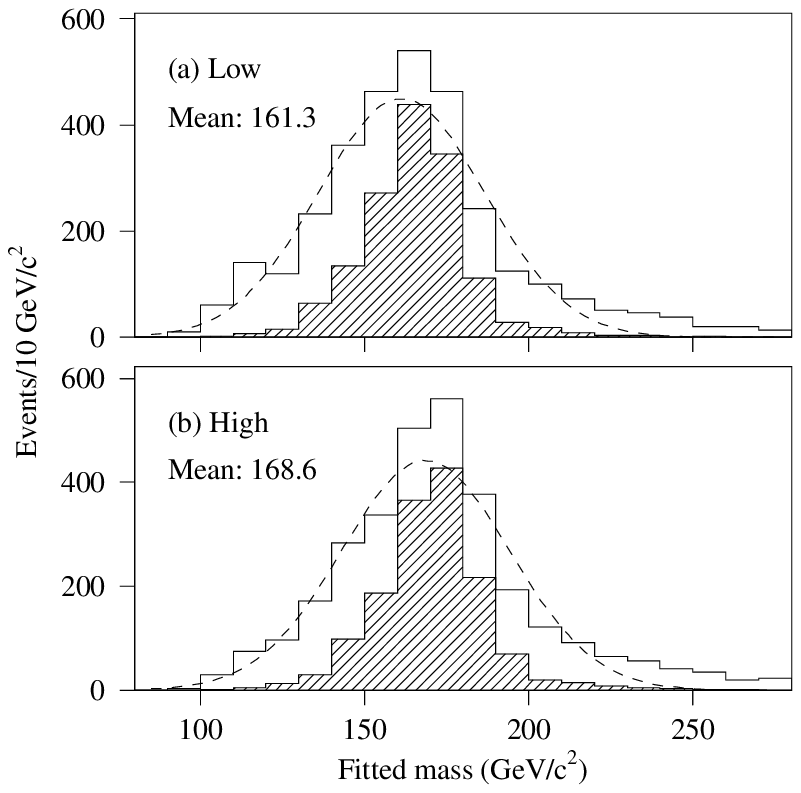,width=\hsize}
\caption{Fitted mass distributions
for $t\tbar$ Monte Carlo samples ($m_t = 170\gevcc$, $e+\jets$
channel).  With jets scaled (a) down and (b) up by $2.5\%+0.5\gev$.
Hatched histograms show the results for the correct jet permutation
(regardless of whether or not it has the lowest $\chisq$).
Displayed means 
are from a Gaussian fit, shown by the dashed curve.}
\label{fg:jet-scalecomp}
\end{figure}

\Figref{fg:jet-scalecomp} shows the distributions which result after
the jets in each Monte Carlo event are scaled up or down by the
per-jet systematic error of $2.5\%+0.5\gev$.  This
shifts the fitted mass by approximately $\pm 3.7\gevcc$.

\begin{figure}
\psfig{file=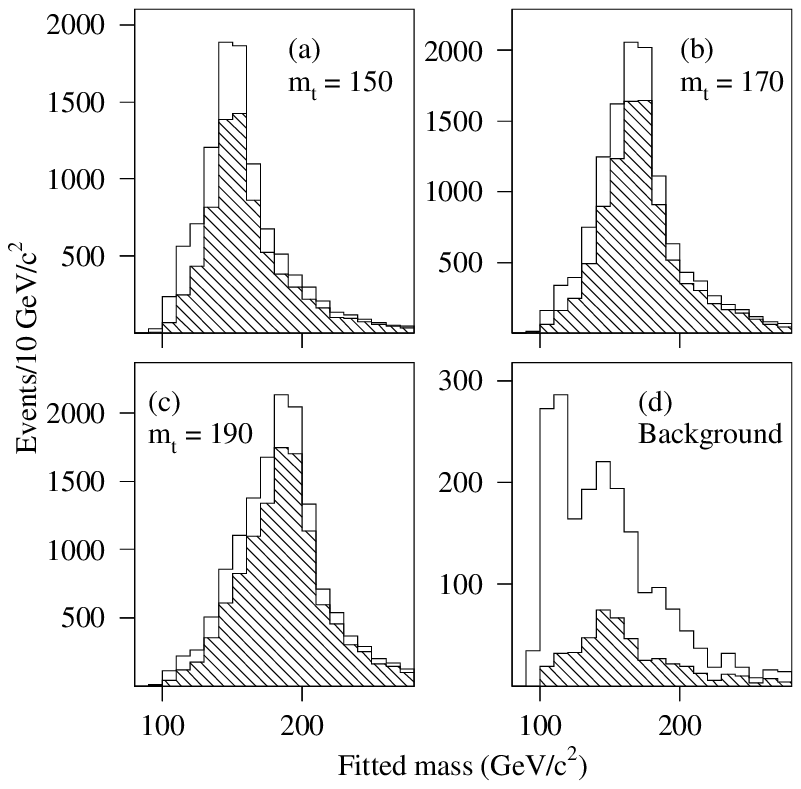,width=\hsize}
\caption{Fitted mass distributions, all channels combined.
Shown is $t\tbar$ Monte Carlo with (a) $m_t = 150\gevcc$,
(b) $m_t = 170\gevcc$, and (c) $m_t = 190\gevcc$ and (d) background.
The hatched distributions are after the LB selection is applied.}
\label{fg:mt-vs-mass}
\end{figure}

\Figref{fg:mt-vs-mass} shows the fitted mass distribution for several
top quark masses and for the background.

\begin{figure}
\psfig{file=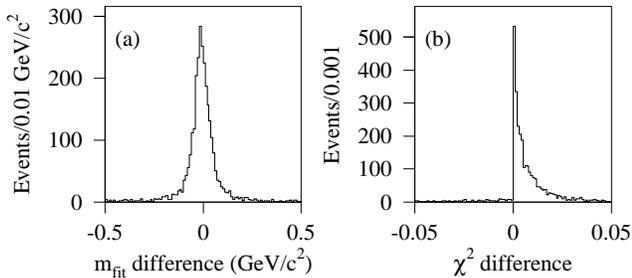,width=\hsize}
\caption{Differences between the results obtained from
the \progname{minuit}-based fitter 
and 
the Lagrange-multiplier based fitter 
for (a) $\mfit$
and (b) $\chisq$.  (For $t\tbar$ Monte
Carlo with $m_t = 170\gevcc$, $e+\jets$ channel.)}
\label{fg:minuitcomp}
\end{figure}

A possible objection to the fit method described here is that it does
not take into account the intrinsic widths of the $W$~boson and
top~quark decays.  To investigate this, an alternate fitting
method was tried which explicitly incorporates these widths.  This
method is based on a standard unconstrained minimization package
(\progname{minuit}~\cite{minuit}).  The quantity minimized is
the $\chisq$ as defined in \eqref{eq:chisq} with three
Breit-Wigner constraint terms added: two for the two $W$~bosons,
and one for the top quark mass difference:
\begin{eqnarray}
{\chisq}' = \chisq &-& 2 \ln {\Gamma_W^2/4\over \Gamma_W^2/4 +(m(l\nu)-M_W)^2} 
                                                                   \nonumber \\
                   &-& 2 \ln {\Gamma_W^2/4\over \Gamma_W^2/4 + (m(q\qbar)-M_W)^2}\\
                   &-& 2 \ln {\Gamma_t^2\over \Gamma_t^2 +
                            (m(l\nu b)-m(q\qbar\bbar))^2}. \nonumber
\end{eqnarray}
(The factor of~4 difference in the last term comes from convoluting two
Breit-Wigner functions centered on $m(l\nu b)$ and $m(q\qbar\bbar)$.)
The $W$~boson width is taken to be $2\gevcc$.  The top quark width is
taken to depend on the mass as $\Gamma_t = (\alpha m_t)^3$; the
proportionality constant $\alpha$ is set so that $\Gamma_t = 0.6\gevcc$
at $m_t = 140\gevcc$.  (Here, $m_t = (m(l\nu b) + m(q\qbar\bbar))/2$.)
These widths are small compared to the experimental resolutions.
The results of this procedure are compared to those from the
Lagrange-multiplier based fitter in \figref{fg:minuitcomp}.  In most
cases, the results are nearly identical, implying that neglecting the
widths is not a serious problem.  Since this
algorithm takes several times longer to execute, it is not used further.

\subsection{Likelihood fit}

The next problem to be solved is the extraction of the top quark mass from
the data sample, which is a mixture of signal and background.  This is
done using a binned Poisson-statistics maximum-likelihood fit at
discrete top quark masses.  (The method is described in more detail
in Ref.~\cite{bayespl}.)

We bin the data according to some
characteristics of the events.  (For this analysis, we will be using
$\mfit$ and either $\DLB$ or $\DNN$.)  Call the number of bins $M$,
the total
number of events $N$, and the number of events in each bin $N_j$.

We also know the distribution expected for different values of the
top quark mass, and also for the background.  (This is from Monte Carlo
except for the QCD multijet background.)
For both the signal and background, we
have a distribution of events among the $M$ bins; call the
numbers of events in each bin of these distributions $A^s_j$
and $A^b_j$.

We regard these distributions as drawn
from ``true'' distributions $a^s_j$ and $a^b_j$, and write
the probability for seeing the observed data set $D$ given these parameters
as a Poisson likelihood
\begin{equation}
L(D|A, a, p) = \prod_{j=1}^M q(N_j, p_s a^s_j + p_b a^b_j)
                    \, q(A^s_j, a^s_j) \, q(A^b_j, a^b_j),
\label{eq:likedef}
\end{equation}
where $q$ is the Poisson distribution
$q(N, a) \equiv e^{-a} a^N / N!$
and $p_s$ and $p_b$ are the signal and background
strengths.  These strengths can be related to the number of expected
events
$n_s$ and $n_b$ by $p_s = n_s / (M + \sum_j A^s_j)$, and similarly for
$n_b$.
(The $M$ term in the denominator ensures that the sum of the
maximum likelihood estimates for $n_s$ and $n_b$ equals $N$.
See Ref.~\cite{bayespl} for further discussion.
Note that usually $M \ll \sum_j A_j$.)
The total number of events expected is thus
$n_j = p_s a^s_j + p_b a^b_j$.
We eliminate the $a_j$'s from this likelihood by integrating over
them; the result is
\begin{eqnarray}
L(D|A, p) &=& \prod_{j=1}^M \sum_{k=0}^{N_j}
  {p_s^{k} \over (1 + p_s)^{A^s_j + k + 1}}
  {p_b^{N_j - k} \over (1 + p_b)^{A^b_j + N_j - k + 1}} \nonumber \\
  &&\times
  {A^s_j + k \choose k}
  {A^b_j + N_j - k \choose N_j - k}.
\end{eqnarray}
Following Ref.~\cite{baker}, we then modify the likelihood by dividing by the
constant factor
\begin{equation}
\prod_j q(N_j, N_j) .
\end{equation}
This has the effect of making the quantity $- 2 \ln L$
behave asymptotically like a $\chisq$ distribution.
(Note, however, that for our
experiment, the sample size is too small for this asymptotic behavior
to be accurately realized.)

We now have a set of signal models,
each corresponding to a different top quark mass $m_t$.  For each
signal model, we fit it plus the background to the data, yielding
$n_s$ and $n_b$.  A maximum likelihood fit is used, based on
\progname{minuit}~\cite{minuit}.  The minimum value of $- \ln L$
is retained; call this $\lnlmin$.  The resulting values of
$(m_t, \lnlmin)$ then define a likelihood curve as a function of top quark
mass.

We also define a statistical error on $\lnlmin$ due to the finite
Monte Carlo statistics.  This is done by the simple method of taking
in turn each bin $j$ in the input Monte Carlo histograms, varying the
contents up or down by $\sqrt{A_j}$, and re-evaluating the likelihood.
(To save time, the fit for $n_s$ and $n_b$ is not redone for each
variation; early testing showed it to make very little difference.)
The resulting variations in $\lnlmin$
for each bin are then added in quadrature.
This error is calculated separately for the signal and background
samples; however, any effects from fluctuations in the background
sample will be highly correlated from mass point to mass point.  Thus,
the errors shown on the plots and used in the fit below come from the
signal samples only.

The final step is to extract a mass value from this set of $(m_t,\lnlmin)$
points.  This is done by fitting a quadratic function
to the smallest $\lnlmin$ and the four closest points on each side.
The points are
weighted by the statistical errors assigned to the $\lnlmin$ values.  The
position of the minimum of this quadratic defines the mass estimate,
and its width (where the curve has risen by~$0.5$) gives an error
estimate.  We also want estimates for $n_s$ and $n_b$.  For each mass
$m_t$, we have a separate estimate for $n_s$ and $n_b$ returned from
\progname{minuit}.  The final estimates of these values are determined
by a linear interpolation between the two points bracketing the final
$m_t$ estimate.  The errors are found in the same manner.

For comparison, some results are also given using 11~points instead of 9
for the polynomial fit, and using a cubic function instead of a quadratic one.

\subsection{Fitting variables and binning}
\label{vbls-and-binning}

From each event, we derive two variables: the fitted mass
$\mfit$ and a discriminant $\calD$.  We use these variables to
bin the data into a two-dimensional histogram.  The top quark mass is
then extracted from a fit to the expectations from Monte Carlo, as
described in the previous section.

\begin{figure}
\psfig{file=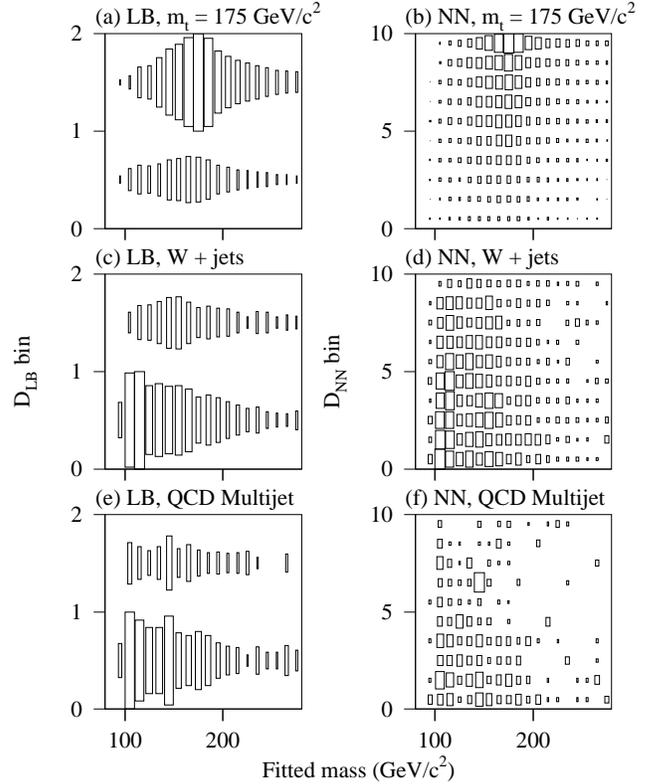,width=\hsize}
\caption{Monte Carlo histograms for LB and NN analyses
for $t\tbar$ Monte Carlo with $m_t = 175\gevcc$, \progname{vecbos}
$W+\jets$ background, and QCD multijet background.  More top~quark-like
events are towards the top of the plots.}
\label{fg:resfunc2d}
\end{figure}

Two different discriminants and histogram binnings are used.  For
both binnings, the fitted mass axis has twenty bins of width $10\gevcc$ over
the range $80$ to $280\gevcc$.  They differ in the definition of the
discriminant axis.  For the ``LB'' analysis, the discriminant axis is
divided into two bins, the first bin containing events which fail the
LB selection (as defined in \secref{lb-discrim}), and the second containing
events which pass it.  (Recall that all tagged events pass the LB selection.)
For the ``NN'' analysis, the
discriminant axis is the NN variable $\DNN$.
(Note that tagging information is not used in forming $\DNN$.)
There are ten unevenly
spaced bins, as defined in \tabref{tb:nnbins}.  These bin boundaries were
chosen so that the expected signal + background distribution populates
the bins approximately uniformly.
There are thus 40 bins in the LB binning, and 200 bins in the NN binning.
Examples of the resulting histograms are shown 
in \figref{fg:resfunc2d}.

\begin{table}
\caption{NN bin definitions.}
\begin{tabular}{rc}
Bin    &  $\DNN$ range \\
\tableline
 1&  $0.000$ -- $0.105$ \\
 2&  $0.105$ -- $0.166$ \\
 3&  $0.166$ -- $0.257$ \\
 4&  $0.257$ -- $0.373$ \\
 5&  $0.373$ -- $0.488$ \\
 6&  $0.488$ -- $0.595$ \\
 7&  $0.595$ -- $0.687$ \\
 8&  $0.687$ -- $0.766$ \\
 9&  $0.766$ -- $0.846$ \\
10&  $0.846$ -- $1.000$ \\
\end{tabular}
\label{tb:nnbins}
\end{table}

These histograms are generated separately for each of the four channels.
They are then combined using the set of fixed weights given in
\tabref{tb:chanweights}.  We derive these numbers by calculating
the expected signal and background in each channel using the same
techniques as used for the cross section measurement~\cite{xsecprl}
(except that only the precuts are applied).
We also combine the histograms for
\progname{vecbos} $W+\jets$ background and the QCD multijet background
using a fixed QCD fraction of $(22\pm 5)\%$, derived in the same manner.

\mediumtext
\begin{table*}
\caption{Fraction of events expected in each channel after the precuts.}
\begin{tabular}{lcccc}
  & $e+\jets$        & $e+\jets/\mu$    & $\mu+\jets$      & $\mu+\jets/\mu$\\
\tableline
\progname{herwig} $\ttbar$ &&&&\\
\quad $110\text{--}150\gevcc$
  & $0.376\pm 0.020$ & $0.085\pm 0.013$ & $0.468\pm 0.025$ & $0.071\pm 0.018$\\
\quad $155\text{--}170\gevcc$
  & $0.418\pm 0.018$ & $0.097\pm 0.011$ & $0.425\pm 0.021$ & $0.059\pm 0.015$\\
\quad $172\text{--}190\gevcc$
  & $0.427\pm 0.016$ & $0.093\pm 0.010$ & $0.409\pm 0.019$ & $0.071\pm 0.013$\\
\quad $195\text{--}230\gevcc$
  & $0.416\pm 0.014$ & $0.097\pm 0.009$ & $0.419\pm 0.018$ & $0.068\pm 0.012$\\
\progname{vecbos}
  & $0.531\pm 0.077$ & $0.015\pm 0.017$ & $0.441\pm 0.079$ & $0.013\pm 0.003$\\
QCD
  & $0.443\pm 0.111$ & $0.013\pm 0.030$ & $0.488\pm 0.115$ & $0.056\pm 0.020$\\
\end{tabular}
\label{tb:chanweights}
\end{table*}
\narrowtext

\subsection{Fits to data}
\label{fits-to-data}

\begin{figure}
\psfig{file=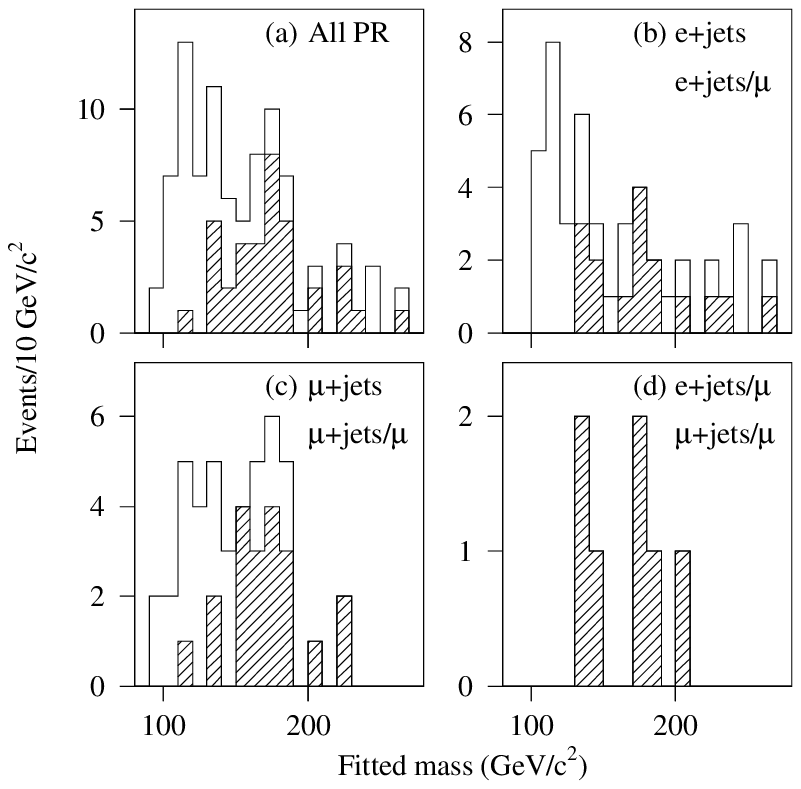,width=\hsize}
\caption{Fitted mass distributions for candidate events.
The hatched histograms show the LB subsample.}
\label{fg:datanocut}
\end{figure}

\begin{figure}
\psfig{file=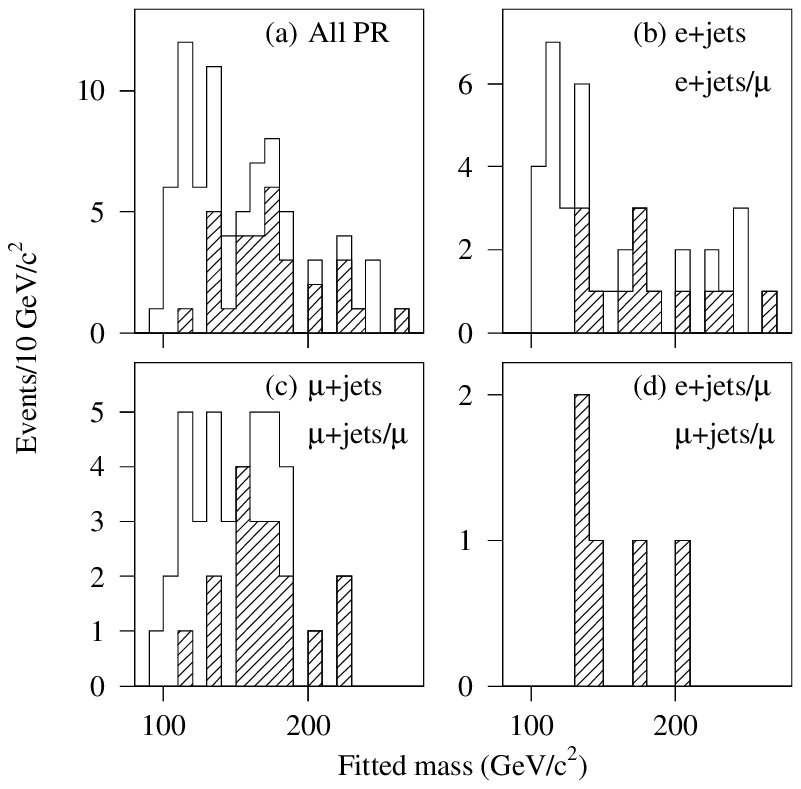,width=\hsize}
\caption{Fitted mass distributions for candidate events with $\chisq < 10$.
The hatched histograms show the LB subsample.}
\label{fg:datacut}
\end{figure}

\begin{figure}
\psfig{file=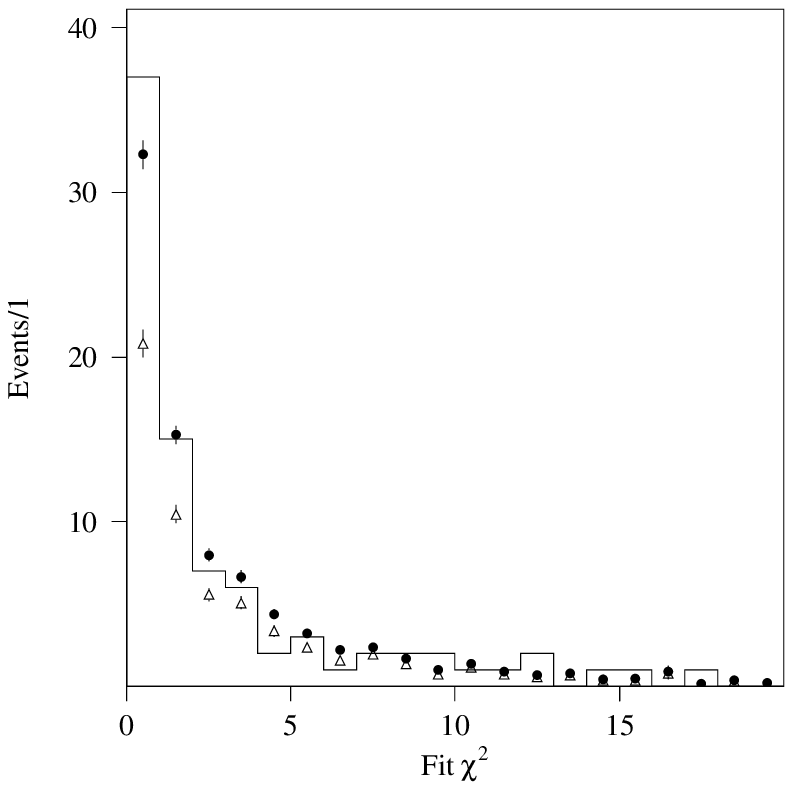,width=\hsize}
\caption{Fit $\chisq$ distribution from data (histogram), the
expected $\ttbar$ signal + background (filled circles), and background
alone (open triangles).}
\label{fg:chisq-comp}
\end{figure}

The results of the kinematic fit for the candidate events are given in
\tabsref{tb:candtable-ej} through~\ref{tb:candtable-mjm}.  (Complete
details of the candidate events are available in Ref.~\cite{cand-details}.)
There
are $91$ events passing the precuts (PR).  One of these, however, had no
successful fits, and is not considered further.  Thirty-six of these events
then pass the LB selection.  The distributions of the fitted masses of these
candidates are shown in \figref{fg:datanocut}.  When the $\chisq < 10$
cut is imposed, there are $77$~PR events and $31$~LB events.
Distributions of their fitted masses are shown in \figref{fg:datacut}.  The
$\chisq$ distribution of the 90~events is shown in
\figref{fg:chisq-comp}.  It compares well to the
expectation from Monte Carlo.

\begin{table}
\caption{Kinematic fit results and top quark discriminants
for events in the $e + \jets$ channel for the jet permutation
with the smallest $\chisq$.
The ``Perm'' column gives the assignment of the jets to partons, listed
in order of decreasing jet $\et$: $B_l$ and $B_h$ denote the
$b$~quarks associated with the leptonically and hadronically decaying
top~quarks, respectively, while $W$ denotes the quarks from the
hadronically decaying $W$ boson.  The fitted mass $\mfit$ is in
$\ugevcc$.}
\begin{tabular}{rrrdddd}
Run & Event & Perm. & $\mfit$     & $\chisq$ & $\DLB$ & $\DNN$ \\
\tableline
\tablenotemark[3]
62199&15224&$B_lWWB_h$&265.4&15.9&0.09&0.21\\
\tablenotemark[2]
\tablenotemark[3]
62431&  788&$WB_hB_lW$&241.7&0.23&0.16&0.09\\
\tablenotemark[1]
\tablenotemark[2]
\tablenotemark[3]
63066&13373&$B_lWB_hW$&206.8&1.35&0.85&0.95\\
\tablenotemark[2]
\tablenotemark[3]
64464&21611&$B_hWWB_l$&115.7&0.64&0.22&0.31\\
\tablenotemark[1]
\tablenotemark[2]
\tablenotemark[3]
81949&12380&$B_lB_hWW$&132.7&1.10&0.77&0.82\\
\tablenotemark[2]
\tablenotemark[3]
82024&44002&$WB_lB_hW$&130.2&0.97&0.06&0.31\\
\tablenotemark[2]
\tablenotemark[3]
82220&20012&$B_hWB_lW$&120.8&2.53&0.03&0.06\\
82996&24461&$WB_hWB_l$&166.8&31.8&0.73&0.74\\
84331&13271&$B_hWB_lW$&116.8&14.4&0.25&0.27\\
\tablenotemark[2]
\tablenotemark[3]
84890&28925&$B_hWB_lW$&126.4&0.78&0.06&0.07\\
\tablenotemark[1]
\tablenotemark[2]
\tablenotemark[3]
85917&   22&$B_lWWB_h$&162.3&2.26&0.79&0.81\\
\tablenotemark[2]
\tablenotemark[3]
86518&11716&$B_hWWB_l$&243.5&0.54&0.18&0.29\\
\tablenotemark[1]
\tablenotemark[2]
\tablenotemark[3]
86601&33128&$WB_lB_hW$&179.2&0.39&0.43&0.29\\
\tablenotemark[1]
\tablenotemark[2]
\tablenotemark[3]
87063&39091&$B_hWB_lW$&188.4&0.39&0.58&0.63\\
\tablenotemark[2]
\tablenotemark[3]
87104&25823&$WB_hB_lW$&119.9&2.11&0.06&0.09\\
\tablenotemark[2]
\tablenotemark[3]
87329&13717&$B_hB_lWW$&242.1&1.95&0.39&0.23\\
\tablenotemark[2]
\tablenotemark[3]
87446&14294&$WWB_hB_l$&118.3&1.11&0.59&0.52\\
88038&14829&$WWB_hB_l$&101.0&12.8&0.37&0.28\\
\tablenotemark[3]
88044& 9807&$WB_lWB_h$&145.2&34.0&0.09&0.11\\
\tablenotemark[1]
\tablenotemark[2]
\tablenotemark[3]
88045&35311&$WB_hWB_l$&178.2&2.71&0.83&0.81\\
\tablenotemark[2]
\tablenotemark[3]
88125&15437&$WB_hWB_l$&115.9&0.16&0.78&0.74\\
\tablenotemark[2]
\tablenotemark[3]
88463& 3627&$WB_hWB_l$&111.7&9.93&0.16&0.46\\
\tablenotemark[2]
88588&15993&$WWB_hB_l$&103.4&7.44&0.29&0.30\\
\tablenotemark[1]
\tablenotemark[2]
\tablenotemark[3]
89484&11741&$B_hB_lWW$&135.0&0.76&0.53&0.58\\
\tablenotemark[2]
\tablenotemark[3]
89550&18042&$WWB_hB_l$&103.5&0.07&0.30&0.27\\
\tablenotemark[1]
89708&24871&$WB_hB_lW$&144.6&20.1&0.62&0.74\\
\tablenotemark[1]
\tablenotemark[2]
\tablenotemark[3]
89936& 6306&$WB_hB_lW$&220.4&1.29&0.50&0.68\\
\tablenotemark[1]
\tablenotemark[2]
\tablenotemark[3]
89972&13657&$WB_hB_lW$&176.7&9.08&0.65&0.77\\
\tablenotemark[2]
\tablenotemark[3]
90108&31611&$WB_hWB_l$&137.4&0.41&0.21&0.21\\
\tablenotemark[2]
\tablenotemark[3]
90435&32258&$B_hB_lWW$&154.1&1.05&0.27&0.62\\
\tablenotemark[2]
\tablenotemark[3]
90496&28296&$B_hWB_lW$&112.9&0.28&0.23&0.19\\
\tablenotemark[2]
90693& 8678&$B_hB_lWW$&105.5&8.98&0.51&0.27\\
\tablenotemark[3]
90795&14246&$B_hWB_lW$&193.9&12.8&0.09&0.07\\
\tablenotemark[2]
\tablenotemark[3]
90804& 6474&$WWB_lB_h$&114.2&0.64&0.34&0.59\\
\tablenotemark[2]
\tablenotemark[3]
91923&  502&$WB_hWB_l$&162.1&0.14&0.09&0.15\\
\tablenotemark[2]
\tablenotemark[3]
92013&11825&$WB_hB_lW$&134.1&3.68&0.11&0.15\\
\tablenotemark[2]
\tablenotemark[3]
92217&  109&$WWB_hB_l$&107.8&0.58&0.77&0.82\\
\tablenotemark[2]
\tablenotemark[3]
92278&21744&$WB_hWB_l$&125.9&7.26&0.17&0.31\\
\tablenotemark[1]
\tablenotemark[2]
\tablenotemark[3]
92673& 4679&$B_lB_hWW$&267.7&1.85&0.92&0.97\\
\tablenotemark[2]
\tablenotemark[3]
94750& 4683&$B_hWWB_l$&201.5&3.63&0.32&0.49\\
\tablenotemark[1]
96329&13811&&---&          &0.54&0.79\\
\tablenotemark[2]
\tablenotemark[3]
96676&79957&$WB_lB_hW$&224.1&0.47&0.36&0.46\\
\tablenotemark[1]
\tablenotemark[2]
\tablenotemark[3]
96738&27592&$B_lWWB_h$&236.6&5.68&0.60&0.83\\
\end{tabular}
\tablenotetext[1]{Passes LB selection.}
\tablenotetext[2]{Used in variable-mass analysis.}
\tablenotetext[3]{Used in pseudolikelihood analysis.}
\label{tb:candtable-ej}
\end{table}

\begin{table}
\caption{Same as \tabref{tb:candtable-ej} for the $\mu + \jets$ channel.}
\begin{tabular}{rrrdddd}
Run & Event & Perm. & $\mfit$     & $\chisq$ & $\DLB$ & $\DNN$ \\
\tableline
\tablenotemark[2]
\tablenotemark[3]
61514& 4537&$B_hWB_lW$&120.8&3.40&0.26&0.59\\
\tablenotemark[1]
\tablenotemark[2]
\tablenotemark[3]
63183&13926&$WWB_hB_l$&133.7&1.26&0.84&0.83\\
\tablenotemark[1]
\tablenotemark[2]
\tablenotemark[3]
63740&14197&$B_lWB_hW$&185.3&2.56&0.94&0.96\\
\tablenotemark[2]
\tablenotemark[3]
80703&31477&$WB_hB_lW$&167.2&0.54&0.24&0.40\\
\tablenotemark[1]
\tablenotemark[2]
\tablenotemark[3]
81909&11966&$B_hWB_lW$&162.9&1.11&0.67&0.66\\
\tablenotemark[2]
81949&13778&$WB_hWB_l$&109.2&8.25&0.27&0.25\\
\tablenotemark[2]
\tablenotemark[3]
82639&11573&$WB_lWB_h$&117.3&2.24&0.35&0.47\\
\tablenotemark[1]
\tablenotemark[2]
\tablenotemark[3]
82694&25595&$WB_lWB_h$&114.0&2.03&0.56&0.53\\
\tablenotemark[1]
\tablenotemark[2]
\tablenotemark[3]
84696&29253&$WB_hB_lW$&221.0&1.05&0.74&0.89\\
\tablenotemark[2]
\tablenotemark[3]
84728&18171&$B_hB_lWW$&136.0&3.65&0.40&0.38\\
\tablenotemark[2]
\tablenotemark[3]
85888&28599&$B_hWWB_l$&189.6&5.78&0.18&0.09\\
\tablenotemark[1]
\tablenotemark[2]
\tablenotemark[3]
87063&14368&$WWB_hB_l$&182.1&0.02&0.50&0.72\\
87604&14282&$B_lWWB_h$& 90.6&40.6&0.14&0.38\\
\tablenotemark[1]
\tablenotemark[3]
87820& 6196&$B_hB_lWW$&178.0&17.8&0.87&0.97\\
\tablenotemark[1]
\tablenotemark[2]
\tablenotemark[3]
88464& 2832&$B_hWB_lW$&154.1&0.14&0.87&0.93\\
\tablenotemark[1]
\tablenotemark[2]
\tablenotemark[3]
88530& 7800&$WB_lB_hW$&151.2&0.08&0.62&0.60\\
88597& 1145&$WWB_hB_l$&124.6&10.2&0.20&0.42\\
\tablenotemark[2]
\tablenotemark[3]
88603& 2131&$WB_lWB_h$&123.7&0.66&0.13&0.17\\
\tablenotemark[2]
\tablenotemark[3]
89751&27345&$B_hWWB_l$&132.4&1.14&0.15&0.14\\
\tablenotemark[1]
\tablenotemark[2]
\tablenotemark[3]
89943&19016&$WB_hB_lW$&163.7&0.03&0.65&0.74\\
\tablenotemark[2]
\tablenotemark[3]
90133&14110&$WB_hWB_l$&169.4&4.88&0.26&0.28\\
\tablenotemark[1]
\tablenotemark[2]
\tablenotemark[3]
90660&20166&$WB_lB_hW$&222.6&1.28&0.70&0.90\\
\tablenotemark[1]
\tablenotemark[2]
\tablenotemark[3]
90690&12392&$B_hWB_lW$&153.3&0.58&0.70&0.78\\
\tablenotemark[2]
\tablenotemark[3]
90836&14924&$WB_lWB_h$&147.4&3.13&0.07&0.08\\
\tablenotemark[2]
\tablenotemark[3]
90864&17697&$WB_hWB_l$& 96.6&0.81&0.44&0.62\\
\tablenotemark[2]
\tablenotemark[3]
91359&15030&$WB_hWB_l$&118.9&1.81&0.54&0.60\\
\tablenotemark[2]
\tablenotemark[3]
92081& 3825&$WB_hB_lW$&117.7&3.72&0.07&0.40\\
\tablenotemark[2]
\tablenotemark[3]
92082&34466&$WB_hB_lW$&176.2&0.30&0.31&0.49\\
\tablenotemark[1]
92114& 1243&$WB_lB_hW$&187.0&11.7&0.96&0.96\\
\tablenotemark[1]
\tablenotemark[2]
\tablenotemark[3]
92126&21544&$B_lWWB_h$&157.2&0.02&0.82&0.91\\
\tablenotemark[2]
\tablenotemark[3]
92142&27042&$WB_lB_hW$&148.7&4.71&0.24&0.21\\
\tablenotemark[2]
\tablenotemark[3]
92226&34133&$WB_hB_lW$&140.3&0.49&0.41&0.66\\
\tablenotemark[2]
\tablenotemark[3]
92714& 4141&$WWB_hB_l$&106.4&6.28&0.43&0.59\\
\tablenotemark[1]
\tablenotemark[2]
\tablenotemark[3]
92714&12581&$B_lB_hWW$&166.3&1.66&0.57&0.66\\
\tablenotemark[2]
\tablenotemark[3]
94750& 1147&$WWB_lB_h$&126.9&0.82&0.32&0.23\\
\tablenotemark[2]
\tablenotemark[3]
96258& 2707&$B_lWB_hW$&171.2&1.02&0.49&0.28\\
\tablenotemark[2]
\tablenotemark[3]
96264&93611&$B_hWWB_l$&111.7&0.41&0.06&0.14\\
\tablenotemark[2]
\tablenotemark[3]
96280&14555&$WB_hB_lW$&133.8&0.07&0.69&0.68\\
\tablenotemark[2]
\tablenotemark[3]
96287&20104&$WB_hB_lW$&182.5&5.64&0.16&0.14\\
\tablenotemark[1]
\tablenotemark[2]
\tablenotemark[3]
96399&32921&$B_lB_hWW$&172.8&0.28&0.68&0.83\\
\tablenotemark[1]
\tablenotemark[2]
\tablenotemark[3]
96591&39318&$B_hB_lWW$&174.3&0.94&0.55&0.75\\
\end{tabular}
\tablenotetext[1]{Passes LB selection.}
\tablenotetext[2]{Used in variable-mass analysis.}
\tablenotetext[3]{Used in pseudolikelihood analysis.}
\label{tb:candtable-mj}
\end{table}

\begin{table}
\caption{Same as \tabref{tb:candtable-ej} for the $e + \jets/\mu$ channel.}
\begin{tabular}{rrrdddd}
Run & Event & Perm. & $\mfit$     & $\chisq$ & $\DLB$ & $\DNN$ \\
\tableline
\tablenotemark[1]
62199&13305&$B_lB_hWW$&173.2&40.0&0.55&0.61\\
\tablenotemark[1]
\tablenotemark[2]
\tablenotemark[3]
85129&19079&$WB_lB_hW$&137.0&0.93&0.81&0.85\\
\tablenotemark[1]
\tablenotemark[2]
\tablenotemark[3]
86570& 8642&$B_hWWB_l$&144.5&0.66&0.74&0.29\\
\tablenotemark[1]
89372&12467&$B_hWWB_l$&186.6&22.1&0.23&0.25\\
\end{tabular}
\tablenotetext[1]{Passes LB selection.}
\tablenotetext[2]{Used in variable-mass analysis.}
\tablenotetext[3]{Used in pseudolikelihood analysis.}
\label{tb:candtable-ejm}
\end{table}

\begin{table}
\caption{Same as \tabref{tb:candtable-ej} for the $\mu + \jets/\mu$ channel.}
\begin{tabular}{rrrdddd}
Run & Event & Perm. & $\mfit$     & $\chisq$ & $\DLB$ & $\DNN$ \\
\tableline
\tablenotemark[1]
\tablenotemark[2]
\tablenotemark[3]
58203& 4980&$WB_hB_lW$&138.3&0.25&0.56&0.62\\
\tablenotemark[1]
\tablenotemark[2]
\tablenotemark[3]
91712&   22&$B_hWB_lW$&203.3&0.44&0.51&0.44\\
\tablenotemark[1]
\tablenotemark[2]
\tablenotemark[3]
92704&14022&$WB_hB_lW$&175.8&0.11&0.79&0.88\\
\end{tabular}
\tablenotetext[1]{Passes LB selection.}
\tablenotetext[2]{Used in variable-mass analysis.}
\tablenotetext[3]{Used in pseudolikelihood analysis.}
\label{tb:candtable-mjm}
\end{table}

Results of likelihood fits to the data sample are shown in
\tabref{tb:fitres}.  Several methods of extracting the final top~quark mass
are tabulated.
The labels ``quad$N$'' and ``cub$N$'' denote, respectively, $N$-point
quadratic and cubic fits to the negative log likelihood values.  The
reported central value is the minimum of the fit curve, and the error
indicated is the width of the curve where it has risen by $0.5$ from the
minimum.  For the ``avg'' fits, the central value is the mean of the
likelihood curve (calculated using trapezoidal-rule
integration), and the reported error on the mass
is the symmetric interval around
the mean containing $68\%$ of the likelihood.
\Tabref{tb:fitres} also shows the result for the ``NN2'' binning.
This is a variant of the NN binning which uses only two bins
in $\DNN$: both the first six bins and the last four bins are coalesced.
The result is seen to be consistent with the 10-bin NN analysis.

\begin{figure}
\psfig{file=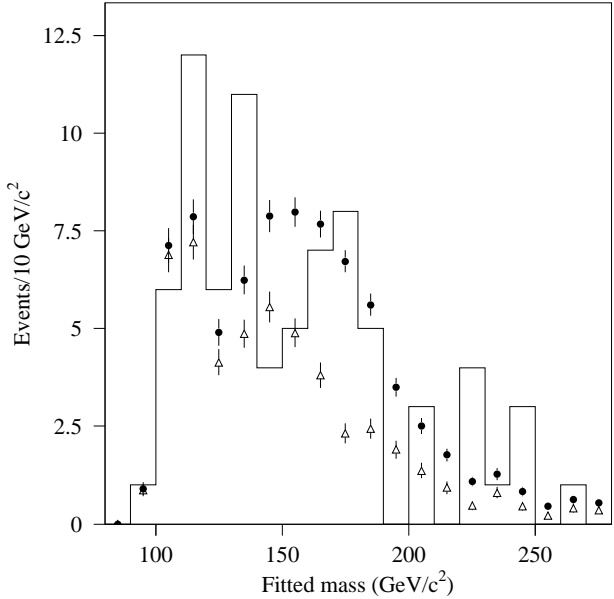,width=\hsize}
\caption{Fitted mass for all
events which pass the precuts and the $\chisq$ cut.
Filled circles are a mixture of $\ttbar$ signal
and background and open triangles are the background only,
both averaged between the results of the LB and NN analyses.}
\label{fg:allcand}
\end{figure}

\begin{figure}
\psfig{file=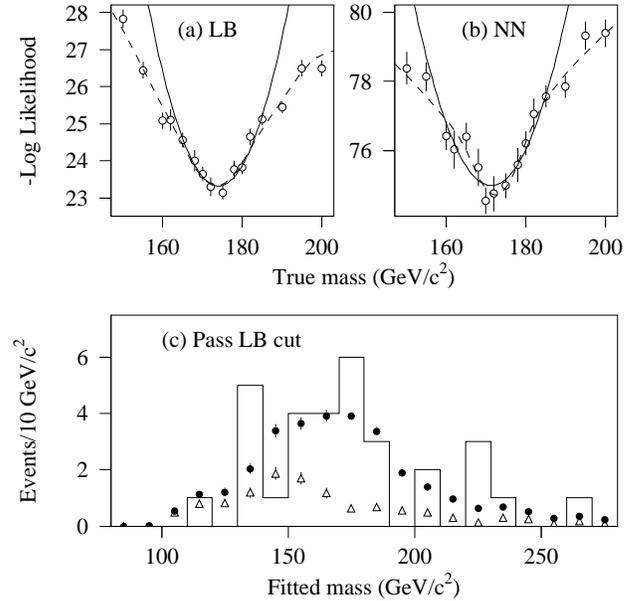,width=\hsize}
\caption{Negative log likelihood for (a) LB and (b) NN analyses.
The solid curve is a quadratic fit to the 9~points around the minimum;
the dashed curve is from fitting \eqref{eq:fdef} to all points in the
range 110--$230\gevcc$.  (c) Results of the LB fit for events passing
the LB selection.  The histogram is data, filled circles are a mixture
of $m_t=175\gevcc$ $\ttbar$ signal and background, normalized using the results
of the LB fit, and open triangles are background only.}
\label{fg:like-and-lb}
\end{figure}

\begin{figure}
\psfig{file=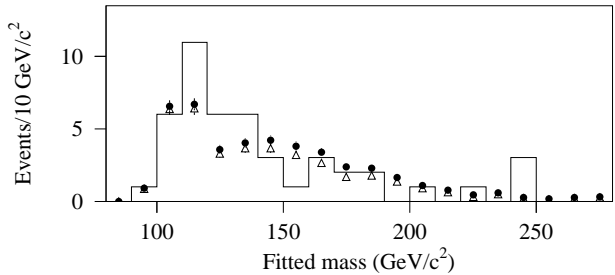,width=\hsize}
\caption{Results of the LB fit for events failing the LB selection.
The histogram is data, filled circles are a mixture
of $m_t=175\gevcc$ $\ttbar$ signal and background, normalized using the results
of the LB fit, and open triangles are background only.}
\label{fg:lb-data-fail}
\end{figure}

\begin{figure}
\psfig{file=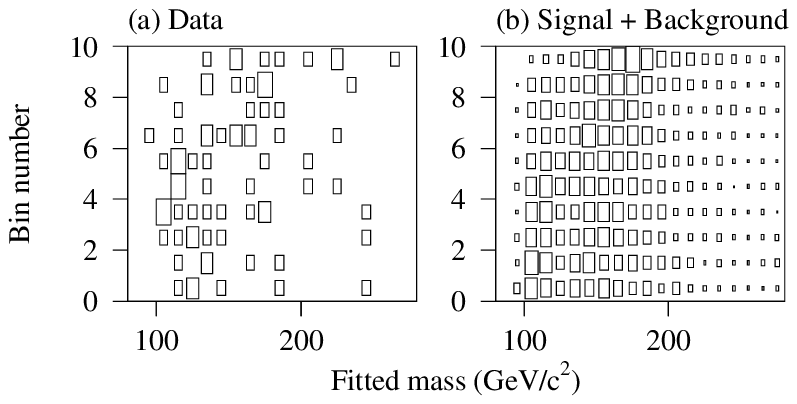,width=\hsize}
\caption{Results of NN fit: (a) Data, (b) $m_t=172\gevcc$ $\ttbar$
signal plus background, normalized using the results of the NN fit.}
\label{fg:nn-data}
\end{figure}

For our final result, we use the nine-point quadratic fit.  This choice is
motivated by a desire to use a simple functional form; furthermore,
it will be seen in the next section that among the polynomial fits
considered, it gives the slope closest to unity when one plots extracted
mass versus Monte Carlo input mass.  The resulting mass is then
$174.0\pm 5.6\gevcc$ for the LB binning, and $171.3\pm 6.0\gevcc$ for NN.
These fit results are
exhibited in \figsref{fg:allcand}--\ref{fg:nn-data}.

Note in \figref{fg:like-and-lb} that $-\ln L$ tends to
flatten out away from the minimum.  Due to this, we limit the
polynomial fit to the central region, where $-\ln L$ is most
nearly quadratic.  This flattening is related to the fact that we do
not impose an external constraint on the number of signal or
background events in the likelihood fit.  If such a constraint is
imposed, as was done in Ref.~\cite{d0discovery}, the $-\ln L$ curve shows
less tendency to flatten.

To use more likelihood points in the fit, a
functional form which can model this flattening is needed.  One such function
which we investigated is
\begin{eqnarray}
\label{eq:fdef}
F(x) = -\ln (P_1 &+& P_2 g (x-P_5, P_8) \nonumber \\
                 &+& P_3 g (x-P_6, 2P_8) \\
                 &+& P_4 g (x-P_7, 4P_8)), \nonumber
\end{eqnarray}
where $g$ is the Gaussian form
$g(x,\sigma) = \exp (-(x/\sigma)^2 / 2)$.  We determine the parameters
$P_1$--$P_8$ by fitting this function 
(using \progname{minuit}) to the
likelihood points over the entire range of 110--$230\gevcc$; the
results are plotted in \figref{fg:like-and-lb}.  If we extract
from these curves the positions of the minima,
the results are
$173.6^{+5.6}_{-5.5}\gevcc$ for LB and $172.4^{+4.1}_{-4.2}\gevcc$ for NN
(taking the error from where the curve rises by 0.5).
From this, we conclude that the procedure of fitting a quadratic in
the central region does not seriously underestimate the width.
In addition, in
Monte Carlo studies, $F(x)$ did not perform better
on average than the simple quadratic fit; thus,
we do not use $F(x)$ for the final mass extraction.

\mediumtext
\begin{table*}
\caption{Results of fits to the candidate sample, showing
the top~quark mass $m_t$ and the number of signal and background events
$n_s$ and $n_b$.  The labels
``quad$N$''  and ``cub$N$'' denote $N$-point quadratic and cubic fits, while
``avg'' denotes the mean value of the posterior mass probability distribution.
``$\lnlmin$'' is the minimum $-\ln L$ point; $\chisq_{\text{poly}}$
is for the polynomial fit to the likelihood points.}
\begin{tabular}{ldlcccd}
      Binning & $\lnlmin$ & Method &
      $m_t$                   &$n_s$                &$n_b$  &
      $\chisq_{\text{poly}}$ \\
 &&& ($\ugevcc$) &&\\
\tableline
      LB  %
      & 23.1
      & quad9 &
      $174.0^{+5.6}_{-5.6}$ &$23.8^{+8.3}_{-7.8}$ &$53.2^{+10.7}_{-9.3}$ & 
      4.7 \\
    && quad11 &
      $174.3^{+7.5}_{-7.5}$ &$23.8^{+8.5}_{-9.1}$ &$53.2^{+12.2}_{-9.4}$ & 
      29.7 \\
    && cub9 &
      $173.7^{+5.7}_{-5.4}$ &$23.8^{+8.3}_{-7.8}$ &$53.2^{+10.7}_{-9.3}$ & 
      4.5 \\
    && cub11 &
      $172.4^{+6.4}_{-5.4}$ &$23.8^{+8.5}_{-7.8}$ &$53.2^{+10.7}_{-9.4}$ & 
      14.7 \\
    && avg &
      $175.4^{+7.7}_{-7.7}$ &$23.7^{+8.5}_{-9.2}$ &$53.3^{+12.4}_{-9.5}$ & 
        \\
\tableline
      NN  %
      & 74.5
      & quad9 &
      $171.3^{+6.0}_{-6.0}$ &$28.8^{+8.4}_{-9.1}$ &$48.2^{+11.4}_{-8.7}$ & 
      8.4 \\
    && quad11 &
      $170.8^{+6.1}_{-6.1}$ &$29.1^{+8.2}_{-9.3}$ &$47.9^{+11.6}_{-8.5}$ & 
      9.9 \\
    && cub9 &
      $173.7^{+3.7}_{-4.7}$ &$28.0^{+9.7}_{-8.4}$ &$49.0^{+10.7}_{-10.0}$ & 
      3.9 \\
    && cub11 &
      $172.5^{+4.8}_{-5.5}$ &$28.3^{+9.0}_{-8.6}$ &$48.7^{+10.9}_{-9.4}$ & 
      6.3 \\
    && avg &
      $170.7^{+6.7}_{-6.7}$ &$29.2^{+8.4}_{-9.5}$ &$47.8^{+11.8}_{-8.7}$ & 
        \\
\tableline
      NN2  %
      & 29.8
      & quad9 &
      $172.0^{+5.5}_{-5.5}$ &$28.4^{+9.6}_{-9.0}$ &$48.6^{+11.3}_{-9.9}$ & 
      5.7 \\
\end{tabular}
\label{tb:fitres}
\end{table*}
\narrowtext

We have explored some additional variations in the definition
of the likelihood function.  The algorithm of
\progname{hmcmll}~\cite{hmcmll} starts with the same
likelihood as \eqref{eq:likedef}, but eliminates the nuisance parameters
$a_j^s$ and $a_j^b$ using a maximum likelihood estimate rather than
integration.  To be able to compare likelihoods from
different Monte Carlo samples, though, we modify the likelihood
following the prescription of Ref.~\cite{baker}:
\begin{equation}
L_\lambda = {L \over \prod_j q(N_j, N_j) q(A_j^s, A_j^s) q(A_j^b, A_j^b)}.
\end{equation}
The results of this procedure are given in \tabref{tb:addfits}.
Alternatively, we can eliminate $n_s$ and $n_b$ by integrating over
them, rather than by using a maximum likelihood estimate.  The results
of this are also given in \tabref{tb:addfits}.  These variations do not
have a large effect on the final result.

To further test the stability of these results, we repeat the fits
using samples in which one candidate event is removed, for a total
of 77 distinct fits.  For the LB~case, the RMS of the resulting
distribution of fits was $0.3\gevcc$; the smallest result seen was
$173.0\gevcc$, and the largest was $174.7\gevcc$.  For the NN~case,
the RMS was $0.5\gevcc$, the smallest result was $170.1\gevcc$, and
the largest was $172.5\gevcc$.

To summarize the main results of this section, the LB analysis yields
$m_t = 174.0\pm 5.6\gevcc$,
and the NN analysis yields $m_t = 171.3\pm 6.0\gevcc$.

\begin{table}
\caption{Additional fit results.}
\begin{tabular}{cccccc}
Method & Binning
      & $-\ln L_{\text{min}}$ &
      $m_t$                   &$n_s$                &$n_b$  \\
 &&& ($\ugevcc$) &&\\
\tableline
\progname{hmcmll}
      & LB  %
      & $22.7$ &
      $174.1^{+5.6}_{-5.6}$ &$23.6^{+7.9}_{-8.2}$ &$53.4^{+10.3}_{-9.7}$ \\

      & NN  %
      & $73.1$ &
      $172.0^{+5.1}_{-5.1}$ &$34.0^{+1.9}_{-14.3}$ &$42.6^{+16.0}_{-2.7}$ \\
Integration
      & LB  %
      & $17.2$ &
      $174.5^{+7.6}_{-7.6}$ &$24.9^{+8.1}_{-9.7}$ &$54.2^{+12.0}_{-9.9}$ \\

      & NN %
      & $68.5$ &
      $169.8^{+7.3}_{-7.3}$ &$30.6^{+8.4}_{-10.2}$ &$48.5^{+11.7}_{-9.4}$ \\
\end{tabular}
\label{tb:addfits}
\end{table}

\subsection{Tests with Monte Carlo samples}

We test the mass extraction procedure by performing fits
to ensembles of Monte Carlo experiments of known composition.
The size of the experiments is fixed;  the number of 
background events in each is chosen from a binomial
distribution with a fixed mean.

For the first set of tests, the ensembles consist of 1000
experiments with a composition of
$\brocket{n_s} = 26$ and $\brocket{n_b} = 52$, for an experiment
size of $N=78$ events with a 1:2
signal/background ratio.  Results for the LB and NN
analyses are shown in \tabsref{tb:lb-ensembles} and~\ref{tb:nn-ensembles}.
For these tests, the tabulated mean value is from a Gaussian fit
to the extracted mass distribution, and the width is the
symmetric interval around the mean which contains $68\%$ of the
entries.
(We estimate the statistical errors on these means and widths to be
in the range $0.5$--$1.0\gevcc$.)
Note that the 9-point quadratic fit gives the slope closest
to unity.  Some results for ensembles containing signal only are given
in \tabsref{tb:lb-ensembles-sig} and~\ref{tb:nn-ensembles-sig}.

\mediumtext
\begin{table*}
\caption{Ensemble tests for the LB analysis with 1:2 signal/background, 
showing means and $68\%$ widths.  ``Slope'' is from a linear fit to the means.}
\begin{tabular}{cdddddddd}
Input
   & \multicolumn{2}{c}{quad9} 
   & \multicolumn{2}{c}{quad11}
   & \multicolumn{2}{c}{cub9}
   & \multicolumn{2}{c}{cub11} \\
Mass
   & mean & width & mean & width & mean & width & mean & width \\
($\ugevcc$)
   & \multicolumn{2}{c}{($\ugevcc$)} 
   & \multicolumn{2}{c}{($\ugevcc$)} 
   & \multicolumn{2}{c}{($\ugevcc$)} 
   & \multicolumn{2}{c}{($\ugevcc$)} \\
\tableline
150&   150.4& 10.7&   150.8& 11.1&   151.5& 10.3&  151.9& 10.9   \\
155&   155.2&  9.1&   155.3&  9.8&   155.3&  9.0&  156.5&  8.4   \\
160&   160.7&  9.2&   160.9&  9.1&   160.9&  9.3&  161.4&  8.3   \\
162&   162.6&  8.5&   162.8&  8.5&   162.8&  9.0&  162.9&  8.3   \\
165&   165.1&  9.0&   165.3&  9.0&   165.2&  8.7&  165.3&  8.7   \\
168&   168.2&  9.3&   168.3&  9.3&   168.1&  9.0&  168.1&  9.0   \\
170&   168.9&  7.6&   169.0&  7.7&   169.2&  7.2&  169.1&  7.4   \\
172&   172.2&  7.4&   172.2&  7.8&   172.0&  7.4&  172.1&  7.5   \\
175&   174.9&  8.4&   174.9&  8.5&   174.9&  8.4&  174.7&  8.3   \\
178&   177.6&  8.5&   177.5&  8.5&   177.4&  8.0&  177.2&  8.0   \\
180&   179.7&  8.7&   179.6&  8.6&   179.4&  8.2&  179.2&  8.1   \\
182&   181.8&  8.1&   182.1&  8.2&   181.3&  7.8&  181.1&  7.5   \\
185&   183.9&  8.9&   183.9&  9.1&   183.3&  8.2&  183.2&  8.1   \\
190&   190.5&  9.7&   191.1& 10.0&   189.0&  9.0&  189.0&  8.9   \\
\tableline
Slope
  & 0.98&
  & 0.98&
  & 0.94&
  & 0.91& \\
\end{tabular}
\label{tb:lb-ensembles}
\end{table*}
\narrowtext

\mediumtext
\begin{table*}
\caption{Same as \tabref{tb:lb-ensembles} for the NN analysis.}
\begin{tabular}{cdddddddd}
Input
   & \multicolumn{2}{c}{quad9} 
   & \multicolumn{2}{c}{quad11}
   & \multicolumn{2}{c}{cub9}
   & \multicolumn{2}{c}{cub11} \\
Mass
   & mean & width & mean & width & mean & width & mean & width \\
($\ugevcc$)
   & \multicolumn{2}{c}{($\ugevcc$)} 
   & \multicolumn{2}{c}{($\ugevcc$)} 
   & \multicolumn{2}{c}{($\ugevcc$)} 
   & \multicolumn{2}{c}{($\ugevcc$)} \\
\tableline
150&   149.0&  9.8&   150.1& 10.8&   150.0&  8.9&  150.8&  9.9 \\
155&   154.6&  9.6&   154.6& 10.0&   155.1&  8.6&  155.5&  8.2 \\
160&   159.6&  9.5&   159.8&  9.7&   159.6&  9.4&  160.1&  8.7 \\
162&   161.8&  9.2&   162.1&  9.0&   161.9&  9.1&  162.3&  8.3 \\
165&   163.9&  9.2&   164.4&  9.4&   163.7&  9.2&  164.0&  8.6 \\
168&   167.2&  9.7&   167.6& 10.0&   166.9&  9.8&  167.0&  9.8 \\
170&   168.3&  8.8&   168.3&  8.2&   168.4&  8.0&  168.3&  8.0 \\
172&   171.6&  8.8&   171.5&  8.3&   171.7&  8.4&  171.7&  8.3 \\
175&   174.6&  9.3&   174.6&  9.1&   174.5&  9.0&  174.3&  9.0 \\
178&   176.6&  8.7&   176.6&  8.8&   176.6&  8.6&  176.6&  8.4 \\
180&   179.0&  9.0&   178.9&  8.9&   178.6&  8.7&  179.0&  8.5 \\
182&   181.1&  8.9&   180.9&  9.0&   180.8&  8.4&  180.9&  7.8 \\
185&   183.0&  8.9&   182.8&  9.1&   182.8&  8.6&  182.5&  8.4 \\
190&   189.0&  9.1&   189.0&  9.8&   188.4&  8.5&  188.2&  8.1 \\
\tableline
Slope
  & 0.98&
  & 0.96&
  & 0.95&
  & 0.93& \\
\end{tabular}
\label{tb:nn-ensembles}
\end{table*}
\narrowtext

\mediumtext
\begin{table*}
\caption{Ensemble tests for the LB analysis with $n_s = 26$ events
and $n_b = 0$.}
\begin{tabular}{cdddddddd}
Input
   & \multicolumn{2}{c}{quad9} 
   & \multicolumn{2}{c}{quad11}
   & \multicolumn{2}{c}{cub9}
   & \multicolumn{2}{c}{cub11} \\
Mass
   & mean & width & mean & width & mean & width & mean & width \\
($\ugevcc$)
   & \multicolumn{2}{c}{($\ugevcc$)} 
   & \multicolumn{2}{c}{($\ugevcc$)} 
   & \multicolumn{2}{c}{($\ugevcc$)} 
   & \multicolumn{2}{c}{($\ugevcc$)} \\
\tableline
168&  168.3&  6.7&  168.2&  6.7&  168.4&  6.3&  168.2&  6.5  \\
170&  168.9&  5.9&  168.9&  6.2&  169.1&  5.7&  168.9&  5.8  \\
172&  172.2&  6.2&  172.2&  6.0&  172.1&  5.9&  172.1&  5.9  \\
175&  175.6&  6.6&  175.7&  6.8&  175.5&  6.2&  175.5&  6.4  \\
\end{tabular}
\label{tb:lb-ensembles-sig}
\end{table*}
\narrowtext

\mediumtext
\begin{table*}
\caption{Same as \tabref{tb:lb-ensembles-sig} for the NN analysis.}
\begin{tabular}{cdddddddd}
Input
   & \multicolumn{2}{c}{quad9} 
   & \multicolumn{2}{c}{quad11}
   & \multicolumn{2}{c}{cub9}
   & \multicolumn{2}{c}{cub11} \\
Mass
   & mean & width & mean & width & mean & width & mean & width \\
($\ugevcc$)
   & \multicolumn{2}{c}{($\ugevcc$)} 
   & \multicolumn{2}{c}{($\ugevcc$)} 
   & \multicolumn{2}{c}{($\ugevcc$)} 
   & \multicolumn{2}{c}{($\ugevcc$)} \\
\tableline
168&  167.7&  6.3&  168.1&  6.8&  168.0&  5.8&  167.9&  6.4  \\
170&  168.9&  6.1&  169.0&  6.0&  169.0&  5.6&  168.8&  5.7  \\
172&  172.0&  6.1&  172.3&  6.2&  172.0&  5.5&  172.0&  5.9  \\
175&  175.6&  6.5&  175.6&  6.7&  175.2&  6.0&  175.3&  6.4  \\
\end{tabular}
\label{tb:nn-ensembles-sig}
\end{table*}
\narrowtext

There are several competing factors which contribute to the mass
dependence of the width of the ensemble mass distributions
$\sigma(m_t)$ observed in \tabsref{tb:lb-ensembles}
and~\ref{tb:nn-ensembles}.  As $m_t$ increases, the widths of the $\mfit$
distributions slowly increase.  From this one would expect the
$\sigma(m_t)$ to increase with increasing top quark mass.  However,
we rely on the difference between the signal and background $\mfit$
distributions to set the background normalization.  This difference
is smallest for $m_t$ around $140$--$150\gevcc$; thus, one would
expect $\sigma(m_t)$ to be larger in that region.  Finally, the spacing
of the generated Monte Carlo points is finer in the region near
$170\gevcc$; the available statistics are also larger there.
This permits a more accurate determination of the top quark mass in
that region, leading to a smaller $\sigma(m_t)$.

\begin{figure}
\psfig{file=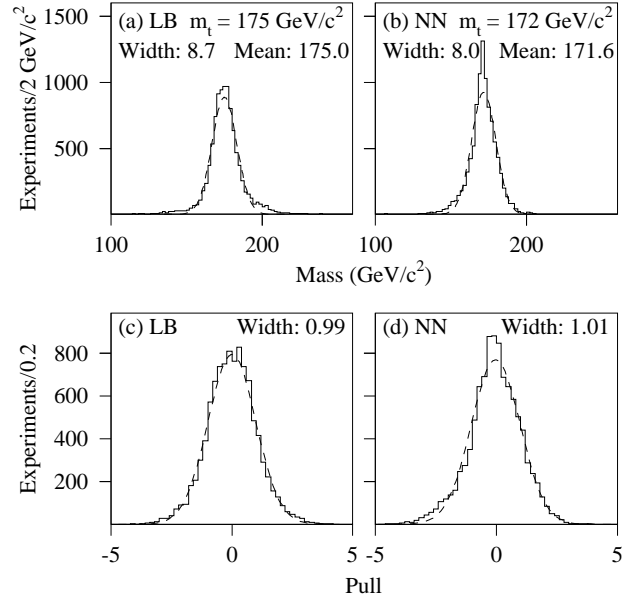,width=\hsize}
\caption{Mass and pull distributions for 10,000 MC experiment ensembles
with compositions matching the fit results.  The dashed curves are
Gaussian fits.  For the mass distributions, the width is
the symmetric interval containing $68\%$ of the entries; for the pull
distributions, it is from the Gaussian fit.}
\label{fg:ensembles}
\end{figure}

Next, we try ensembles with compositions that match the results
of the likelihood fit.  The results are given in \tabref{tb:ens-fitresults}.
(These and all subsequent results use the ``quad9'' prescription.)
Plots of the mass distributions from these ensembles are shown in
\figref{fg:ensembles}.  Also shown are the distributions of the pull
quantity
\begin{equation}
\text{pull} = {m_t(\text{measured}) - m_t(\text{true}) \over \sigma(m_t)}.
\label{eq:pull}
\end{equation}
If the errors produced by the mass extraction procedure are correct,
these distributions should have unit width, as is indeed observed.
In addition, $70\%$ of the $1\sigma$ error intervals
from the LB ensemble include $175\gevcc$, and $69\%$ of those from
the NN ensemble include $172\gevcc$, as expected.

\begin{table}
\caption{Results of mass fits to ensembles of Monte Carlo events.
The ensembles consisted of 10,000 experiments of 77~events each, with
the compositions indicated below.}
\begin{tabular}{lcdd|cc}
   & Input       &&&&\\
   & Mass        & $\brocket{n_s}$ & $\brocket{n_b}$ & Mean      & Width \\
   & ($\ugevcc$) &             &             & ($\ugevcc$) & ($\ugevcc$) \\
\tableline
LB & $175$       & 23.8      & 53.2      & $175.0$ & $8.7$ \\
NN & $172$       & 28.8      & 48.2      & $171.6$ & $8.0$ \\
\end{tabular}
\label{tb:ens-fitresults}
\end{table}

\begin{figure}
\psfig{file=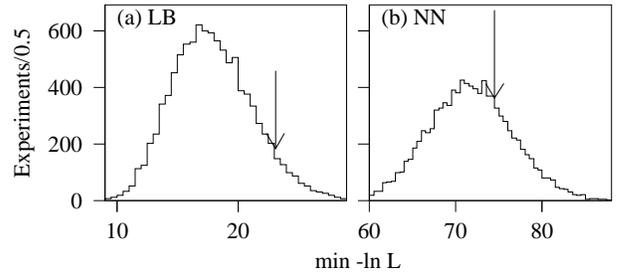,width=\hsize}
\caption{Minimum $-\ln L$ distributions from the LB and NN ensembles.
The arrows show the values corresponding to the data fits.}
\label{fg:minlnl}
\end{figure}

The minimum $-\ln L$ value for the LB fit was $23.1$; for the NN fit,
it was $74.5$.
(A smaller value of $-\ln L$ corresponds to a better fit to the
expected distributions.)
This quantity is plotted for the LB and NN ensembles
in \figref{fg:minlnl}.  A $-\ln L$ value larger than that of the data
is seen in about $7\%$ of LB experiments and in about $28\%$ of NN
experiments.

\begin{figure}
\psfig{file=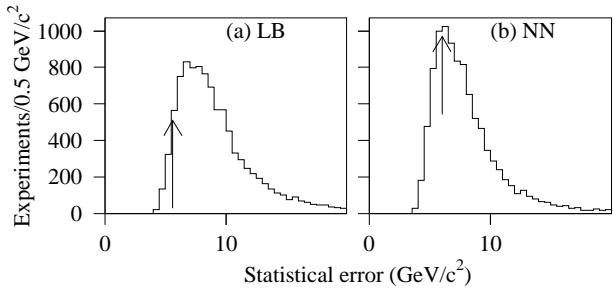,width=\hsize}
\caption{Statistical error distributions from the LB and NN ensembles.
The arrows show the values corresponding to the data fits.}
\label{fg:ensemble-err}
\end{figure}

\begin{figure}
\psfig{file=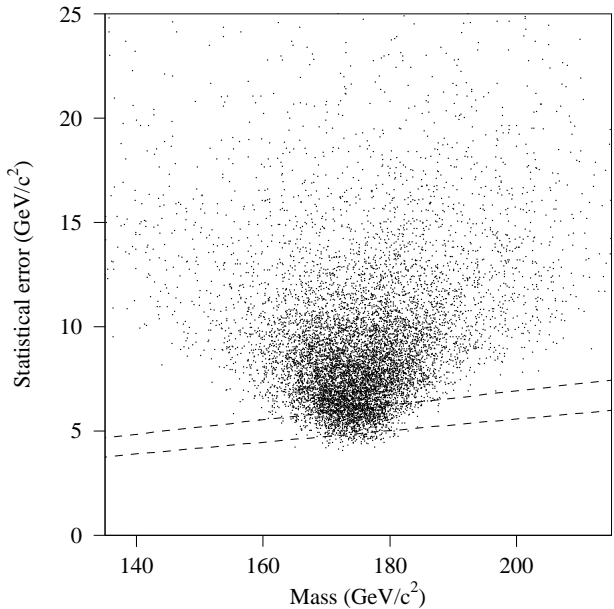,width=\hsize}
\caption{Scatter plot of masses and statistical errors
from the LB ensemble.  The dashed lines of constant relative error
delimit the ``accurate subset'' (see text).}
\label{fg:likescatter}
\end{figure}

One can also look at the distribution of statistical errors from
ensemble tests.  For the data, the statistical error is $5.6\gevcc$
for the LB analysis, and $6.0\gevcc$ for the NN analysis.
Plots of the statistical
error for the ensemble fits are shown in \figref{fg:ensemble-err}.  An
error smaller than that for the data is seen in about~$6\%$ of LB
experiments and in about~$25\%$ of NN experiments.  The correlation
between the mass and the error for the LB ensemble is exhibited
in \figref{fg:likescatter}.  This shows that experiments
with a small error typically yield masses closer to the true value.

\begin{figure}
\psfig{file=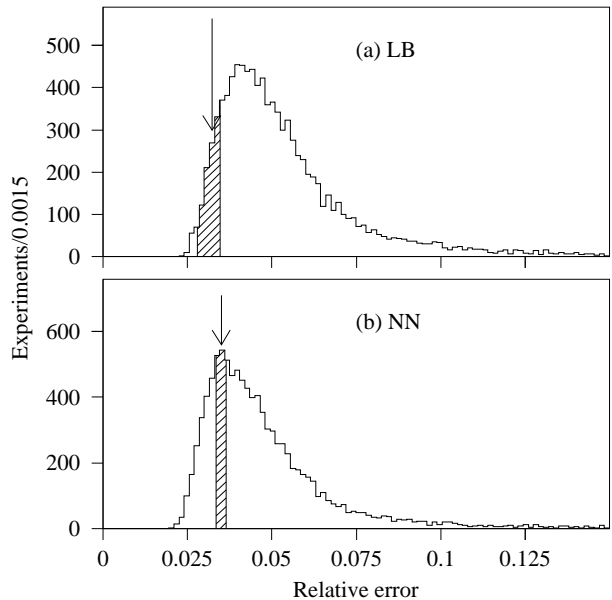,width=\hsize}
\caption{Relative error ($\sigma(m_t)/m_t$)
distributions from the LB and NN ensembles.
The arrows show the value corresponding to the data fits, and the
hatched regions show the definitions of the accurate subsets.}
\label{fg:ensemble-relerr}
\end{figure}

It is interesting to examine the ensemble results for that subset of
experiments where the extracted statistical error is similar to that actually
obtained.  We define this ``accurate subset'' as follows.  First, find
the relative error ($\sigma(m_t)/m_t$) for the result.  For LB, this
is $0.0322$; for NN, it is $0.0350$.  Then convert these numbers to
a percentile in the relative error distribution.  These are $6.0\%$ and
$24.9\%$ for LB and NN, respectively.  For any ensemble, we then
define the accurate subset by looking at its relative error
distribution and selecting those experiments which lie within a range
of $\pm 5\%$ around the above percentiles.  This is illustrated in
\figsref{fg:likescatter}--\ref{fg:ensemble-relerr}.
This procedure thus selects $10\%$ of
the total sample.  (The relative error is used because the statistical
error tends to increase slightly with increasing mass; therefore, cutting on
relative rather than absolute error results in a less biased
subsample.)

\begin{figure}
\psfig{file=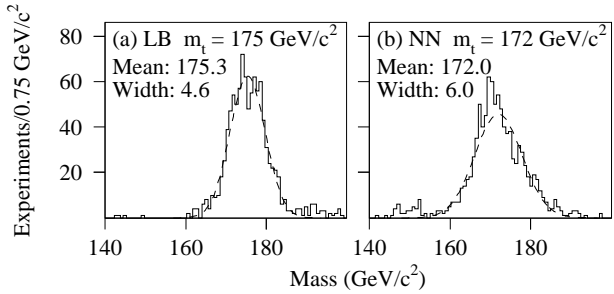,width=\hsize}
\caption{Mass distributions for accurate subsets of ensembles.
The dashed curves are Gaussian fits.}
\label{fg:ensemble-acc}
\end{figure}

There is an additional complication which arises when a cut is made
on the statistical error.  The spacing of the generated mass points is finer
around $m_t = 175\gevcc$.  This permits a more accurate determination
of the top quark mass in that range.  However, this implies that if
a small error is required, the masses of the selected
events will be biased towards the region with finer spacing.  (Note,
however, that as long as a cut on the error is not made, the uneven MC
spacing does not bias the mass.  
Studies of an even but coarser MC spacing show that adding extra points
reduces the statistical error in the region where the
extra points are added, but does not, on average, shift the extracted
mass distribution.)
Thus, for the accurate subset fits we changed
the procedure slightly, adding Monte Carlo points at
intervals of $2.5\gevcc$ between 130~and $160\gevcc$ and also between
185~and $210\gevcc$.  These additional mass points were constructed by
interpolating between the existing MC histograms on either side.  The
results of these fits with the accurate subset cuts are shown in
\figref{fg:ensemble-acc}.  The widths are $4.6\gevcc$ and
$6.0\gevcc$ for LB and NN, respectively.  This is a further indication
that the error estimates from the likelihood fit are reliable.

The results of the LB and NN analyses can be compared
experiment-by-experiment, provided that the ensemble definitions are
the same.  We use the same ensemble definition as for the first set of
tests ($N=78$ events and a 1:2 signal/background ratio) with
$m_t = 175\gevcc$.  The results for 10,000 experiments 
are given in \tabref{tb:ensemble-comp}.  It is seen that given
the observed statistical errors, a difference between the two analyses
of the magnitude seen is expected $\sim 20\%$ of the time.

\begin{table}
\caption{Comparisons of LB and NN ensembles for $m_t=175\gevcc$ and
         a 1:2 signal/background ratio.  The first line is the
         mean difference
         between the results; the second and third lines give the fraction
         of experiments for which the difference exceeds the
         observed difference of $2.7\gevcc$.
         (Numbers are in $\ugevcc$.)}
\begin{tabular}{lccc}
 & Full     & LB & NN \\
 & ensemble & acc. subset & acc. subset \\
\tableline
$\brocket{\hbox{LB}-\hbox{NN}}$     &  $0.78\pm 0.05$
                                    &  $0.34\pm 0.06$
                                    &  $0.51\pm 0.09$  \\
$(\hbox{LB} - \hbox{NN}) > 2.7$     & $29\%$ & $11\%$ & $18\%$ \\
$|\hbox{LB} - \hbox{NN}| > 2.7$     & $45\%$ & $16\%$ & $28\%$ \\
\end{tabular}
\label{tb:ensemble-comp}
\end{table}

It is also interesting to look at the correlation between the LB and NN
measurements.  This can be defined using the ensemble mass
distributions of $\mlb$ and $\mnn$ as
\begin{equation}
\rho = {\brocket{\left(\mlb - \brocket{\mlb}\right)
                 \left(\mnn - \brocket{\mnn}\right)} \over \slb\snn}.
\end{equation}
This is appropriate for Gaussian distributions; however, our
distributions typically have a small number of non-Gaussian outliers.
To explore the sensitivity of this quantity to these outliers,
the following procedure is used.
\begin{itemize}
\item For the cuts of interest, plot $\mlb$ and $\mnn$.  Record the means
      and RMS widths of these distributions ($\brocket{\mlb}$, $\slb$,
      $\brocket{\mnn}$, $\snn$).

\item Reject experiments which are more than $K\sigma$ from the mean.
      Specifically, make the additional cut that
\begin{eqnarray}
|\mlb-\brocket{\mlb}| &<& K\slb\quad\text{and}    \\ 
|\mnn-\brocket{\mnn}| &<& K\snn. \nonumber 
\end{eqnarray}

\item Replot $\mlb$ and $\mnn$ with this additional cut, and record the
      new means and RMS widths ($\brocket{\mlb}'$, $\slb'$,
      $\brocket{\mnn}'$, $\snn'$).

\item Plot (with all cuts) the distribution of
\begin{equation}
\left(\mlb - \brocket{\mlb}'\right) \cdot \left(\mnn - \brocket{\mnn}'\right) .
\end{equation}

\item Find the mean of this distribution.  $\rho$ is then calculated by
      dividing this mean by $\slb'\snn'$.
\end{itemize}

The results are tabulated for the full sample and for the LB and NN
accurate subsets in \tabref{tb:correlation}.  This is done using the
same $m_t = 175\gevcc$ ensembles as for the previous comparisons.
They do not depend strongly on $K$ within reasonable ranges.
To get a single number, we average the $K=5$ results for the two
accurate subset results, giving $0.88$.  This appears to be a
reasonable representation of the accurate subset numbers (within a few percent)
for $K \ge 2$.  Propagating statistical errors through this
calculation gives $\rho = 0.88\pm 0.04$.

\begin{table}
\caption{Values of correlation parameter $\rho$.}
\begin{tabular}{rddd}
  K  &   Full   &  LB      &   NN      \\
     &   Sample &  acc. subset & acc. subset \\
\tableline
100  &   0.62   &   0.89   &   0.77   \\
  5  &   0.65   &   0.89   &   0.88   \\
  4  &   0.67   &   0.89   &   0.89   \\
  3  &   0.70   &   0.89   &   0.89   \\
  2  &   0.77   &   0.87   &   0.88   \\
  1  &   0.75   &   0.67   &   0.78   \\
\end{tabular}
\label{tb:correlation}
\end{table}

In summary, these ensemble tests show that the masses and errors
obtained from the likelihood fit are reliable, and that our observed
data set is not particularly unlikely.

\subsection{Systematic errors}

\subsubsection{Energy scale errors}
\label{jet-scale}

The first major component of the systematic error is the jet energy scale
uncertainty.  What is relevant here is the uncertainty in the relative
scale between the data and MC, rather than in the absolute scale.
This was estimated to be $\pm(2.5\% + 0.5\gev)$ for each jet (see
\secref{jetcorrections}).

We propagate this per-jet error to the final mass
measurement by performing ensemble tests with all the jets in the
events comprising the ensemble scaled up or down by the per-jet
uncertainty.  For these tests, we used large experiment sizes, with
$N = 1000$.  The results are given in \tabref{tb:ensemble-scaling} and
give an error of about $\pm 4\gevcc$.  Comparing this with the shifts
in the $\mfit$ distributions seen after scaling the jets
(\figref{fg:jet-scalecomp}), we estimate the ratio between a shift in
the final extracted mass and a shift in $\mfit$ to be about 1.1.

\begin{table}
\caption{Ensemble means for determining error due to jet energy scale.
Each experiment consisted of $N = 1000$ events;
the signal/background ratios are the same as in
\tabref{tb:ens-fitresults}.}
\begin{tabular}{cdd}
               &  LB       & NN       \\
\tableline
Input mass     &  175.0$\gevcc$  & 172.0$\gevcc$  \\
Input $\brocket{n_s}$
               &  309.1 events   & 374.0 events  \\
\tableline
$-2.5\%-0.5$   &  170.9$\gevcc$  & 167.6$\gevcc$  \\
Nominal        &  175.4$\gevcc$  & 171.3$\gevcc$  \\
$+2.5\%+0.5$   &  179.4$\gevcc$  & 175.2$\gevcc$  \\
Symmetric\\
Error          &   4.2$\gevcc$   & 3.8$\gevcc$    \\
\end{tabular}
\label{tb:ensemble-scaling}
\end{table}

The systematic uncertainty in the electromagnetic energy scale
is much smaller than that of the jets, and can be neglected.
The systematic uncertainty of the muon momentum measurement is estimated
to be $2.5\%$.
The effect of this uncertainty is found to be negligible
relative to the jet scale uncertainty.

\subsubsection{Generator dependencies}
\label{massfit-gendep}

The next component of the systematic error is that due to
uncertainties in how well the underlying Monte Carlo event generators
model reality.  We separate this into signal and background
components.  Of particular concern is the modeling of QCD~radiation by
the $\ttbar$ signal Monte Carlo.

To estimate the error due to the \progname{herwig} generator, we
characterize \progname{herwig} events using variables which are
sensitive to the amount of initial and final state radiation (ISR and
FSR) in each event.  To do this, we match the direction of
reconstructed jets with \progname{herwig} partons and use the
Monte Carlo parentage information to identify the jets which come
from the $b$ quarks and the hadronically-decaying $W$~boson.
We consider the four jets with
highest $\et$ $j_1,\ldots j_4$, and define the variables:
\begin{itemize}
 \item $x\equiv$ Number of jets in $j_1,\ldots j_4$ which do not
       come from a $b$ quark or the $W$~boson (i.e., jets which are
       likely to be due to ISR).
 \item $y\equiv N_j-4\equiv$ Number of extra jets of any kind in the
       event
       ($N_j\equiv$ number of jets with $\et>15\gev$ and
        $|\eta| < 2.0$).
 \item $z\equiv$ Number of non-ISR jets in $j_1,\ldots j_4$ which have
       the same
       parent as a higher $\et$ jet (i.e., the number of extra jets
       due to FSR among the top four).
\end{itemize}
We take a \progname{herwig} Monte Carlo sample (with $m_t = 170\gevcc$)
and bin it using these variables into a three-dimensional histogram
with ranges $0 \le x,y,z \le 2$ (27~bins).
For each bin $(x,y,z)$, we plot the fitted masses for all events in
that bin, fit them to a Gaussian to form $\brocket{\mfit}(x,y,z)$,
and then fit the resulting values to the empirical function
\begin{equation}
 G(x,y,z) = m_0 + ux + v\max\,(0,y-x-z) + wz,
\end{equation}
for fit parameters $m_0$, $u$, $v$, and $w$.  Here, $u$ describes the
dependence of $\brocket{\mfit}$ on ISR and $v$ and $w$ describe its
dependence on FSR.
In particular, the $v$ term describes the dependence of the mass on
the number of extra jets which cannot be attributed to either an ISR
or FSR jet displacing another jet out of the top four.  Additional low
$\et$ jets affect the mass only if they are FSR; thus we group $v$
with $w$.
We compute a population-weighted average of
$G$ over all bins; this is seen to agree well with
$\brocket{\mfit}$ from the entire sample.  Finally, we
recalculate this average with (a) $u$ (ISR) increased by $50\%$ and
(b) $v$ and $w$ (FSR) increased together by $50\%$.  This gives
excursions of $0.69$ and $1.74\gevcc$, respectively.
Adding these in quadrature yields an error of $1.9\gevcc$.
(Monte Carlo studies of ensembles constructed of events from
individual $(x,y,z)$ bins confirm that, for these variations, the
mass resulting from the likelihood fit approximately tracks
$\brocket{\mfit}$.)

We have performed several additional cross checks to verify that
this is a reasonable estimate of the signal generator error.
The first is simply to compare these results to those from a
different event generator, in this case
\progname{isajet}.  We constructed ensembles from \progname{isajet}
events and analyzed them using the MC histograms derived from
\progname{herwig}.  These are compared to ensembles of
\progname{herwig} events in \tabref{tb:isajet-ens}.  Taking the six
differences in the region 160--$180\gevcc$ gives a mean of
$-0.17\gevcc$ and a RMS of $0.8\gevcc$.

\begin{table}
\caption{Ensemble means for determining the difference between
\progname{isajet} and \progname{herwig}.  (All numbers in $\ugevcc$.)
Each ensemble consisted of $N=1000$ event experiments
with a 1:2 signal/background ratio.}
\begin{tabular}{ldddddd}
      & \multicolumn{3}{c}{LB}
      & \multicolumn{3}{c}{NN} \\
      \cline{2-4} \cline{5-7}
$m_t$ & \progname{herwig} & \progname{isajet} & Diff
      & \progname{herwig} & \progname{isajet} & Diff \\
\tableline
150   & 150.5 & 151.7 & $-$1.2  & 149.4 & 150.4 & $-$1.0 \\
160   & 161.0 & 160.9 &    0.1  & 159.8 & 159.4 &    0.4 \\
170   & 169.3 & 170.8 & $-$1.5  & 168.3 & 169.0 & $-$0.7 \\
180   & 180.1 & 180.1 &    0.0  & 179.6 & 178.9 &    0.7 \\
190   & 190.2 & 190.1 &    0.1  & 189.0 & 188.8 &    0.2 \\
200   & 201.9 & 200.9 &    1.0  & 200.5 & 197.6 &    2.9 \\
\end{tabular}
\label{tb:isajet-ens}
\end{table}

We also vary the QCD coupling strength parameter, $\LQCD$, of
the \progname{herwig} $\ttbar$ Monte Carlo.  The default value of this
parameter in \progname{herwig}~5.7 is $0.18\gev$; the current
experimental value from the Particle Data Group is
$0.21^{+0.04}_{-0.03}\gev$ \cite{pdg}.  Accordingly, we generate
additional $\ttbar$ Monte Carlo with $\LQCD$ set to $0.15$, $0.21$,
and $0.25\gev$, with $m_t=170$ and $175\gevcc$ \cite{hwparms}.
We then construct ensembles from these
samples and process them using the standard analysis.  The results are
given in \tabref{tb:lambda-qcd}.  The size of the resulting deviations
is on the order of $1\gevcc$; they appear to be dominated by Monte
Carlo statistics.

\widetext
\begin{table*}
\caption{Ensemble tests with $\LQCD$ varied.  Ensembles consisted
of experiments with $N=1000$ events and a 1:2 signal/background ratio.}
\squeezetable %
\begin{tabular}{ddddddd}
$\LQCD$ & \multicolumn{2}{c}{$\brocket{\mfit}$ ($\ugevcc$)}
        & \multicolumn{2}{c}{LB ($\ugevcc$)}
        & \multicolumn{2}{c}{NN ($\ugevcc$)} \\
\cline{2-3} \cline{4-5} \cline{6-7}
($\ugev$)  & $\mt=170\gevcc$ & $\mt=175\gevcc$
           & $\mt=170\gevcc$ & $\mt=175\gevcc$
           & $\mt=170\gevcc$ & $\mt=175\gevcc$ \\
\tableline
0.15  & 171.0 & 173.5 & 170.5 & 175.2 & 169.5 & 174.8 \\
0.18  & 168.8 & 173.1 & 169.2 & 175.3 & 168.3 & 174.5 \\
0.21  & 170.8 & 173.6 & 170.2 & 174.5 & 169.5 & 173.3 \\
0.25  & 168.7 & 173.2 & 168.3 & 175.7 & 167.2 & 175.0 \\
\end{tabular}
\label{tb:lambda-qcd}
\end{table*}
\narrowtext

We can make another comparison by using a version of
\progname{herwig}~5.8 in which final state radiation (FSR) in top quark
decays is substantially suppressed.  We compare results from ensembles
made from this version to those from \progname{herwig}~5.8 with normal
radiation.  The results are shown in \tabref{tb:hw58}.  Averaging
over LB and NN, this is seen to give an excursion of about
$2.15\gevcc$.  Note that the $\mfit$ distribution with FSR suppressed
is significantly narrower on the low mass side than distributions
with normal radiation.  This difference in shape is why the relation
between means of $\mfit$ and ensemble results is different here than
described above.

\begin{table}
\caption{Comparison of ensembles constructed using
\progname{herwig}~5.8 both with and without FSR suppressed.
The ensembles consist of $N=77$ event experiments.  For the LB
case, $\brocket{n_s}=23.8$, and for NN, $\brocket{n_s}=28.8$.
For both cases, $m_t = 170\gevcc$.}
\begin{tabular}{lddd}
          & $\brocket{\mfit}$ & LB    & NN \\
          & ($\ugevcc$)       & ($\ugevcc$) & ($\ugevcc$) \\
\tableline	                                
FSR suppressed  &     176.0         & 172.2 & 172.7 \\
Normal FSR      &     170.1         & 170.7 & 169.9 \\
\tableline
Difference&       5.9         &   1.5 &   2.8 \\
\end{tabular}
\label{tb:hw58}
\end{table}

The results of these cross checks confirm that our estimate for the
systematic error due to the signal generator of $1.9\gevcc$ is reasonable.

We also study the effects of varying the \progname{vecbos}
background model.  Besides the sample used for the mass measurement
(which uses a $Q^2$ scale of $\smallbrocket{\pt^{\jet}}^2$
and \progname{herwig}
fragmentation), we have samples with a $Q^2$ scale of $M_W^2$ and with
\progname{isajet} fragmentation.  Results from ensembles made from
these samples are shown in \tabref{tb:vecbos-ens}.  (The ensemble
compositions were the same as for the jet energy scale tests.)  The largest
difference seen is about $2.5\gevcc$ using the $M_W^2$ scale with
\progname{herwig} fragmentation.

\begin{table}
\caption{Ensemble means for determining
\progname{vecbos} differences.  Samples were generated with
a \progname{vecbos} $Q^2$
scale of both $M_W^2$ and $\smallbrocket{\pt^{\jet}}^2$, and using both
\progname{herwig} (HW) and \progname{isajet} (IS) for fragmentation.
Each experiment consisted of $N = 1000$ events;
the signal/background ratios are the same as in
\tabref{tb:ens-fitresults}.}
\begin{tabular}{cdd}
                        &  LB     & NN       \\
\tableline
Input mass     &  175.0$\gevcc$  & 172.0$\gevcc$  \\
Input $\brocket{n_s}$
               &  309.1 events   & 374.0 events  \\
\tableline	                                
$\smallbrocket{\pt^{\jet}}^2$, HW  &  175.4$\gevcc$  & 171.3$\gevcc$  \\
$M_W^2$, HW             &  177.9$\gevcc$  & 173.8$\gevcc$  \\
$\smallbrocket{\pt^{\jet}}^2$, IS  &  175.0$\gevcc$  & 171.2$\gevcc$  \\
$M_W^2$, IS             &  175.8$\gevcc$  & 171.6$\gevcc$  \\
\tableline
Max.\\
difference              &  2.5$\gevcc$    & 2.5$\gevcc$ \\
\end{tabular}
\label{tb:vecbos-ens}
\end{table}

A concern is that the systematic error assigned here to
\progname{vecbos} may not adequately reflect the level of agreement
between \progname{vecbos} and data for $\eta^W$ in the forward region
(\figref{fg:eta_w}).  To check this, we reweight the
\progname{vecbos} events using a smooth function of $\eta^W$ (a
Gaussian) chosen to optimize the agreement between the simulation and
the data.  When we redo the mass extraction with this reweighted
background, the top quark mass shifts by only $0.4$--$0.5\gevcc$, a
value much smaller than the error we attribute to \progname{vecbos}.
This error can therefore be neglected.

We also do the fits with the fraction of QCD multijets contributing to
the background histogram [$(22\pm 5)\%$] varied within its errors.  The
changes to the final extracted mass are $<0.2\gevcc$, well below the
assigned error.

\subsubsection{Noise and multiple interactions}

At the luminosities at which most of our data were collected,
it is likely that during a single beam crossing, there will be
multiple $p\pbar$ inelastic interactions (MI).  (This is expected
about $2/3$ of the time.)  While these extra interactions
rarely give rise to additional high-$\pt$ objects, they do
deposit a small amount of additional energy over the entire calorimeter,
affecting the jet energy calibration.  Additional noise in the
calorimeter is produced by the radioactive decay of the uranium absorber.
The Monte Carlo samples used for this analysis do not include these
effects.  To estimate
them, we generate a small number of additional Monte Carlo events which
include noise, and which are overlaid with one
or two additional interactions.  The means of the $\mfit$ distribution
for these samples are given in \tabref{tb:mierror}.  Based on the
luminosity profile of the collected data, we estimate that in order to
represent the data, these samples should be combined in the ratio
$0.31:0.33:0.36$.  The weighted average of the three means is then
$170.5\pm 0.6\gevcc$; the shift from the zero additional interaction case
is $1.2\pm 0.7\gevcc$.  Scaling this by the factor 1.1 for the
ratio between a shift in final extracted mass and a shift in $\mfit$
(\secref{jet-scale}) gives an estimated shift due to noise and
multiple interactions of $1.3\pm 0.8\gevcc$.  Since this effect is
relatively poorly known and is small compared to other error sources,
we do not attempt to correct the result for this effect, but instead
include it as a systematic error.

\begin{table}
\caption{Means of $\mfit$ distributions of $\ttbar$ Monte Carlo
for multiple interaction error determination.  (For the $e$ + jets channel,
$m_t=170\gevcc$.)}
\begin{tabular}{lcc}
                          & $\brocket{\mfit}$ ($\ugevcc$) & Weight \\
\tableline
0 additional interactions & $169.3\pm 0.4$ & 0.31 \\
1 additional interaction  & $170.5\pm 1.3$ & 0.33 \\
2 additional interactions & $171.6\pm 1.2$ & 0.36 \\
\end{tabular}
\label{tb:mierror}
\end{table}

\subsubsection{Monte Carlo statistics}

We assess the effect of Monte Carlo statistics on the final result
by performing the fit to the data many times, each time smearing the
MC histograms used to calculate the likelihood according to Poisson
statistics.  This is done separately for signal and background. 
The $68\%$ widths of the resulting mass distributions are given in
\tabref{tb:mcstat}.

\begin{table}
\caption{Errors due to Monte Carlo statistics.}
\begin{tabular}{cdd}
              &  LB    & NN     \\
              & ($\ugevcc$) & ($\ugevcc$) \\
\tableline
Signal        &  0.49  &  0.99  \\
Background    &  0.33  &  0.57  \\
\tableline
Total         &  0.6   &  1.1   \\
\end{tabular}
\label{tb:mcstat}
\end{table}

\subsubsection{Systematic error summary}

\Tabref{tb:systerr} gives a summary of the systematic errors.
In addition to the errors already discussed, the mean difference of
$0.8\gevcc$ between the LB and NN ensemble results from
\tabref{tb:ensemble-comp} has been added as a systematic uncertainty, and an
additional error of $1\gevcc$ has been added to cover possible small
biases in the likelihood fitting method (this is approximately the RMS
spread of the different polynomial fits in \tabref{tb:fitres}).
Note that these two components are of the same order as the
estimated error due to Monte Carlo statistics, and that these
small biases are probably due in large part to statistical
fluctuations in the Monte Carlo histograms.  Nevertheless, we retain
these as separate components of the systematic error in lieu of
exploring this further with still larger Monte Carlo samples.

\begin{table}
\caption{Systematic error summary.}
\begin{tabular}{lddd}
                    &  LB       & NN     & Average    \\
                    & ($\ugevcc$)& ($\ugevcc$) & ($\ugevcc$) \\
\tableline
Jet energy scale    &   4.2     &  3.8   &  4.0   \\
Generator&&&\\
\ \ $t\tbar$ signal &   1.9     &  1.9   &  1.9   \\
\ \ \progname{vecbos} flavors
                    &   2.5     &  2.5   &  2.5   \\
Noise/MI            &   1.3     &  1.3   &  1.3   \\
Monte Carlo stat.   &   0.6     &  1.1   &  0.85  \\
LB/NN diff          &   0.8     &  0.8   &  0.8   \\
Likelihood fit      &   1.0     &  1.0   &  1.0   \\
\tableline
Total               &   5.6     &  5.4   &  5.5   \\
\end{tabular}
\label{tb:systerr}
\end{table}

The total systematic errors here are slightly smaller than
those reported in Ref.~\cite{ljmassprl}.  The signal generator error
was $3.3\gevcc$, taken from the difference between
\progname{herwig} and an older version of \progname{isajet},
and the LB/NN difference was $1.35\gevcc$, taken from 
half the difference of the fit results.

\subsection{Summary}

For the final mass result, we combine the results of these two
analyses, taking into account
their correlation $\rho$ determined earlier.  Let $\mlb$ and $\mnn$
be the two results and $\slb$ and $\snn$ be their errors.  Then we
form a $\chisq$ as a function of the combined mass $M$:
\begin{eqnarray}
\chisq(M) &=& {1\over \slb^2\snn^2(1-\rho^2)} \nonumber \\
          & &\times [\snn^2(M-\mlb)^2 \nonumber \\
          & & \quad - 2\rho\slb\snn(M-\mlb)(M-\mnn) \\
          & & \quad + \slb^2(M-\mnn)^2]. \nonumber
\end{eqnarray}
The combined result and its error is then defined by the minimum of
this curve and the points where the curve rises by one unit from the
minimum.  (Monte Carlo studies of this combination give a width of the
pull distribution of 1.11 for the full sample, but 0.76 for the LB
accurate subset and 0.97 for the NN accurate subset.)  Inserting
$\mlb=174.0\gevcc$, $\slb=5.6\gevcc$, $\mnn=171.3\gevcc$,
$\snn=6.0\gevcc$, and $\rho = 0.88$ (for the accurate subsets) gives
\begin{equation}
M = 173.3 \pm 5.6 \gevcc .
\end{equation}
The systematic errors of the two methods are averaged, giving a final
result of
\begin{equation}
m_t = 173.3 \pm 5.6(\text{stat}) \pm 5.5(\text{syst}) \gevcc .
\end{equation}

\section{Pseudolikelihood Analysis}
\label{pseudolikelihood}

\subsection{Introduction}

The pseudolikelihood (PL) analysis is an alternate method of
extracting the top quark mass, with several important differences from
the analyses of the previous section.  It thus serves as a nearly
independent check of the previous result.  In this analysis, we
kinematically fit candidate events at a series of fixed top quark masses
$\mfit$ (3C fits) over the range $100$--$250\gevcc$.
These fits are done using a
different kinematic fitting program
(\progname{squaw}~\cite{squaw1}\mynocite{*squaw2})
than was used in the previous section.
In addition, when looping over jet permutations, we allow the
assignment of jets beyond the fourth (in which case at least one of
the top four jets is treated as ISR).
At each $\mfit$, we choose the jet
permutation yielding the smallest $\chisq$, and interpret the
resulting plot of $\chisq/2$ versus $\mfit$ as
defining a top quark mass ``pseudolikelihood'' $L$ for a
particular event given by
\begin{equation}
L(\mfit) = e^{-\chisq/2 (\mfit)}.
\end{equation}
We then sum this plot over all candidate
events, subtract the expected background contribution, and
fit the remainder to a quadratic function to extract the top quark mass.
This analysis is performed mainly for signal-enriched subsamples of the
entire precut sample.

\begin{figure}
\psfig{file=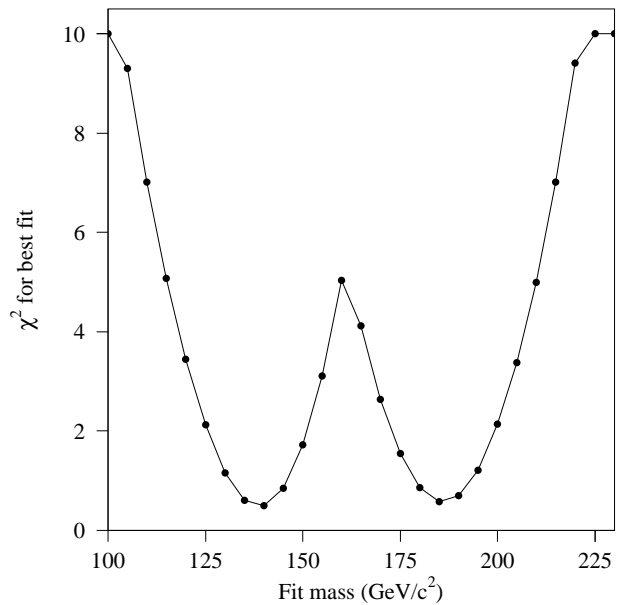,width=\hsize}
\caption{$\chisq$ plot for \progname{SQUAW} fixed-mass fits for event
58203, 4980.}
\label{fg:prdpl1}
\end{figure}

A major motivation for this analysis method is to more fully take into
account the information from different jet permutations.  For example,
the fixed-mass $\chisq$ plot for one top quark candidate is
shown in \figref{fg:prdpl1}.  The information about both minima in
this figure is incorporated directly into the PL
analysis, but is not used in the LB and NN likelihood analyses.

\subsection{PL method}

\begin{figure}
\psfig{file=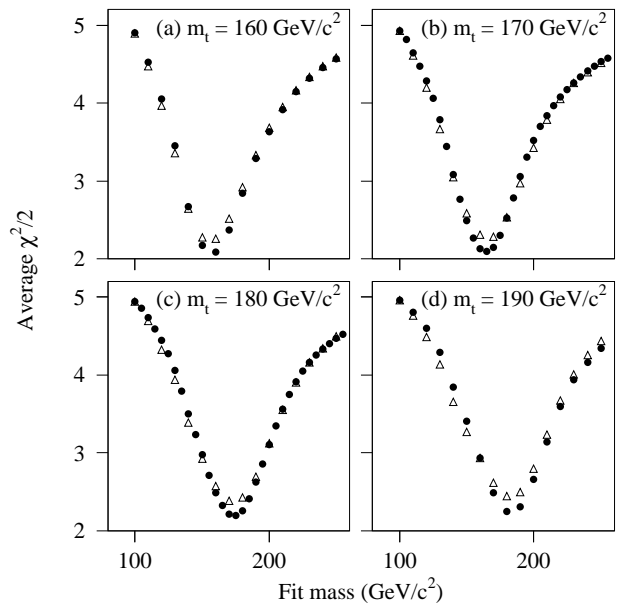,width=\hsize}
\caption{Average $\chisq/2$ plots (after LB selection)
for \progname{HERWIG} 
(filled circles) and
\progname{ISAJET} (open triangles) $\ttbar$ events.}
\label{fg:prdpl2}
\end{figure}

\begin{figure}
\psfig{file=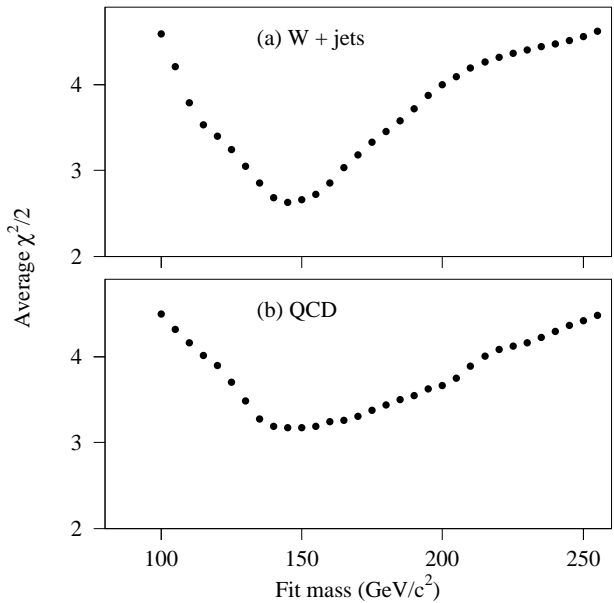,width=\hsize}
\caption{Average $\chisq/2$ plots (after LB selection)
for (a) \progname{VECBOS}
$W+\jets$ and (b) QCD multijet background samples.
}
\label{fg:prdpl3}
\end{figure}

Some examples of $\chisq/2$ plots for $\ttbar$~events are
shown in \figref{fg:prdpl2}.  These are ``average $\chisq/2$''
plots: for each $\mfit$, we average the $\chisq/2$ over all
events in the sample.  The figure shows plots for events
generated with both \progname{herwig} and \progname{isajet} for
top~quark masses from 160 to $190\gevcc$.
The plots from
\progname{isajet} are slightly wider than those from
\progname{herwig}.  We will also need the background shape to
subtract the expected background contribution from the data sample.
It is determined by combining the average $\chisq/2$
plot of the \progname{vecbos} $W+\jets$ sample with that of
the QCD multijet sample.  These plots are shown in
\figref{fg:prdpl3}.  They are broader and have minima at
about $150\gevcc$, lower
than those for $\ttbar$ events (for $m_t > 160\gevcc$).  The
\progname{VECBOS} sample uses the average jet transverse momentum
$Q^2$ scale and \progname{HERWIG} for fragmentation, as in the 
variable-mass analyses.

The next step is to determine the background
normalization.  The nominal background fraction in the precut event
sample is found from the cross section analysis to be
$\approx 2/3$.
One can improve on this nominal background by using properties of the
particular sample being analyzed which are sensitive to the background
fraction.  One such property is the average value of one of the top
quark discriminants (either $\DLB$ or $\DNN$).  The background
fraction can be calculated as
\begin{equation}
\text{BG fraction} = (\calD^T - \calD^D) / (\calD^T - \calD^B),
\end{equation}
where $\calD^T$ is the average value expected for $\ttbar$ events,
$\calD^B$ is that expected for background events, and $\calD^D$ is
that of the sample being analyzed.

We can do an analogous calculation using the $\chisq/2$ plot.
There is, however, a complication, due to the fact that the $\chisq/2$
plots depend on the top quark mass to a much greater extent
than do the likelihood discriminants.  Therefore, to get a
background from this method, we need a rough estimate of the top quark mass.
We find this as follows.  For each sample, we construct
the average $\chisq/2$ plot.
We compare the plot from data
to that predicted from MC signal plus background, with the MC top quark
mass varied in 10 steps from~140 to~$210\gevcc$.  We pick the mass
which yields the smallest RMS difference with the data.

An additional complication is that, in general, the average $\chisq/2$
plots for signal and background will cross at some $\mfit$.
We thus define the variable
\begin{equation}
C = \sum_{\mfit > m_{\text{cross}}} \chisq/2 \, (\mfit) -
    \sum_{\mfit < m_{\text{cross}}} \chisq/2 \, (\mfit),
\end{equation}
where $m_{\text{cross}}$ is the point at which the plots
cross.
($m_{\text{cross}}$ is near $150\gevcc$ for top~quark masses above
$160\gevcc$.)
We then estimate the background in the same manner as
before, using
\begin{equation}
\text{BG fraction} = (C^T - C^D) / (C^T - C^B),
\end{equation}
where $C^T$, $C^B$, and $C^D$ are the values of $C$ from MC signal,
background, and the data sample, respectively.

The background fraction for the full precut sample is taken
to be the average of three values:    
the nominal value, the value determined from the top quark discriminants,
and the value from the $\chisq/2$ plot.  They are weighted by the
squared inverses of their errors.

When analyzing subsets of the precut sample, we determine the nominal
background for the subset by scaling down the background determined from the
full precut sample.  The subset background fraction is then
the weighted average of this nominal background fraction and the
fraction estimated from the $\chisq/2$ plots.  The background
estimate from the top quark discriminants is not used in this case,
as the subset selections tend to make the distributions of
these discriminants similar for signal and background.
The precut and LB subset background fractions determined from the data
are 0.60 and 0.32, respectively.

For each $\mfit$, we subtract the $\chisq/2$ contribution
expected for the background from the total.
This is evaluated over the range 100--$250\gevcc$ with a
distance between points $\Delta\mfit = 10\gevcc$.   
We then extract
the top~quark mass and error using a quadratic fit near the
minimum of this background-subtracted $\chisq/2$ plot.
The
extracted mass $\mmin$ is the value at which the fit function has its
minimum, and its error is the deviation that corresponds to an
increase of 0.5~units above the minimum.  We try to use as
many points as possible in the fit provided that the plot
remains parabolic over the fit range.  The algorithm used to select
the fit range is determined empirically by fitting the average
$\chisq/2$ plots for $\ttbar$~Monte Carlo events.
With $\Delta\mfit = 10\gevcc$, at least three
points below and two points above the minimum are required; thus, the
mass range covered is at least $50\gevcc$.  If necessary, we add
points at the extremes until the value of $\chisq/2$ exceeds that
at the minimum
by an amount equal to the number of events in the plot.
However, we add points on the high side only if the $\chi^2$/2
values change at an increasing rate, as expected for a parabola.
We also do some fits with $\Delta\mfit=5\gevcc$ over the range
100--$255\gevcc$.  In that case, we use at least five points on each
side of the minimum.

\subsection{Results of fits to Monte Carlo events}

\Tabref{tab:table1} contains results of fits to average
$\chisq/2$ plots from MC samples.  The mass $\mmin$ 
(from a quadratic fit near the minimum) for $\ttbar$
Monte Carlo is slightly different from the MC input mass.  It has a
roughly linear dependence on the input top quark mass, with a slope
that is only slightly smaller than that determined from fits with the
correct jet assignment.  A linear fit to these points gives the
following prescription for a ``corrected'' mass $\mcorr$:
\begin{equation}
\mcorr = (\mmin - 27.0 \gevcc) / 0.815 .
\end {equation}
This relation is used to correct the masses $\mmin$ obtained from
fits.

\mediumtext
\begin{table*}
\caption{%
Results of fits to average $\chisq/2$ plots from MC.
$\mmin$ is the minimum of a quadratic fit to the points, ``width'' is
the width where the fit curve rises by 0.5, and $\brocket{\mfit}$ is
the weighted average of the $\mfit$ values, where the weights are
$e^{-\chisq/2}$.
Entries labeled ``jet high'' and ``jet low'' are after scaling jet energies
by $\pm(2.5\%+0.5\gev)$.}
    \label{tab:table1}
    \begin{tabular}{lcccccc}
            &\multicolumn{3}{c}{\progname{HERWIG}}  
            &\multicolumn{3}{c}{\progname{ISAJET}} \\
\cline{2-7}
 Sample     & $\mmin$ &width &$\brocket{\mfit}$ 
            & $\mmin$ &width &$\brocket{\mfit}$ \\
            & ($\ugevcc$) & ($\ugevcc$) & ($\ugevcc$)
            & ($\ugevcc$) & ($\ugevcc$) & ($\ugevcc$) \\
\hline
\progname{herwig}\\
\quad $m_t=150\gevcc$ &149.5  &16.4    &159.5     & ~      & ~        & ~   \\
\quad $m_t=160\gevcc$ &157.7  &17.1    &165.5     &158.9   &20.4      &165.4\\
\quad $m_t=165\gevcc$ &161.7  &18.4    &167.5     &~       &~         &~    \\
\quad $m_t=170\gevcc$ &164.7  &18.9    &170.0     &166.3   &22.0      &170.6\\
\quad $m_t=175\gevcc$ &169.9  &19.6    &173.1     &~       &~         &~    \\
\quad\quad  jet high  &172.3  &19.7    &175.3     & ~      & ~        &~    \\
\quad\quad  jet low   &166.5  &18.4    &171.1     & ~      & ~        &~    \\
\quad $m_t=180\gevcc$ &173.2  &20.5    &175.9     &172.4   &23.9      &175.2\\
\quad $m_t=190\gevcc$ &182.5  &21.2    &182.3     &180.4   &25.7      &180.4\\
\quad $m_t=200\gevcc$ &191.3  &21.9    &188.0     &188.7   &26.9      &185.8\\
\hline
\progname{vecbos}\\
 \quad $M_W^2$ scale
            &156.4   &29.9     &166.2      &152.8   &28.0      &164.0\\
 \quad $\smallbrocket{\pt^{\jet}}^2$ scale
            &147.1   &24.5     &160.4      &142.2   &23.1      &157.1\\
 QCD (data) &158.0   &33.6     &169.4      & ~      & ~        & ~   \\
    \end{tabular}
\end{table*}
\narrowtext

\subsection{Ensemble studies}

\mediumtext
\begin{table*}
    \caption{%
Ensembles with $N=78$ and a 1:2 signal/background ratio.
Entries labeled ``jet high'' and ``jet low'' are after scaling jet energies
by $\pm(2.5\%+0.5\gev)$.
``Slope'' is from a linear fit to the masses.
The LB discriminant is used in the background determination for
analyses of the precut samples.}
    \label{tab:table4}
    \begin{tabular}{lcdcddd}

  ~              &\multicolumn{2}{c}{$\mmin$}
                 &\multicolumn{2}{c}{$\mcorr$}
                 &\multicolumn{2}{c}{Width containing} \\
\cline{2-7}
  ~              &avg. mass&RMS &avg. mass  &RMS    &68.27\%   &95.45\%\\
                    &($\ugevcc$) &($\ugevcc$) &($\ugevcc$)
                    &($\ugevcc$) &($\ugevcc$) &($\ugevcc$) \\
\tableline
Precut sample, \progname{herwig}\\
\quad $m_t = 165\gevcc$ &160.0  &8.5   &163.2    &10.4    &8.99     &22.13\\
\quad $m_t = 170\gevcc$ &163.5  &8.4   &167.5    &10.3    &8.85     &21.88\\
\quad $m_t = 175\gevcc$ &168.1  &8.4   &173.1    &10.4    &9.04     &20.98\\
\quad $m_t = 180\gevcc$ &171.8  &9.5   &177.7    &11.7    &10.00    &22.77\\
Slope                   &0.80   &      &0.98     &        &         &        \\
\tableline
LB subset, \progname{herwig}\\
\quad $m_t = 150\gevcc$ &150.6  &7.3   &151.7    &8.9     &7.68     &16.84   \\
\quad $m_t = 160\gevcc$ &158.8  &7.4   &161.7    &9.0     &7.82     &18.07   \\
\quad $m_t = 165\gevcc$ &161.6  &7.1   &165.2    &8.7     &7.34     &17.27   \\
\quad $m_t = 170\gevcc$ &165.2  &7.0   &169.6    &8.6     &7.51     &17.22   \\
\quad $m_t = 175\gevcc$ &169.6  &6.7   &175.0    &8.2     &7.93     &16.83   \\
\quad\quad  jet high    &172.6  &7.5   &178.7    &9.2     &8.22     &18.32   \\
\quad\quad  jet low     &167.0  &8.0   &171.7    &9.9     &8.35     &19.73   \\
\quad $m_t = 180\gevcc$ &173.3  &7.5   &179.5    &9.2     &8.47     &18.28   \\
\quad $m_t = 190\gevcc$ &182.4  &7.7   &190.7    &9.5     &8.61     &19.54   \\
Slope                   &0.78   &      &0.96     &        &         &        \\
\tableline
LB subset, \progname{isajet}\\
\quad $m_t = 160\gevcc$ &158.6  &8.9   &161.5    &10.9    &9.23     &21.02   \\
\quad $m_t = 170\gevcc$ &166.0  &8.6   &170.5    &10.6    &9.59     &21.57   \\
\quad $m_t = 180\gevcc$ &173.0  &9.2   &179.1    &11.3    &10.38    &22.44   \\
\quad $m_t = 190\gevcc$ &180.6  &10.0  &188.5    &12.2    &11.38    &24.93   \\
Slope                   &0.73   &      &0.90     &        &         &        \\
\tableline
NN subset, \progname{herwig}\\
\quad $m_t = 150\gevcc$ &149.4  &8.3   &150.2    &10.2    &8.55     &19.03   \\
\quad $m_t = 160\gevcc$ &158.1  &8.3   &160.8    &10.2    &8.75     &20.21   \\
\quad $m_t = 165\gevcc$ &161.1  &8.5   &164.6    &10.4    &8.44     &19.87   \\
\quad $m_t = 170\gevcc$ &164.8  &7.8   &169.1    &9.6     &8.41     &19.10   \\
\quad $m_t = 175\gevcc$ &169.5  &7.8   &174.8    &9.6     &8.45     &20.50   \\
\quad $m_t = 180\gevcc$ &173.3  &8.5   &179.5    &10.5    &9.53     &21.30   \\
\quad $m_t = 190\gevcc$ &182.4  &8.7   &190.6    &10.7    &9.67     &21.78   \\
Slope                   &0.81   &      &1.00     &        &         &        \\
\end{tabular}
\end{table*}
\narrowtext

We study the performance of the PL method by forming ensembles of
simulated experiments consisting of MC events which pass the precuts.
These experiments contain $N=78$ events each, with an average of
26~events from signal and the balance from background.  The results
are shown in \tabref{tab:table4}.  (All use $\Delta \mfit =
10\gevcc$.)  
The typical errors on the average ensemble masses are about
$0.5\gevcc$, so the LB and NN subset masses are consistent.
We also show in \tabref{tab:table2} results for ensembles
of experiments consisting of 26 signal events and no background.
The agreement of the corresponding average mass values between
\tabsref{tab:table4} and~\ref{tab:table2} indicates that the
background subtraction does not produce a mass bias.

\mediumtext
\begin{table*}
    \caption{%
Results of fits to LB subsets
using ensembles with $N=26$ and no background.
Entries labeled ``jet high'' and ``jet low'' are after scaling jet energies
by $\pm(2.5\%+0.5\gev)$.
``Slope'' is from a linear fit to the masses.}
    \label{tab:table2}
    \begin{tabular}{lcccccc}

  ~              &\multicolumn{2}{c}{$\mmin$}
                 &\multicolumn{2}{c}{$\mcorr$}
                 &\multicolumn{2}{c}{Width containing}
\\ \cline{2-7}
  Input mass &avg. mass&RMS &avg. mass  &RMS    &68.27\%   &95.45\%\\
  $(\ugevcc$)        &($\ugevcc$) &($\ugevcc$) &($\ugevcc$)
                     &($\ugevcc$) &($\ugevcc$) &($\ugevcc$) \\
\hline
  150    &150.6  &5.0   &151.6    &6.1     &5.96     &12.07\\
  160    &158.6  &5.1   &161.5    &6.2     &6.02     &12.56\\
  165    &161.6  &4.7   &165.2    &5.8     &5.62     &12.18\\
  170    &165.2  &5.0   &169.5    &6.2     &6.15     &12.72\\
  175    &169.8  &5.0   &175.2    &6.2     &6.06     &12.51\\
  \quad jet high
         &172.6  &5.3   &178.7    &6.5     &6.41     &13.27\\
  \quad jet low
         &166.9  &5.5   &171.7    &6.7     &6.40     &13.78\\
  180    &173.5  &5.6   &179.8    &6.9     &6.95     &13.89\\
  190    &182.7  &5.8   &191.0    &7.1     &6.99     &14.40\\
  200    &191.0  &6.6   &201.3    &8.0     &7.88     &16.09\\
Slope    &0.81   &      &1.00     &        &         &     \\
\end{tabular}
\end{table*}
\narrowtext

\begin{figure}
\psfig{file=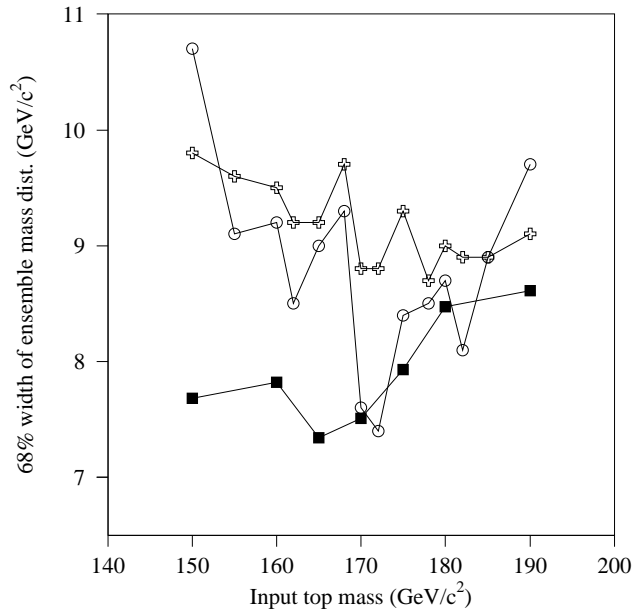,width=\hsize}
\caption{ 68\% widths of ensemble mass distributions for different analyses.
Squares are for PL fits to the LB subset,
circles are for LB variable-mass fits,
and plus symbols are for the NN variable-mass fits.
Typical errors on the plotted values are between 0.5 and $1.0\gevcc$.
}
\label{fg:prdpl4}
\end{figure}

\begin{figure}
\psfig{file=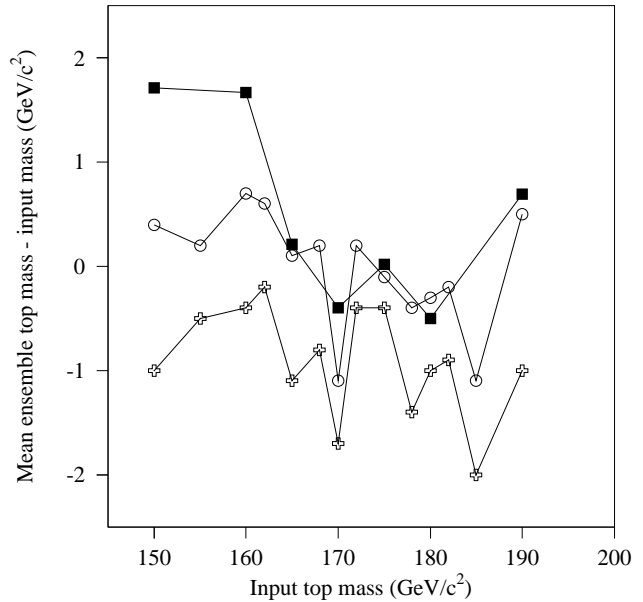,width=\hsize}
\caption{Same as \figref{fg:prdpl4} for mean ensemble mass deviations.}
\label{fg:prdpl5}
\end{figure}

The widths of the $\mcorr$ distributions for the subset analyses are
smaller than those from the
entire sample; further, the widths for LB~subsets are all smaller
than those for the corresponding NN~subsets.  The widths for the LB~subset are
smaller because the background for the LB~subset is smaller than for
the NN~subset: at $m_t = 175\gevcc$, the background fraction for LB is
$35\%$, and for NN, it is $42\%$.  Results will therefore be
based primarily on LB~subset fits.  The widths and shifts from the input
mass are plotted in \figsref{fg:prdpl4} and~\ref{fg:prdpl5} for the
LB subset.

\begin{figure}
\psfig{file=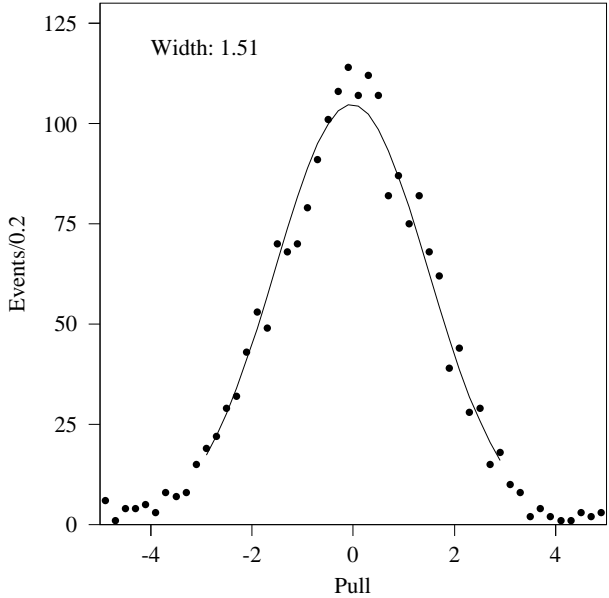,width=\hsize}
\caption{Pull distribution for LB subset fits to precut ensemble samples
with $m_t = 175\gevcc$.  The curve is a
Gaussian fit to the region $-3$ to $+3$.
}
\label{fg:prdpl6}
\end{figure}

\Figref{fg:prdpl6} shows the pull distribution (as defined in
\eqref{eq:pull}) for LB subset fits.  We find the error on $\mcorr$ by
dividing the width of the quadratic fit by the slope of
the mass correction.  A Gaussian fit to the pull distribution for
$m_t = 175\gevcc$ has a width of $1.51$.  Therefore, the
corrected errors from quadratic fits typically underestimate the width of the
ensemble mass distribution and need to be scaled up by an additional
factor of 1.51.

\subsection{Analysis of data sample}

\begin{figure}
\psfig{file=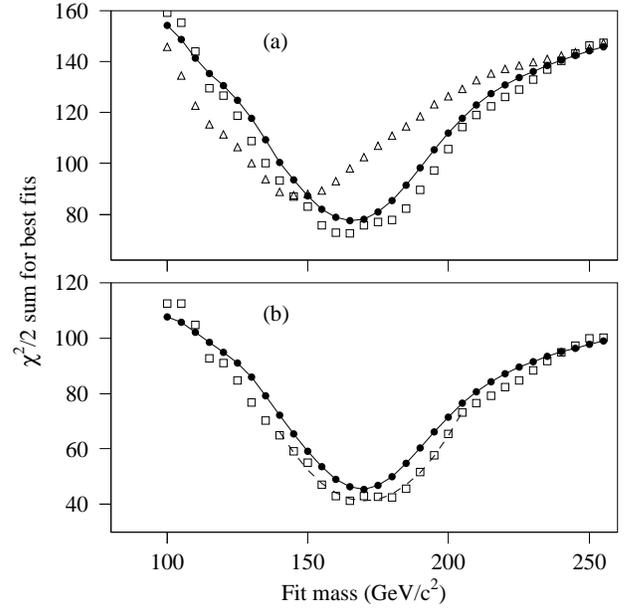,width=\hsize}
\caption{(a) $\chisq/2$ plots for the LB subset of the PR sample.
Data are the open squares, filled circles are the
prediction for a mixture of background and $175\gevcc$ top events,
and open triangles are the prediction for pure background.
The solid line joins the filled circles.
(b) Background-subtracted $\chisq/2$ plot for LB subsets.  Data are
the open squares, and filled circles
are the prediction for $175\gevcc$ top events.  The
dashed curve is a parabola fit near the minimum.}
\label{fg:prdpl7}
\end{figure}

We analyze the data
for the two subsets defined by the LB and NN selections
(see \secref{likelihood}).
These subset selections are about $80\%$ efficient for the $\ttbar$
signal, versus about $30\%$ for background.

We select the data sample for analysis by requiring that each event
have at least one fit with $\chisq < 10$.  This yields a sample of
78~events, 32 of which pass the LB~selection, and 33 of which pass the
NN~selection, with 27~events in common between these two subsets.
(Due to differences in the kinematic fitting, three events in the
variable-mass analysis fail the $\chisq$ cut for 3C \progname{SQUAW}
fits, and four events not in the variable-mass analysis are included
in the PL analysis.)
Results
of fits to these samples are given in \tabref{tab:table5}.  They are
listed for $\Delta \mfit$ values of both 5 and $10\gevcc$.  A
$5\gevcc$ increment gives slightly smaller errors.  The $\chisq/2$
plot for the LB subsample is plotted in \figref{fg:prdpl7}.

The top quark mass from the NN~subset is smaller than that from the
LB~subset, and has a larger error.  This is due to the fact that the
events accepted by the NN~selection but rejected by the LB~selection tend to be
of lower mass than those accepted by LB but rejected by NN.
These low mass events are typically rejected from the LB~subsample by
the $H_{T2} > 90\gev$ cut.

If we look at the subset of events selected by both the PL and
variable-mass analysis, there are 74~events, with 31 events passing the
LB~selection and 32 events passing the NN~selection.
Results of fits to these samples
are also given in \tabref{tab:table5}.

\begin{table}
    \caption{Fits to data samples.}
    \label{tab:table5}
    \begin{tabular}{ccdccdd}
Cut & $N$& $\Delta\mfit$ 
                 &$\mmin$
                 &$\mcorr$
                 &\multicolumn{2}{c}{BG fractions} \\
\cline{6-7}
        &&($\ugevcc$)&($\ugevcc$)
                     &($\ugevcc$)
                     &precut  &subset \\
\tableline
LB & 32 & 10.0  &$171.0 \pm 4.6$   &$176.7 \pm  8.4$     &0.60     &0.32 \\
LB & 32 &  5.0  &$170.4 \pm 4.3$   &$176.0 \pm  7.9$     &0.60     &0.32 \\
NN & 33 & 10.0  &$164.3 \pm 5.5$   &$168.4 \pm  10.1$    &0.65     &0.41 \\
\tableline
\multicolumn{7}{l}{Subset common to both PL and variable-mass}\\
LB & 31 & 10.0  &$169.0 \pm 4.6$   &$174.3 \pm  8.5$     &0.56     &0.29 \\
LB & 31 &  5.0  &$169.8 \pm 4.4$   &$175.2 \pm  8.0$     &0.56     &0.29 \\
NN & 32 & 10.0  &$163.0 \pm 5.4$   &$166.8 \pm  9.9$     &0.60     &0.38 \\
\end{tabular}
\end{table}

\subsection{Systematic errors}

This section gives estimates of the
systematic errors for the PL analysis.  
The uncertainty in the jet energy scale is 
$\pm(2.5\% + 0.5\gev)$ per jet (\secref{jetcorrections}).
To estimate the effect
of this on $\mcorr$, we redo the fits for a $\ttbar$ MC sample with
all jets scaled up or down by this uncertainty.  The results are given
in \tabref{tab:table1}.  After applying the slope correction, this
yields an estimate of $\pm 3.6\gevcc$.  Note that this is only valid in
the limit of a large number of $\ttbar$ events with negligible background.
We can also estimate this error by constructing ensembles with all the
jets in the $\ttbar$ signal sample scaled up or down.  The results are
given in \tabref{tab:table4}; the estimated error is $\pm 3.5\gevcc$.  The
same value for this error would be obtained using the mass shifts
from ensemble studies with no background, as given in
\tabref{tab:table2}.

The differences seen in $\mmin$ between \progname{herwig} events and
\progname{isajet} events are shown in \tabref{tab:table1}.  The
corresponding differences in $\mcorr$ vary from $-1.6$ to $2.6\gevcc$
over the range $m_t = 160$--$200\gevcc$, and have a minimum between
170 and $180\gevcc$.  We then construct ensembles using
\progname{isajet} events and compare these results to those from
\progname{herwig}.  This is done in \tabref{tab:table4}.  The
resulting difference varies from $-0.9$ to $2.2\gevcc$ over the range
$m_t = 160$--$190\gevcc$, so we assign a systematic error of
$2.2\gevcc$ for the signal model.

We estimate the contribution to the systematic error due to the choice of the
\progname{vecbos} $Q^2$ scale and fragmentation method by
examining the four different choices listed in \tabref{tab:table1}.
One can see that our choice of average jet $\pt$ scale and
\progname{herwig} fragmentation represents an intermediate
case.  The resulting uncertainty in $m_t$ is obtained by constructing
ensembles from the different \progname{VECBOS} parameter choices (but still
using the favored choice for background calculation and subtraction).
For ensemble samples with $m_t = 175\gevcc$ events, the average
corrected masses for the four choices range from 174.5 to
$176.4\gevcc$, for a maximum difference of $1.9\gevcc$.

Some of the other systematic error contributions evaluated for the LB
and NN analyses (see \tabref{tb:systerr})
cannot be determined in the same way for
the PL analysis.  The noise and multiple interaction
error is determined from the shift in the
mean fitted mass for the variable-mass fits, which are not used in the
PL analysis.  However, the kinematic fitters used give similar
results, so the size of this effect for the PL analysis should be
similar to that from  the LB and NN variable-mass analyses.
The error due to Monte Carlo statistics is assumed to be negligible.
The LB-NN difference can be calculated from the PL ensemble results in
\tabref{tab:table4}.
For the 170--$180\gevcc$ mass range, the mean LB-NN difference
is $0.23\gevcc$.  Finally, the likelihood fit error contribution can be
calculated from the four LB fit values given in \tabref{tab:table5}.  
The RMS of
the four LB corrected mass values is $0.9\gevcc$.  Combining in quadrature
these error contributions with those for the
energy scale ($3.5\gevcc$), signal
generator ($2.2\gevcc$ from the maximum \progname{HERWIG}-\progname{ISAJET}
difference in the
160--$190\gevcc$ mass range), and \progname{VECBOS}
flavors ($1.9\gevcc$) gives a total 
PL systematic error of $4.8\gevcc$.

\subsection{Summary}

Pseudolikelihood analysis of the LB subset of the data gives a top
quark mass of $176.0 \pm 7.9\ \textrm{(stat)} \pm 4.8\ \textrm{(syst)}\gevcc$.
This is based upon a 14-point
quadratic fit (with a mass increment of $5\gevcc$)
to the background-subtracted
$\chisq/2$ plot over the range $\mfit = 140 \text{--} 205
\gevcc$.

\section{Further Kinematic Studies}
\label{kinematics}

This section presents distributions of additional kinematic
quantities derived from the data.  In these plots, the data sample is
compared to a mixture of $\ttbar$ (generated with \progname{herwig}
with $m_t=175\gevcc$ unless otherwise specified) and
background models.  The distributions are shown for the LB subsample
and are normalized according to the results of the LB analysis.
There are $18.5$ signal events and $12.5$ background events
expected in this subsample.  The error bars shown on these plots
are from signal and background sample statistics only, and do not
include the correlated error in the overall normalization.

To test the compatibility of our predictions with the data, we use a
Kolmogorov-Smirnov (K-S) test~\cite{eadie}.  
The resulting probability is indicated
on each plot.  Note that binning the data induces an upwards bias in
the K-S probabilities.  To mitigate this effect, all such
probabilities for distributions of continuous variables are calculated
using histograms consisting of 10,000 bins.

\begin{figure}
\psfig{file=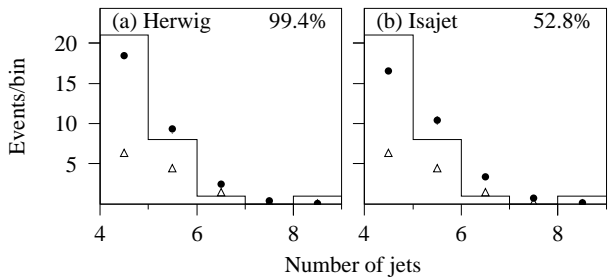,width=\hsize}
\caption{Number of jets in each event with $\et > 15\gev$ and $|\eta| < 2$
for (a) \progname{herwig} ($m_t = 170\gevcc$) and
(b) \progname{isajet} ($m_t = 170\gevcc$).
The histogram is data,
open triangles are expected background, and filled circles are expected
signal plus background.}
\label{fg:njets}
\end{figure}

\begin{figure}
\psfig{file=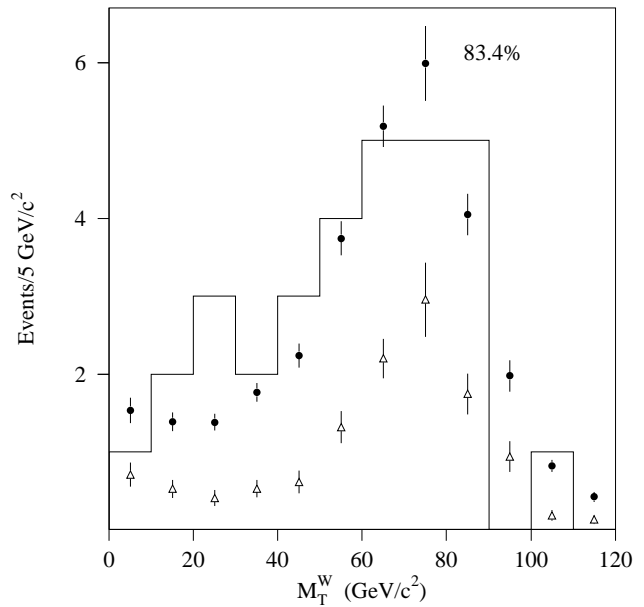,width=\hsize}
\caption{Transverse mass of the lepton and neutrino.
The histogram is data,
open triangles are expected background, and filled circles are expected
signal plus background.}
\label{fg:mtw}
\end{figure}

\begin{figure}
\psfig{file=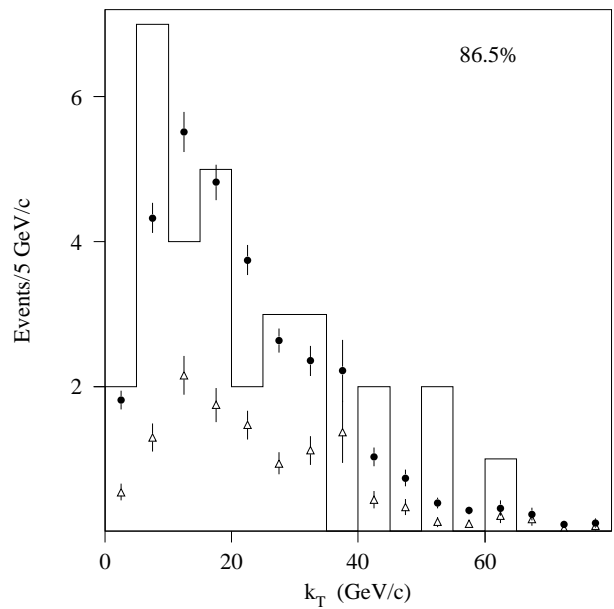,width=\hsize}
\caption{Total transverse momentum $\kt$ of all objects used in the mass fit
(the highest four jets, the lepton, and the $\met$).
This is a vector sum.
The histogram is data,
open triangles are expected background, and filled circles are expected
signal plus background.}
\label{fg:kt}
\end{figure}

\Figref{fg:njets} shows the distribution of the number of jets in
each event in the sample.  For comparison, the prediction of
\progname{isajet} is shown as well as that of \progname{herwig}.  (Note that
since the number of jets is unavoidably a discrete variable, the K-S
probabilities are expected to be biased high.)  \Figref{fg:mtw}
shows the transverse mass of the lepton and neutrino.
The slight rise of the prediction
at low $m_T^W$ is due to the QCD multijet background.
\Figref{fg:kt} shows the total transverse momentum $\kt$ (vector sum)
of all the objects
used in the mass fit.  (The full jet corrections are used; however,
for this plot only, all untagged jets are corrected using the light
quark corrections.)  Note that due to the procedure of using only the
top four jets for the fit, this is not necessarily the actual transverse
momentum of the $\ttbar$ system ($\kt$ tends to be somewhat lower, on average).

\begin{figure}
\psfig{file=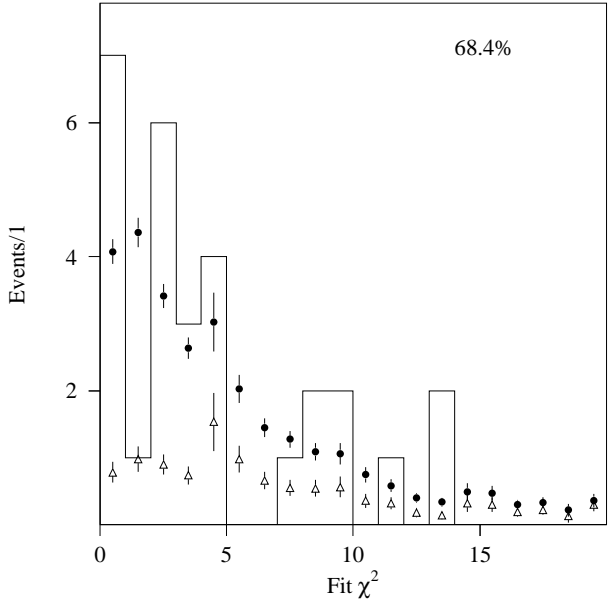,width=\hsize}
\caption{$\chisq$ distributions from the 3C fit.  The histogram is data
(with two overflows),
open triangles are expected background, and filled circles are expected
signal plus background.}
\label{fg:chisqfix}
\end{figure}

The remaining distributions depend on the results of the kinematic
fit.  For these, we plot the result corresponding to the jet
permutation with the smallest $\chisq$.  We also show the
distributions which result if the data and Monte Carlo are refit with
the additional constraint that $m_t = 173.3\gevcc$.
This is now a 3C fit.  Note, however, that when making the
$\chisq$~cut to define the sample, the 2C $\chisq$ is used in all
cases; thus, adding the additional constraint does not change the
sample definition.  The distribution of the 3C fit $\chisq$ is shown in
\figref{fg:chisqfix}.  There are five events with a 3C fit
$\chisq > 10$, compared to $\approx 7$ expected.  They are consistent
with a mixture of background and $\ttbar$ events where the wrong set
of four jets was selected.

\begin{figure}
\psfig{file=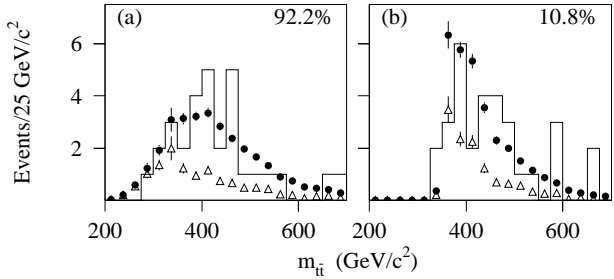,width=\hsize}
\caption{Invariant mass distribution of the $t\tbar$ pair.
The histogram is data,
open triangles are expected background, and filled circles are expected
signal plus background.
(a) 2C fit, (b) 3C fit with $m_t=173.3\gevcc$.}
\label{fg:mtt}
\end{figure}

\begin{figure}
\psfig{file=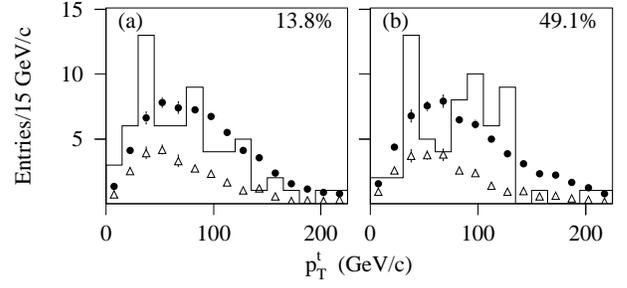,width=\hsize}
\caption{Same as \figref{fg:mtt} for the transverse momenta of
the top quarks (two entries per event).}
\label{fg:ptt}
\end{figure}

\begin{figure}
\psfig{file=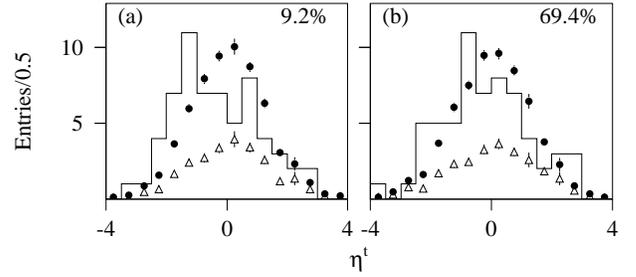,width=\hsize}
\caption{Same as \figref{fg:mtt} for the pseudorapidities
of the top quarks (two entries per event).}
\label{fg:etat}
\end{figure}

\begin{figure}
\psfig{file=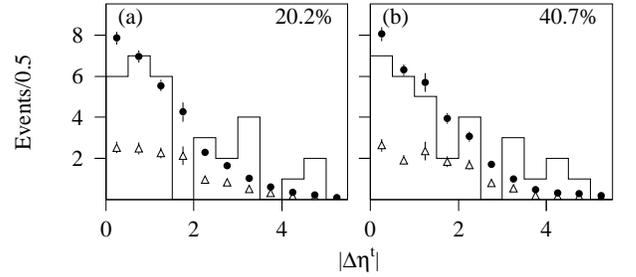,width=\hsize}
\caption{Same as \figref{fg:mtt} for the difference in pseudorapidity
$\eta$ between the two top quarks.}
\label{fg:deta_tt}
\end{figure}

\begin{figure}
\psfig{file=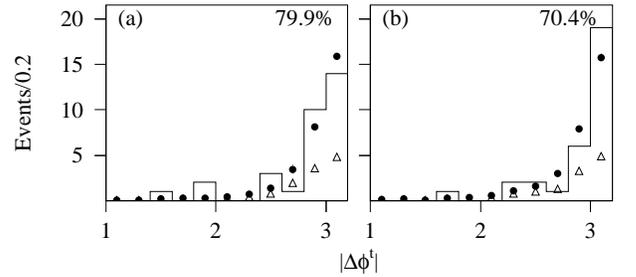,width=\hsize}
\caption{Same as \figref{fg:mtt} for the difference in azimuthal angle
$\phi$ between the two top quarks.}
\label{fg:dphi_tt}
\end{figure}

\Figref{fg:mtt} shows the invariant mass of the $t\tbar$ pair.
\Figref{fg:ptt} shows the transverse momenta of the two top
quarks, and \figref{fg:etat} shows their pseudorapidity.
\Figsref{fg:deta_tt} and~\ref{fg:dphi_tt} show, respectively, the
distance in $\eta$ and $\phi$ between the two top quarks.
The mean of the~13 K-S probabilities we calculate from continuous
distributions is $(53\pm9)\%$, consistent with the hypothesis that our
predictions for $\ttbar$ signal plus background adequately represent
our data.

\section{Conclusions}
\label{conclusion}

In summary, we measure the top quark mass using lepton + jets events
to be
$m_t(lj) = 173.3\pm 5.6\ \textrm{(stat)} \pm 5.5\ \textrm{(syst)}\gevcc$.
We have also measured the top quark mass from
dilepton events \cite{dilmassprl}, yielding $m_t(ll) = 168.4\pm 12.3\
\textrm{(stat)} \pm 3.6\ \textrm{(syst)}\gevcc$.  We combine these two
values, assuming that the systematics for jet energy scale, multiple
interactions, and $\ttbar$ signal generator dependencies are fully
correlated, and that other systematics are uncorrelated.  The result
is
\begin{eqnarray}
m_t &=& 172.1\pm 5.2\ \textrm{(stat)} \pm 4.9\ \textrm{(syst)}\gevcc\\
    &=& 172.1\pm 7.1\gevcc.\nonumber
\end{eqnarray}

\begin{figure}
\psfig{file=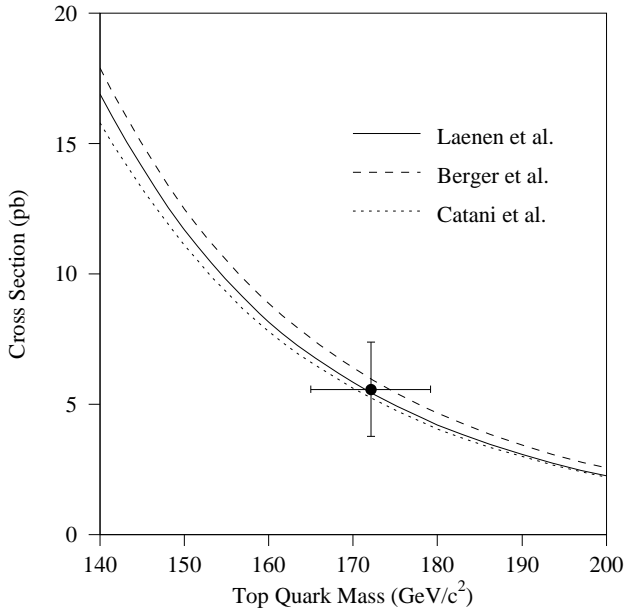,width=\hsize}
\caption{Comparison of the measured top quark mass and production
cross section with theoretical
calculations~\protect\cite{laenen}.}
\label{fg:xsectplot}
\end{figure}

In a separate publication~\cite{xsecprl}, we describe the measurement
of the $p\pbar \rightarrow t\tbar$ production cross section.  The
result for $m_t = 172.1\gevcc$ is
\begin{equation}
\sigma(m_t = 172.1\gevcc) = 5.6\pm 1.8\,\text{pb}.
\end{equation}
Our results are plotted in \figref{fg:xsectplot} and are compared to
several theoretical calculations of the $\ttbar$ production cross
section~\cite{laenen}\mynocite{*berger1,*berger2,*catani}.
The agreement of the standard model expectations with
our measurement is excellent.
We also find agreement between our data and predictions
for distributions of various kinematic variables for $\ttbar$
decays.

An alternate analysis technique using three constraint fits
to fixed top quark masses using the lepton + jets data gives a result of
$m_t(lj) = 176.0\pm 7.9\ \textrm{(stat)}\pm 4.8\ \textrm{(syst)}\gevcc$,
consistent with the above result.

\section*{Acknowledgments}
We thank the staffs at Fermilab and collaborating institutions for their
contributions to this work, and acknowledge support from the 
Department of Energy and National Science Foundation (U.S.A.),  
Commissariat  \` a L'Energie Atomique (France), 
State Committee for Science and Technology and Ministry for Atomic 
   Energy (Russia),
CNPq (Brazil),
Departments of Atomic Energy and Science and Education (India),
Colciencias (Colombia),
CONACyT (Mexico),
Ministry of Education and KOSEF (Korea),
CONICET and UBACyT (Argentina),
and CAPES (Brazil).


\begin{thebibliography}{10}
\bibitem{ev10822-note}
Note that one event which would have otherwise passed the cuts, event
  $(95653,10822)$, was removed from this analysis because it was selected by
  the dilepton mass analysis~\cite{dilmassprl}. That event contained two
  electron candidates. However, one of the electrons just grazed the edge of
  the outer drift chamber, passing through the inner layers but not the outer
  ones. Thus, although there are hits in the inner layers, a complete track was
  not reconstructed. This event also contains a muon near one of its jets,
  consistent with $b$ quark decay. If this event is treated as a $l+\jets$
  candidate, it has a fit $\chisq$ of 0.92 and $\mfit=138.7\gevcc$.
\bibitem{cafix-note}
The current version of the standard corrections (which was not available in
  time for this analysis) is better behaved in this regard.
\bibitem{vbl-note}
Some additional transformations are performed on the $x_i$'s before the fit so
  that the functions can be better approximated by low order polynomials:
  \begin{eqnarray} x_1' &=& \exp{[-(\sqrt{\met/(1\gev)}-A)/B]} \nonumber \\ & &
  \hbox{($A,B = 5,1.5$ for electrons and $4.47,3$ for muons)},\nonumber\\ x_2'
  &=& \exp{(-11 \aplan)}, \nonumber \\ x_3' &=& \ln x_3, \nonumber \\ x_4' &=&
  \sqrt{x_4}. \nonumber \end{eqnarray}
\bibitem{hwparms}
When we increased $\LQCD$ from its nominal value of $0.18\gev$, we also had to
  adjust the \progname{herwig} quark and gluon virtuality cutoffs {\tt vqcut}
  and {\tt vgcut} as follows: \begin{center}\begin{tabular}{ddd} $\LQCD$ & {\tt
  vqcut} & {\tt vgcut} \\ ($\ugev$) & ($\ugevcc$) & ($\ugevcc$) \\ \tableline
  0.18 & 0.48 & 0.10 \\ 0.21 & 0.58 & 0.10 \\ 0.25 & 0.68 & 0.30 \\
  \end{tabular}\end{center}
\end{thebibliography}
\end{document}